\documentclass[fleqn,usenatbib]{mnras}

\usepackage{newtxtext,newtxmath}

\usepackage[T1]{fontenc}
\usepackage{ae,aecompl}

\usepackage{graphicx}	
\usepackage{amsmath}	
\usepackage{amssymb}	
\usepackage{rotating}
\usepackage{pdflscape}
\usepackage{afterpage}

\newcommand{\chandra}{{\it Chandra}}

\newcommand{\suzaku}{{\it Suzaku}}
\newcommand{\xmm}{{\it XMM-Newton}}

\newcommand{\nustar}{\textit{NuSTAR}}

\newcommand{\ms}{$M_{\odot}$}

\title[Searching for UFOs with variability]{Searching for Ultra-fast Outflows in AGN using Variability Spectra}

\author[Z. Igo et al.]{
Z. Igo,$^{1}$\thanks{E-mail: zsofi.igo@durham.ac.uk}
M. L. Parker,$^{1}$\thanks{E-mail: mparker@sciops.esa.int}
G. A. Matzeu,$^{1}$
W. Alston,$^{2}$
N. Alvarez Crespo,$^{1}$\newauthor
D. J. K. Buisson,$^{3}$
F. F\"{u}rst,$^{1}$
A. M. Joyce,$^{1}$
L. Mallick,$^{4}$
N. Schartel,$^{1}$
M. Santos-Lle\'{o}$^{1}$
\\ 
$^{1}$European Space Agency (ESA), European Space Astronomy Centre (ESAC), E-28691\\
$^{2}$Institute of Astronomy, University of Cambridge, Madingley Road, Cambridge, CB3 0HA, UK\\
$^{3}$Department of Physics and Astronomy, University of Southampton, Highfield, Southampton, SO17 1BJ, UK\\
$^{4}$Department of Astronomy and Astrophysics, Pennsylvania State University, 525 Davey Laboratory, University Park, PA 16802, USA\\
}

\date{Accepted XXX. Received YYY; in original form ZZZ}

\pubyear{2020}

\begin{document}
\label{firstpage}
\pagerange{\pageref{firstpage}--\pageref{lastpage}}
\maketitle

\begin{abstract}

We present a qualitative search for ultra-fast outflows (UFOs) in excess variance spectra of radio-quiet active galactic nuclei (AGN). We analyse 42 sources from the \citet{Tombesi10} spectroscopic UFO detection sample, and an additional 22 different sources from the \citet{Kara16} variability sample. A total of 58 sources have sufficient observational data from \xmm\ EPIC-pn and variability for an excess variance spectrum to be calculated. We examine these spectra for peaks corresponding to variable blue-shifted H- and He-like ion absorption lines from UFOs. We find good evidence for such outflows in 28\% of the AGN sample and weak evidence in a further 31\%, meaning that $\sim$ 30--60\% of the AGN sample hosts such UFOs. The mean and median blue-shifted velocity is found to be $\sim$ 0.14c and 0.12c, respectively. Current variability methods allow for a fast, model-independent determination of UFOs, however, further work needs to be undertaken to better characterize the statistical significance of the peaks in these spectra by more rigorous modelling. Detecting good evidence for variable UFO lines in a large number of sources also lays the groundwork for detailed analysis of the variability timescales of the absorbers. This will allow us to probe their densities and hence distances from the central super-massive black hole.

\end{abstract}




\section{Introduction}

The detection of ultra-fast outflows (UFOs) in AGN has been hotly debated over the years. While there are a handful of sources where absorption lines from highly ionized gas are seen consistently every time they are observed \citep[most famously PDS 456:][]{Reeves03, Nardini15, Matzeu17}, there are many more claimed detections where a line feature is only seen in a single observation, and is absent or found at a different energy when the source is re-observed or re-analysed (for example, \citealp{Laha14} were unable to reproduce the UFO detections of six sources from the sample of \citealp{Tombesi10}, although \citealp{Tombesi14} argue that this is due to difference in the data reduction).

In order to cleanly differentiate UFOs from the (colder, slower) warm absorbers (WA), a lower limit on their outflow velocity of $>$10,000~km~s$^{-1}$ ($\sim$ 0.033c) is typically adopted in the literature \citep{Tombesi10}. They are usually highly ionized winds with high column densities of $N_H \sim $ 10$^{23}$ cm$^{-2}$. The high velocities indicate that these winds originate from within a few hundred gravitational radii from the central source and are closely related to the accretion processes of the AGN \citep{Parker17_nature}. 

The importance of a comprehensive study on UFOs is highlighted by their rather ubiquitous nature, appearing in over 40\% of AGN \citep[including radio-loud AGN][]{Tombesi14}, alluding to a geometry comprised of a wide solid angle \citep{Tombesi10, Gofford13}. Furthermore, studies have found that outflows possessing mechanical energies of $>$ 0.5--5\% of $L_\mathrm{bol}$ have enough power to drive out gas and dust from the host galaxy, thus quenching star formation and impacting the growth of the central engine as they remove angular momentum from the accretion disk \citep{Hopkins2010}. Therefore, obtaining a better understanding of the statistical presence and dynamic range of these UFOs is vital in order to better understand the complex feedback processes of AGN. For example, outflows could serve as a natural explanation of the M--$\sigma$ relation, seeing as they may relate to the AGN accretion activity \citep{Ferrarese&Meritt, Kormendy&Ho}. 

Due to these features being predominantly found at high energies (7--10 keV), towards the edge of the detector bandpass for \xmm, \suzaku, and \chandra, there is a great deal of noise in the spectrum. There is also possible background contamination, for example from the Cu K-alpha line found in the \xmm\ EPIC-pn instrumental background. Additionally, the claimed detection significance of these features can depend strongly on the model used for the continuum. For example, some high energy absorption lines that are found to be significant assuming a power-law continuum with optional Gaussian emission lines are not significant when a more complex continuum model is used \citep{Zoghbi15, Lobban2016}.

The net result of this is that the true prevalence of UFOs in AGN is still unknown and some past UFO claims could potentially have been skewed by publication bias \citep{Vaughan08}. One way of helping this is to use a variety of different methods and aim to produce consistent results. For example, there are a small number of AGN where UFOs have been observed in both X-ray and UV spectra \citep[e.g.][]{Danehkar18, Hamann18}. However, this requires high quality UV spectra to be taken in addition to X-rays, which is not always feasible, and UFOs can only be detected in the UV when the gas ionization is relatively low.

An alternative method for detecting the presence of UFOs was developed in \citet{Parker17_irasvariability, Parker18_pds456}, relying on measuring the variability of UFO lines. \citet{Parker17_nature} showed that the strength of the absorption lines in IRAS~13224-3809 is strongly anti-correlated with the source flux \citep[consistent with the gas being more highly ionized at high fluxes][]{Pinto18}, so the lines are stronger when the flux is low and weaker when the flux is high.
This increases the variability amplitude, so the fractional variance is higher in energy bands where UFO lines are present. In practice, this means that UFO lines appear as positive spikes in variability spectra, relating to absorption features from highly ionized atoms. This method of UFO detection has several advantages, which complement conventional spectral fitting: it is fast, less biased by the shape of the continuum and the features appear more pronounced than in traditional energy spectra across the whole bandpass.

The main drawback of this technique is that it cannot be used to rule out the presence of UFOs, if the UFO line strengths are not anti-correlated to the continuum. The effectiveness of the technique therefore depends on how general this correlation is.
We have employed this method in two AGN so far, the narrow-line Seyfert 1 (NLSy1) IRAS~13224-3809 and the low redshift quasar PDS~456, and cleanly detected the UFOs in both cases \citep{Parker17_irasvariability, Parker18_pds456}.

In this work, we use the variability UFO detection method to revisit the samples of \citet{Tombesi10} and \citet{Kara16}, in order to investigate what fraction of the sources found to have UFOs with spectroscopy also show signatures of UFOs in their variability. Our work is the first search for AGN winds with variability to date. Moreover, we aim to motivate further investigation into the different X-ray sources in order to build a more robust way to characterize the preliminary statistical significances of UFOs presented in this paper.

\vspace{-0.5em}
\section{Data Reduction}

Archival data for 64 AGN were obtained in the standard manner from the \xmm\ Science Archive. This sample was based on the spectroscopic study of nearby radio-quiet AGN by \citet{Tombesi10} and the X-ray reverberation study on Seyfert galaxies by \citet{Kara16} as they had already passed certain cuts relating to having longer exposure times, numerous observations and nominal rms variability. We also include the proto-typical UFO source PDS~456 in our sample as a reference/comparison source. After further cuts, where we excluded sources with inadequate signal for a complete spectrum or sources where the noise was larger than any possible line feature (thus rendering it inherently undetectable), a total of 58 sources remained. A full list of Observation Data Files (ODFs), start dates and which sample each source was taken from is available in the online supplementary data. This list is comprised of a total of 385 ODFs. There are a total of 16 NLSy1, 2 quasars (namely PDS 456 and PG 1211+143), 9 obscured Sy 1.9-2.0 galaxies and 31 unobscured Sy 1.0-1.8, all of which are radio-quiet.

We used only the high signal EPIC-pn instrument as it is more sensitive and less background contaminated than MOS data. We reduce the data using \textsc{epproc} from the XMM-Newton Science Analysis Software (SAS, version 18.0.0). We extract only single and double events by specifying PATTERN<=4, and filter for background flares. We extract source photons from 20$^{\prime\prime}$ radius circular regions centered on the AGN, and background photons from an equal or larger sized background region. The background region was chosen far enough away from the source to avoid contamination but close enough (same chip region) to avoid the instrumental background Cu K-alpha line at 8.05~keV. We use a small source extraction region to reduce the level of background contamination. While a small extraction region results in the loss of some source photons, it reduces the number of background photons much more rapidly. For most of our sources reducing the background contribution means that we obtain a higher signal-to-noise  above 7~keV for a smaller extraction region.

For each source, we extract lightcurves in 200 logarithmic-spaced energy bins with 100~s time bins. Additionally, we calculate stacked energy spectra from all available obsIDs for each source by running \textsc{ADDSPEC}, a HEASOFT tool, and binning the EPIC-pn spectra to a signal to noise ratio (SNR) of 6, followed by oversampling the spectral resolution by a factor of 3.  No models were fit to the energy spectra as this paper aims to provide an efficient model-independent study of UFOs and the spectra are for comparison purposes only.

\section{Methods}

For each source, we calculate the fractional excess variance ($F_\mathrm{var}$) spectrum, following Eq. 8--11 from \citet{Vaughan03_variability}. This is a way to quantify the intrinsic variance of the source, taking into account the extra variance from the Poisson noise of X-ray photons. We re-bin the lightcurves in energy and time for each source to find an optimal balance between energy resolution and SNR.  The optimal energy and time binning were generally found to be $\Delta E/E = 0.026$ and 1000~s, respectively, meaning that many rows/columns of the stacked lightcurve were combined into one. Some sources had to be less finely binned in energy to account for slightly shorter exposure time.

Eq. 1 was used on each X-ray lightcurve, for each energy bin, to obtain $F_\mathrm{var}$ over the \xmm\ bandpass (0.3--10~keV). 

\begin{equation}
F_\mathrm{var}= \sqrt{ \frac{S^2-\overline{\sigma_\mathrm{err}^2}}{\bar{x}^2}}.
\end{equation}
The sample variance ($S^2$) and mean ($\bar{x}^2$) are calculated in the conventional manner and $\overline{\sigma_\mathrm{err}^2}$ is defined as, 

\begin{equation}
\overline{\sigma_\mathrm{err}^2}= \frac{1}{N} \sum_{i=1}^N \sigma_{\mathrm{err}, i}^2,
\end{equation}
where N represents the number of data points and $\sigma_{\mathrm{err}, i}^2$ is the Poisson error on the $x_i^\mathrm{th}$ data point. Monte Carlo methods, taken from \citet{Vaughan03_variability}, were used to obtain error estimates on the excess variance.

To quantify the velocity of the potential outflows, the relativistically blue-shifted hydrogen-like \ion{Fe}{xxvi} Ly-$\alpha$ ($E_\mathrm{rest}$ = 6.97 keV), helium-like \ion{Fe}{xxv} ($E_\mathrm{rest}$ = 6.7~keV) and the 6.4~keV fluorescent iron lines were aligned with the peaks of the $F_\mathrm{var}$ spectrum. Errors on velocity were calculated by the width of the energy bin in which the \ion{Fe}{xxvi}/\ion{Fe}{xxv} absorption line was matched to an $F_\mathrm{var}$ peak. However, due to the energy re-binning for the variance spectra, \ion{Fe}{xxv} and \ion{Fe}{xxvi} are usually unresolvable, which often resulted in an added error on the outflow velocity from line identification uncertainty. This error was quantified by the outflow velocity needed to shift the best-fitting line velocity to half the energy range between the \ion{Fe}{xxv} and \ion{Fe}{xxvi} lines.

Lastly, as a preliminary quantification of the significance of the UFO features, the data points closest to the matched absorption lines (the peaks) were removed and a spline was fit through the remaining data points. We then take the residuals to this spline fit, in units of standard deviations. Assuming Gaussian statistics, we calculate a probability of finding $N$ features with that strength by chance, and multiply by the number of resolution elements to calculate a combined false-alarm probability. We subtract this from 1 to give a detection significance ($\sigma_{net}$). This is very approximate, and susceptible to over-fitting from the spline model. We will address this in future work, following the approach detailed in \citet{Parker19_RMS}. We will fit the RMS spectra with physically motivated variance models, and establish the statistical improvement given by adding in UFO models. However, we cannot implement this yet, as we have not yet developed the models needed to fit a large sample of AGN.

\begin{figure}
    \centering
    \includegraphics[width=84mm]{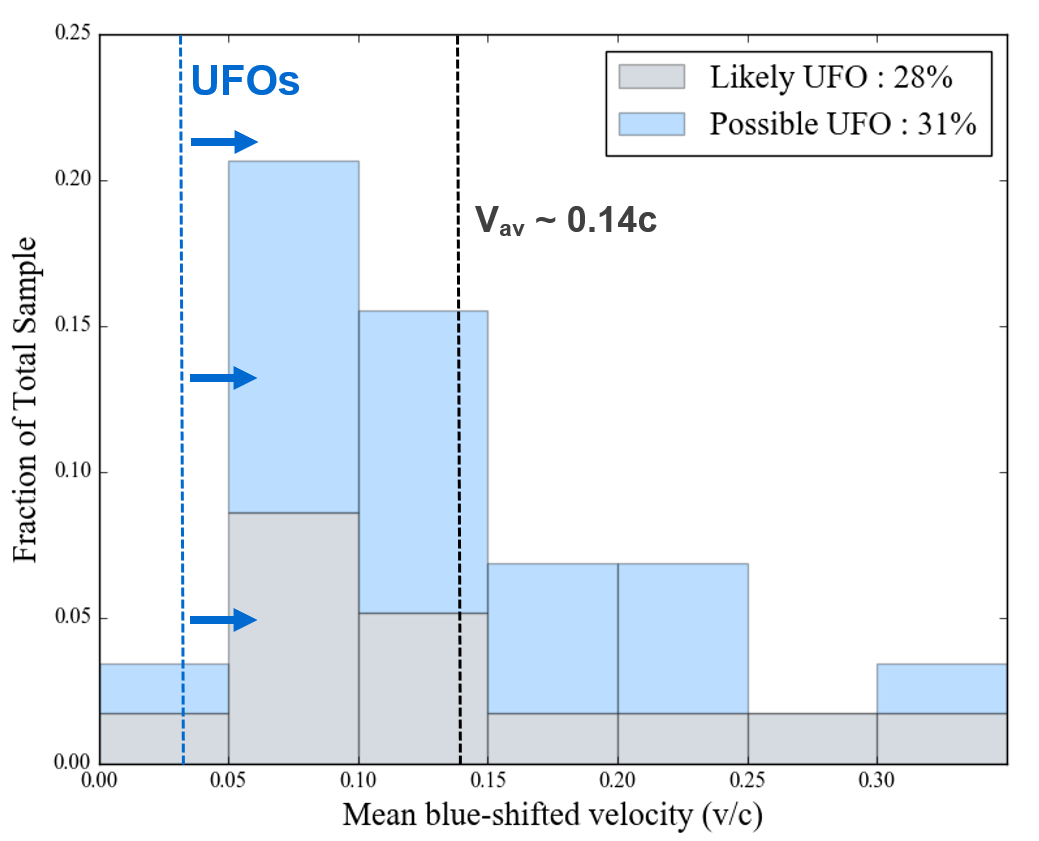}
    \caption{Histogram showing the distribution of mean blue-shifted outflow velocities as a fraction of the total sample. The outflows considered ultra-fast, having v$_\mathrm{out}$ $\geq$ 0.033c are found to the right of the (blue) line with arrows and the mean outflow velocity is shown to be $\sim$ 0.14c.}
\end{figure}

\section{Results}

For each AGN, $F_\mathrm{var}$ as a function of energy in the rest frame of the source was plotted, along with the energy spectra and the preliminary significances of the features.

Table 1 shows a summary of the final results, detailing the properties of the observations, the outflow velocities found by fractional variance methods, the literature values for comparison (if they exist), and $\sigma_{net}$. The table is split into three parts: likely outflows, possible UFOs and no outflows, depending on what was shown by the preliminary statistics, as well as the general alignment of the absorption lines to the peaks in both the variance and energy spectra. Strong evidence from \textit{each} of these three points of interest had to be present to claim a UFO feature. All sources marked with strong evidence for UFOs have at least a $\sigma_{net} > 2$.

Overall, we find that 28\% of the AGN sample strongly support the presence of ultra-fast outflows, whilst a further 31\% may also have UFOs albeit with weaker evidence. In general, this means that $\sim$ 30--60\% of the AGN sample hosts outflows. Figure 1 shows a histogram of the distribution of mean outflow velocities in the AGN sample, ranging from 0.038c to 0.35c. The mean and median blue-shifted velocity is $\sim$ 0.14c and 0.12c, respectively. 

Figure 2 shows the positive correlations seen between the fractional excess variance and $\sigma_{net}$ as well as total exposure and $\sigma_{net}$. This is done as a way of quantifying the reliability of our UFO line detections (discussed in more detail later).

The results for a subset of sources are shown in Figures 3--5, which were chosen specifically to showcase the wide-ranging nature of the UFOs found in this study and highlight interesting points of discussion. Each graph is marked with the corresponding best-fitting outflow velocity, found by comparing the excess variance and count spectra, along with the overall significance of the matched peaks.

\begin{landscape}
\begin{table}
\renewcommand\thetable{1}
\centering
\caption{Outflow velocity results table for AGN sample and literature comparison values.}

\begin{tabular}{lllcllll}

\hline
\hline
Source             & Exposure time (ks) & Total $F_\mathrm{var}$ & Av. Count Rate (s$^{-1}$) & v$_\mathrm{out}$ (c)         & Literature Outflow Velocities (c)                                                                             & $\sigma_{net}$ & References$^{*}$     \\
\hline

\\
                   & \multicolumn{3}{c}{Likely Outflows}                                     &                              &                                                                                                               &                &                      \\
1H 0707-495        & 954.9              & 0.574                  & 3.26                      & $0.165^{+ 0.002}_{- 0.051}$  & 0.11-0.18; 0.13                                                                                               & 6.17           & D12; K19             \\
ESO 323-G77        & 1081.0             & 0.538                  & 0.35                      & $0.085^{+ 0.006}_{- 0.048}$  & $\sim$ 0.007                                                                                                  & 8.14           & T10                  \\
IC 4329A           & 113.4              & 0.042                  & 16.84                     & $0.095^{+ 0.052}_{- 0.020}$  & 0.098 $\pm$ 0.004, $\sim$ 0.1                                                                                 & 3.14           & T11; AM06            \\
IRAS 13224-3809    & 1735.1             & 0.748           & 2.11                          & $0.238^{+ 0.003}_{- 0.049}$  & 0.236 $\pm$ 0.006 ; $0.267^{+0.04}_{-0.03}$c \& 0.225$\pm$0.002c              & 8.16            & P17; J18             \\
IRAS 13349+2438    & 154.4              & 0.191                  & 2.25                      & $0.14^{+ 0.02}_{ - 0.07}$    & 0.13 $\pm$ 0.01                                                                                               & 2.45           & P18                  \\
MCG-6-30-15        & 591.7              & 0.368                  & 16.17                     & $0.08^{+0.02}_{ - 0.05}$     & 0.007 $\pm$ 0.002                                                                                             & 5.82           & G15                  \\
Mrk 205            & 112                & 0.394                  & 7.18                      & $0.14^{+ 0.02}_{- 0.05}$     & $\sim$ 0.1                                                                                                    & 3.67           & T10                  \\
Mrk 766            & 634.1              & 0.543                  & 8.53                      & $0.08^{+ 0.02}_{- 0.05}$     & $\sim$ 0.067; 0.082 $\pm$ 0.006; 0.039 $\pm$ 0.006                                                            & 4.94           & T10; T11; G13        \\
NGC 4051           & 628.7              & 0.821                  & 9.57                      & $0.061^{+ 0.022}_{- 0.033}$  & $\sim$ 0.084; 0.202 $\pm$ 0.006; 0.018 $\pm$ 0.001                                                            & 4.83           & T10; T11; G13        \\
NGC 4395           & 185                & 0.875                  & 0.63                      & $0.13^{+ 0.05}_{- 0.02}$     & $<$ 0.001                                                                                                     & 6.14           & G15                  \\
NGC 7314           & 425                & 0.463                  & 3.28                      & $0.04^{+ 0.02}_{- 0.03}$     &                                                                                                               & 5.31           &                      \\
PDS456             & 664                & 0.353                  & 1.79                      & $0.255^{+ 0.049}_{- 0.020}$  & 0.278 $\pm$ 0.003; 0.273 $\pm$ 0.006; 0.26-0.31                                                               & 2.62           & M17 ; G13 ; R09      \\
TON S180           & 209.6              & 0.243                  & 5.15                      & $0.35^{+ 0.05}_{- 0.02}$     &                                                                                                               & 4.18           &                      \\
\\
                   & \multicolumn{3}{c}{Possible Outflows}                                   &                              &                                                                                                               &                &                      \\
Ark 564            & 571.3              & 0.290                  & 26.42                     & $0.16^{+ 0.02}_{- 0.05}$     &                                                                                                               & 4.09           &                      \\
IRAS 18325-5926    & 208.7              & 0.282                  & 3.48                      & $0.15^{+ 0.02}_{- 0.05}$     & $\sim$ 0.2                                                                                                    & 3.85           & I16                  \\
IZW1 (UGC00545)    & 344.1              & 0.264                  & 4.45                      & $0.21^{+ 0.07}_{ - 0.02}$    & $>$ 0.25                                                                                                      & 3.25           & R19                  \\
MCG-02-14-009      & 61.1               & 0.193                  & 2.17                      & $0.115^{+ 0.020}_{ - 0.056}$ &                                                                                                               & 3.04           &                      \\
Mrk 1040           & 228                & 0.133                  & 6.57                      & $0.06^{+ 0.02}_{- 0.05}$     &                                                                                                               & 1.49           &                      \\
Mrk 335            & 227.4              & 0.593                  & 11.03                     & $0.051^{+ 0.053}_{- 0.020}$  & $0.12^{+ 0.08}_{- 0.04}$                                                                                      & 3.04           & Gallo19              \\
Mrk 509            & 886.5              & 0.184                  & 23.22                     & $0.12^{+ 0.02}_{- 0.06}$     & $\sim$ 0.17; 0.039 $\pm$ 0.03                                                                                 & 3.19           & T10; G13             \\
Mrk 79             & 142.5              & 0.754                  & 4.59                      & $0.093^{+ 0.021}_{- 0.053}$  & $\sim$ 0.091; 0.092 $\pm$ 0.004                                                                               & 2.48           & T10; T11             \\
Mrk 841            & 111.5              & 0.491                  & 6.46                      & $0.051^{+ 0.022}_{- 0.050}$  & $\sim$ 0.034; 0.055 $\pm$ 0.025                                                                               & 2.97           & T10; T11             \\
NGC 3227           & 573.8              & 0.384                  & 8.08                      & $0.038^{+ 0.020}_{- 0.031}$  & 0.005 $\pm$ 0.004                                                                                             & 4.12           & G13                  \\
NGC 4151           & 619.7              & 0.519                  & 6.44                      & $0.105^{+ 0.032}_{- 0.032}$  & $\sim$ 0.105; 0.055 $\pm$ 0.023; 0.106 $\pm$ 0.007                                                             & 8.16           & T10; G13; T11        \\
NGC 4507           & 55.5               & 0.187                  & 0.74                      & $0.1^{+ 0.07}_{- 0.02}$      & $\sim$ 0.18                                                                                                   & 4.89           & T10                  \\
NGC 4593           & 260.2              & 0.428                  & 8.30                      & $0.15^{+0.05}_{- 0.02}$      &                                                                                                               & 3.45           &                      \\
NGC 5506           & 402.6              & 0.291                  & 5.94                      & $0.23^{+ 0.02}_{- 0.05}$     & 0.246 $\pm$ 0.006                                                                                             & 4.11           & G13                  \\
NGC 6860           & 16.0                 & 0.112                  & 6.71                      & $0.19^{+ 0.02}_{- 0.07}$     &                                                                                                               & 3.82           &                      \\
NGC 7213           & 166.9              & 0.309                  & 5.32                      & $0.18^{+ 0.022}_{- 0.065}$   &                                                                                                               & 3.32           &                      \\
PG 1211+143        & 736.4              & 0.253                  & 3.71                      & $0.09^{+ 0.02}_{- 0.07}$     & $\sim$ 0.128; 0.0598 $\pm$ 0.00069, 0.095$\pm$0.005c, none                                                    & 2.27           & T10; KP16, KP03, Z15 \\
PG 1244+026        & 660.7              & 0.274                  & 4.37                      & $0.08^{+ 0.02}_{- 0.05}$     &                                                                                                               & 5.24           &                      \\
PKS 0558-504       & 692.9              & 0.315                  & 10.35                     & $0.30^{+ 0.02}_{- 0.05}$     &                                                                                                               & 2.92           &                      \\
RE J1034+398       & 247.6              & 0.208                  & 3.75                      & $0.22^{+ 0.02}_{- 0.07}$     &                                                                                                               & 2.78           &                      \\
SWIFT J2127.4+5654 & 458.0                & 0.300                  & 3.66                      & $0.142^{+ 0.020}_{- 0.057}$  & 0.231 $\pm$ 0.006                                                                                             & 3.14           & G13                  \\

\hline
\end{tabular}
\end{table}
\end{landscape}
\newpage
\begin{landscape}
\begin{table}
\centering
\renewcommand\thetable{1 (continued)}
\caption{Outflow velocity results table for AGN sample continued. Table shows AGNs with no detected UFO features.}

\begin{tabular}{lllcllll}

\\
\hline
Source             & Exposure time (ks) & Total $F_\mathrm{var}$ & Av. Count Rate (s$^{-1}$) & v$_\mathrm{out}$ (c)         & Literature Outflow Velocities (c)                                                                             & $\sigma_{net}$ & References$^{*}$     \\
\hline
\\
                   & \multicolumn{3}{c}{No Outflows}                                         &                              &                                                                                                               &                &                      \\
1H0419-577         & 207.3              & 0.201                  & 6.77                      & -                            & $\sim$ 0.037; 0.079 $\pm$ 0.007                                                                               & $<$ 1          & T10; T11             \\
Ark 120            & 707.7              & 0.226                  & 15.29                     & -                            & $\sim$ 0.269; 0.29 $\pm$ 0.02                                                                                 & 2.21           & T10; T11             \\
ESO 113-G010       & 83.8               & 0.180                  & 2.71                      & -                            &                                                                                                               & 0.28           &                      \\
ESO 198-G024       & 146.6              & 0.228                  & 4.00                      & -                            &                                                                                                               & 4.11           &                      \\
ESO 362-G18        & 204.5              & 0.219                  & 2.14                      & -                            &                                                                                                               & 4.23           &                      \\
Fairall 9          & 386.2              & 0.369                  & 9.09                      & -                            &                                                                                                               & 1.78           &                      \\
H 0557-385         & 231.6              & 0.167                  & 0.34                      & -                            &                                                                                                               & 2.93           &                      \\
IRAS 17020+4544    & 197.4              & 0.128                  & 4.51                      & -                            &                                                                                                               & 1.19           &                      \\
MCG-5-23-16        & 377.4              & 0.149                  & 11.08                     & -                            & $\sim$ 0.118; 0.116 $\pm$ 0.004                                                                               & 2.50           & T10; T11             \\
MCG+8-11-11        & 37.5               & 0.040                  & 9.52                      & -                            &                                                                                                               & 4.29           &                      \\
Mrk 279            & 148.9              & 0.105                  & 12.35                     & -                            & $\sim -$0.001 (inflow); 0.220 $\pm$ 0.006                                                                     & 1.37           & T10; G13             \\
Mrk 290            & 79.4               & 0.177                  & 3.45                      & -                            & $\sim$ 0.141                                                                                                  & $<$ 1          & T10                  \\
Mrk 590            & 112.6              & 0.157                  & 2.24                      & -                            &                                                                                                               & 1.61           &                      \\
Mrk 704            & 113.0              & 0.342                  & 4.45                      & -                            &                                                                                                               & 2.17           &                      \\
MS22549-3712       & 127.1              & 0.325                  & 1.78                      & -                            &                                                                                                               & 0.93           &                      \\
NGC 1365           & 577.7              & 0.842                  & 1.37                      & -                            & $<$ 0.014                                                                                                     & 8.14           & G13                  \\
NGC 3516           & 445.4              & 0.730                  & 8.44                      & -                            & $\sim$ 0.008; 0.004 $\pm$ 0.002                                                                               & 3.67           & T10; G13             \\
NGC 3783           & 430.8              & 0.499                  & 7.40                      & -                            & $\sim -$0.013 (inflow); $<$ 0.007                                                                             & 4.90           & T10; G13             \\
NGC 4748           & 26.5               & 0.201                  & 8.77                      & -                            &                                                                                                               & 3.90           &                      \\
NGC 526A           & 140.2              & 0.154                  & 3.07                      & -                            &                                                                                                               & 3.58           &                      \\
NGC 5548           & 962.4              & 0.970                  & 4.40                      & -                            &                                                                                                               & 4.73           &                      \\
NGC 7172           & 27.3               & 0.324                  & 2.08                      & -                            &                                                                                                               & 3.68           &                      \\
NGC 7469           & 765.0              & 0.140                  & 16.94                     & -                            &                                                                                                               & 2.52           &                      \\
NGC 7582           & 114.4              & 0.288                  & 0.42                      & -                            & $\sim$ 0.255                                                                                                  & 2.17           & T10      \\
\hline
\hline 

\end{tabular}
\end{table}
$^*$Abbreviated references explained: D12: \citet{Dauser12}, T10: \citet{Tombesi10}, T11: \citet{Tombesi11}, G13: \citet{Gofford13}, G15: \citet{Gofford15}, KP16: \citet{Pounds2016}, M17: \citet{Matzeu17}, I16: \citet{Iwasawa16}, P17: \citet{Parker17_irasvariability}, P18: \citet{ParkerIRAS13349}, R09: \citet{Reeves09}, J18: \citet{Jiang18}, R19: \citet{Reeves19}, Gallo19: \citet{Gallo19}, Z15: \citet{Zoghbi15}, KP03: \citet{Pounds03}, AM06: \citet{Markowitz06}.
\end{landscape}

The rest of the sources are presented and commented on in Appendices A and B.

Almost all of the sources had an additional uncertainty added to the final outflow velocity measurement originating from the \ion{Fe}{xxv}/\ion{Fe}{xxvi} lines being unresolvable. Additionally, the splines used to fit the $F_\mathrm{var}$ spectra were found to over-fit the data, having $\chi^2_{\nu} << 1$, causing overestimated $\sigma_{net}$ values. The implications of this will be further discussed below.

\vspace{+1em}
\section{Discussion}

This study is in overall good agreement with \citet{Tombesi10} and \citet{Gofford13}, who claim that $\sim$ 40\% of the AGN population could host ultra-fast outflows. In comparison to their mean outflow velocity values of $\sim$ 0.1c, we find a marginally higher mean blue-shifted velocity at $\sim$ 0.14c, as shown on Figure 1. This could be a result of the excess variance spectra being effective at accentuating the peaks even at high energies, where the sensitivity of the EPIC-pn is lower. Therefore, absorption features in the count spectra past 8--9~keV, which are usually used to detect UFO lines, may be lost in the low SNR data. It could also be due to \citet{Gofford13} using data obtained from \suzaku, resulting in different AGN samples being analysed to the one presented here (for example one which included 6 radio-loud BLRGs and 5 QSOs).

\begin{figure}
    \includegraphics[width=85mm]{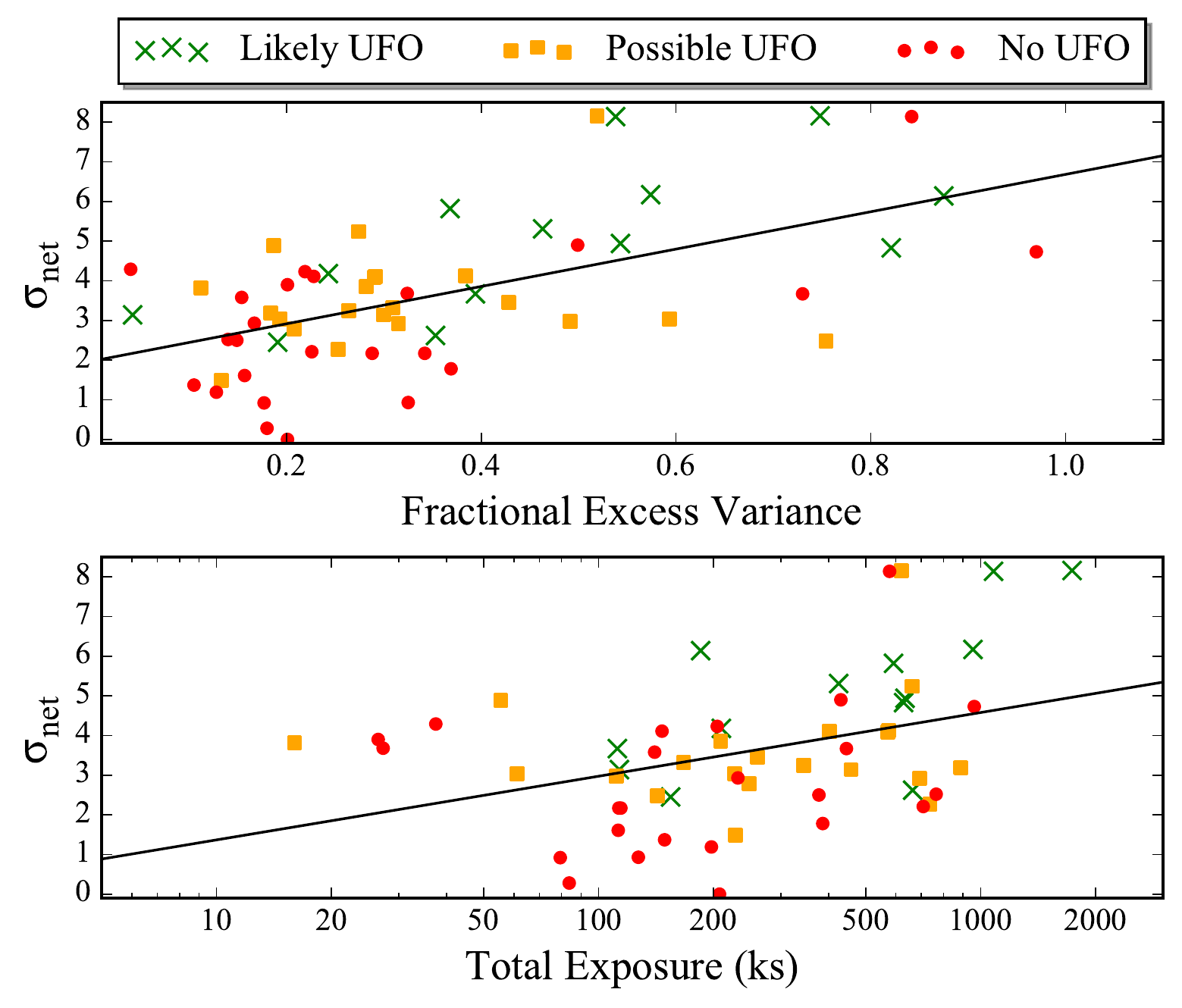}
    \caption{Graph showing how the detection significance of the different classes of UFOs (likely: green crosses; possible: orange squares; no UFO: red dots) compare to total exposure time and fractional excess variance for all sample AGN. A best-fitting trend through all data sets is plotted as the black line.}
\end{figure}

\begin{figure*}     
    \centering
    \includegraphics[width=85mm]{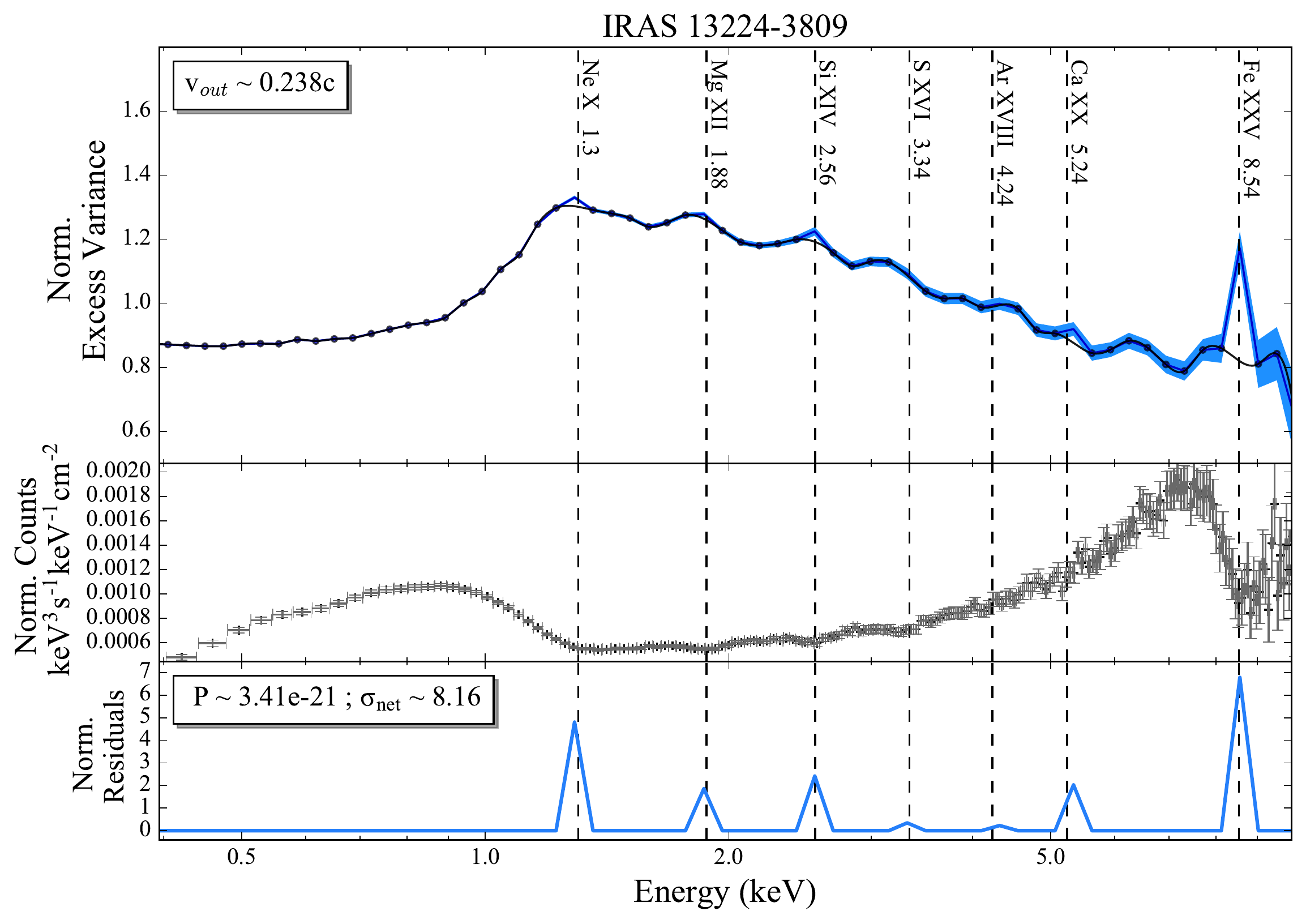}
    \includegraphics[width=85mm]{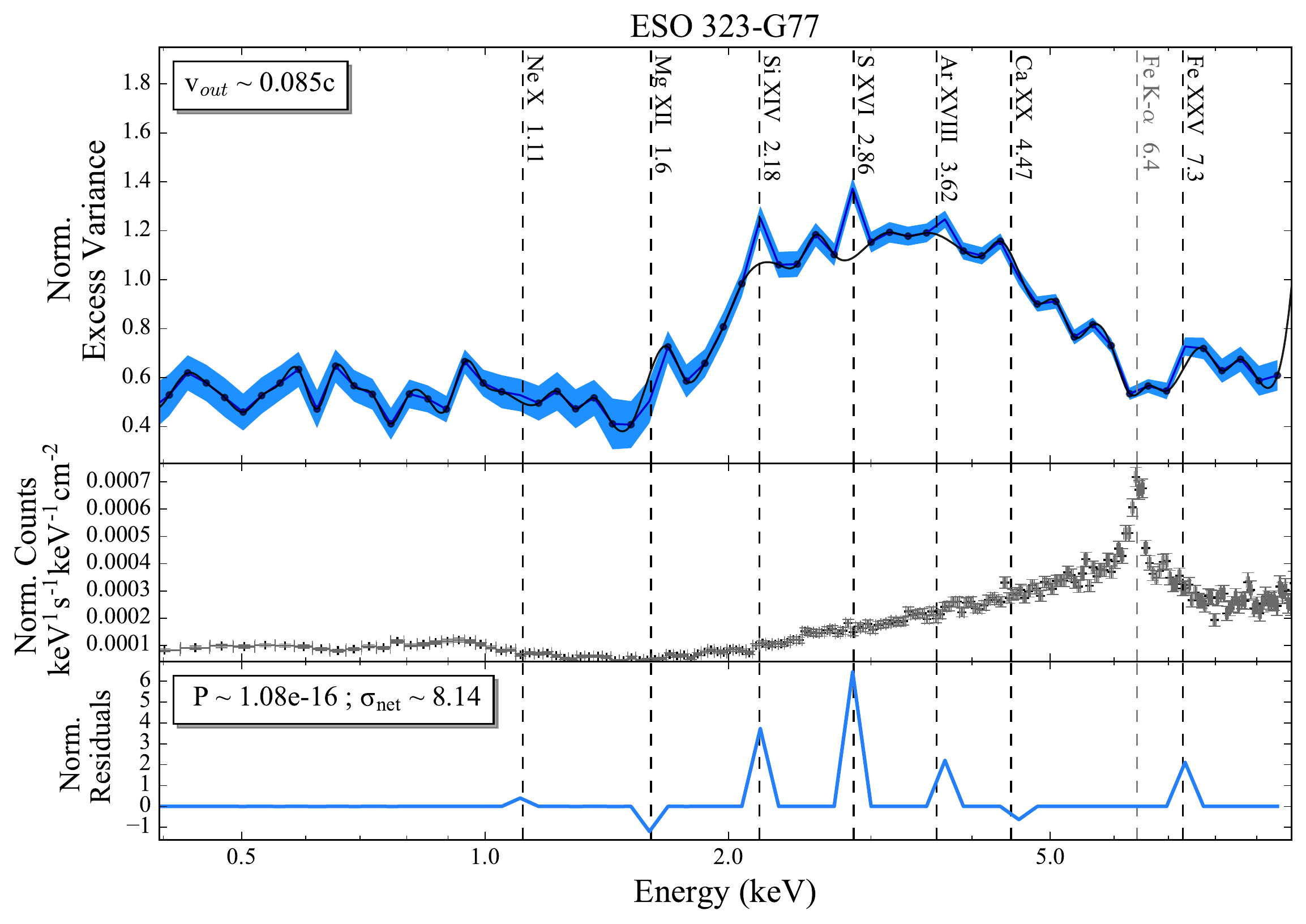}
    \includegraphics[width=85mm]{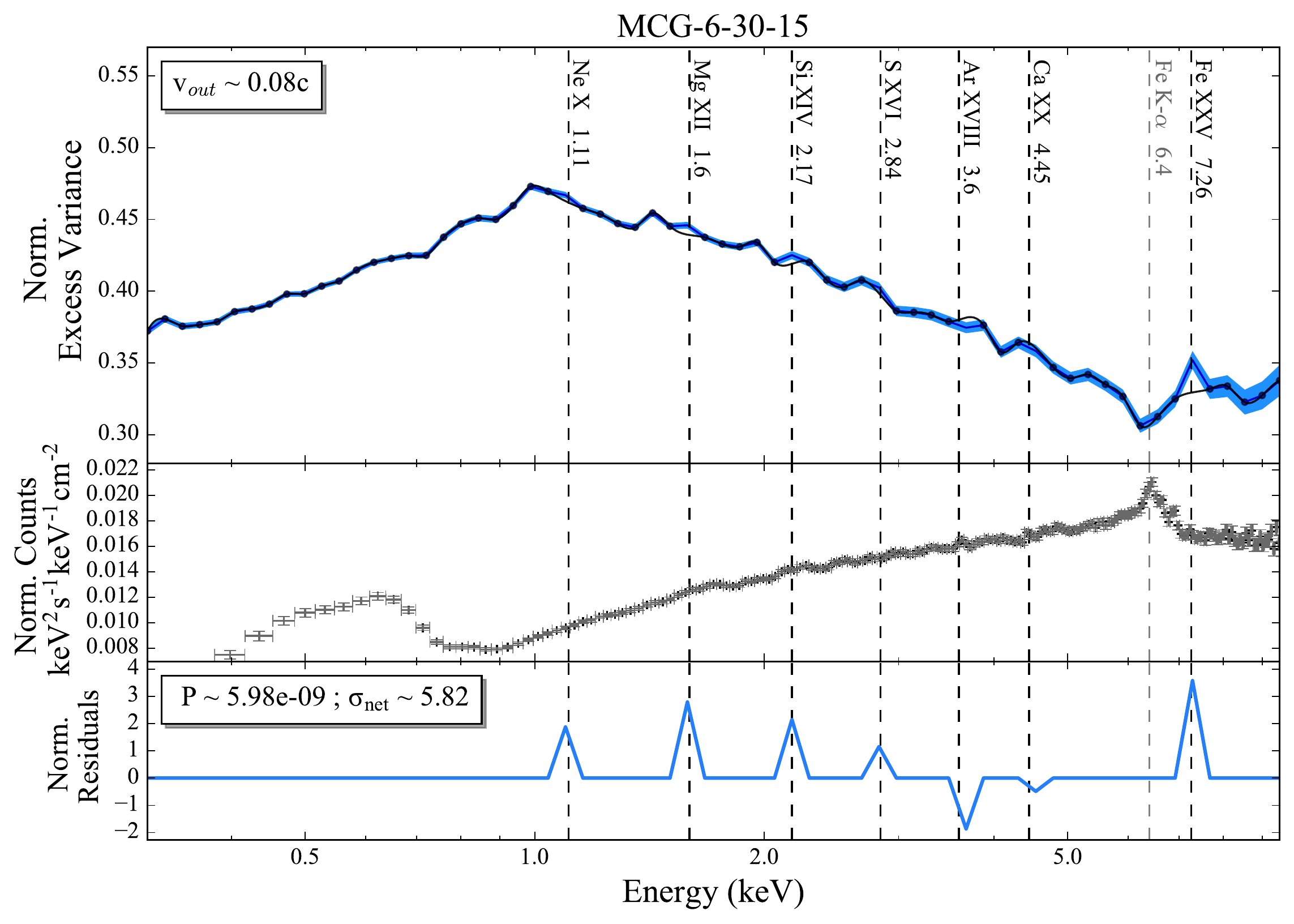}
    \includegraphics[width=85mm]{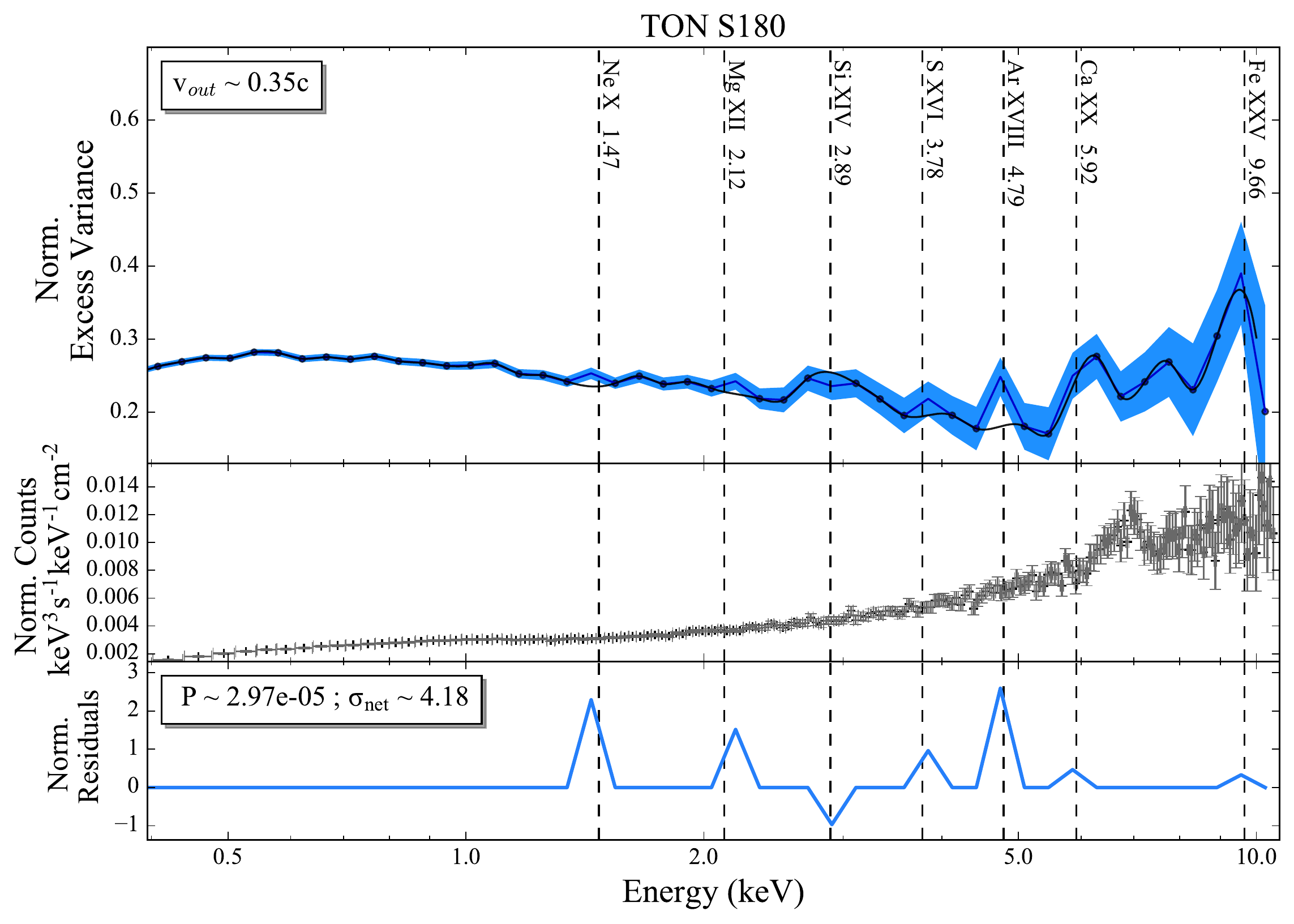}
    \caption{IRAS 13224-3809 (top left), ESO 323-G77 (top right), MCG-6-30-15 (bottom left) TON S180 (bottom right) are examples of ultra-fast outflows with good supporting evidence in the variability and count spectra. The top panel shows the calculated $F_\mathrm{var}$, the middle panel depicts the energy spectrum and the bottom panel shows the normalized residuals with respect to a solid black spline plotted over the $F_\mathrm{var}$ data.}
\end{figure*}

\begin{figure*}
    \centering
    \includegraphics[width=85mm]{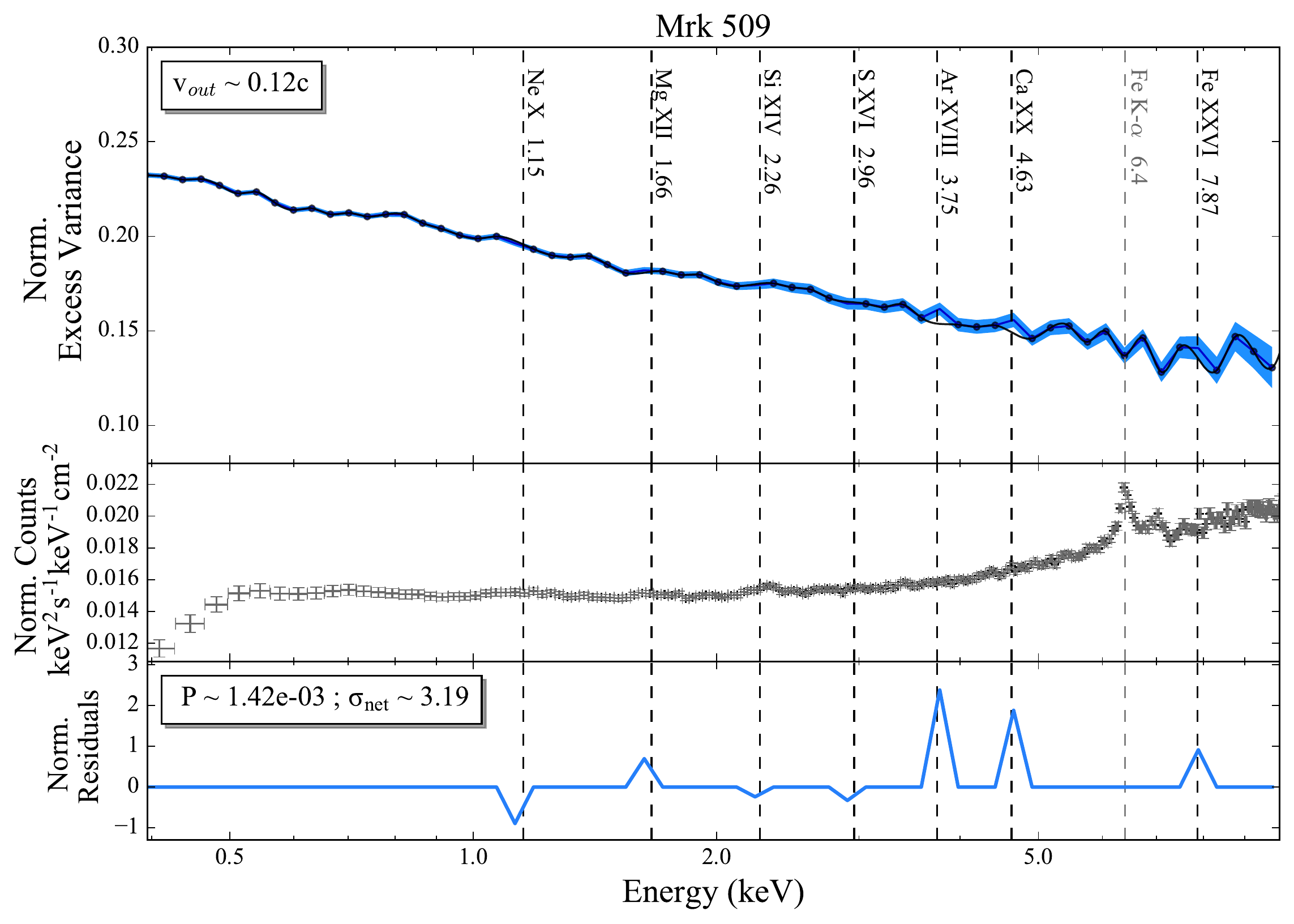}
    \includegraphics[width=85mm]{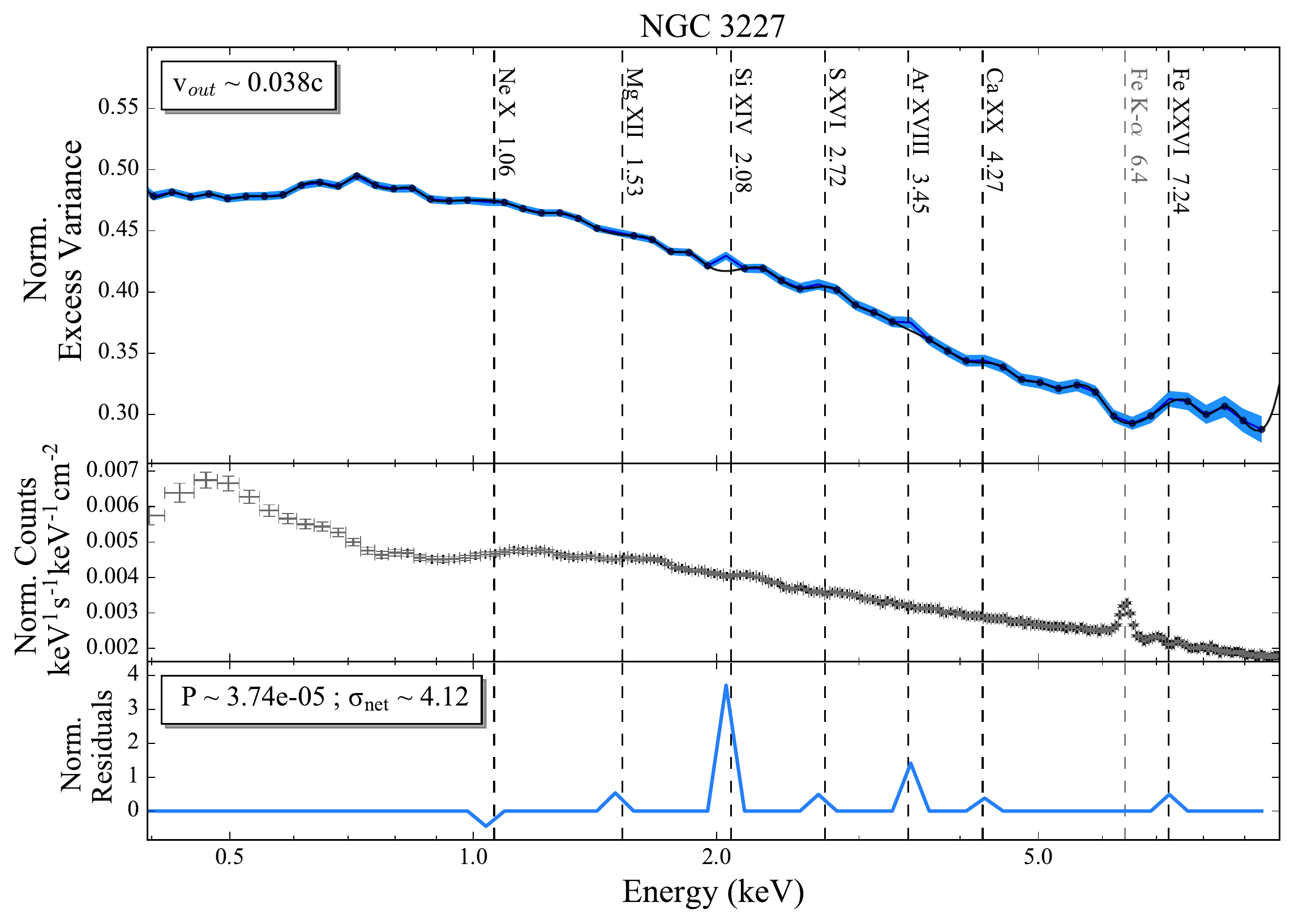}
    \includegraphics[width=85mm]{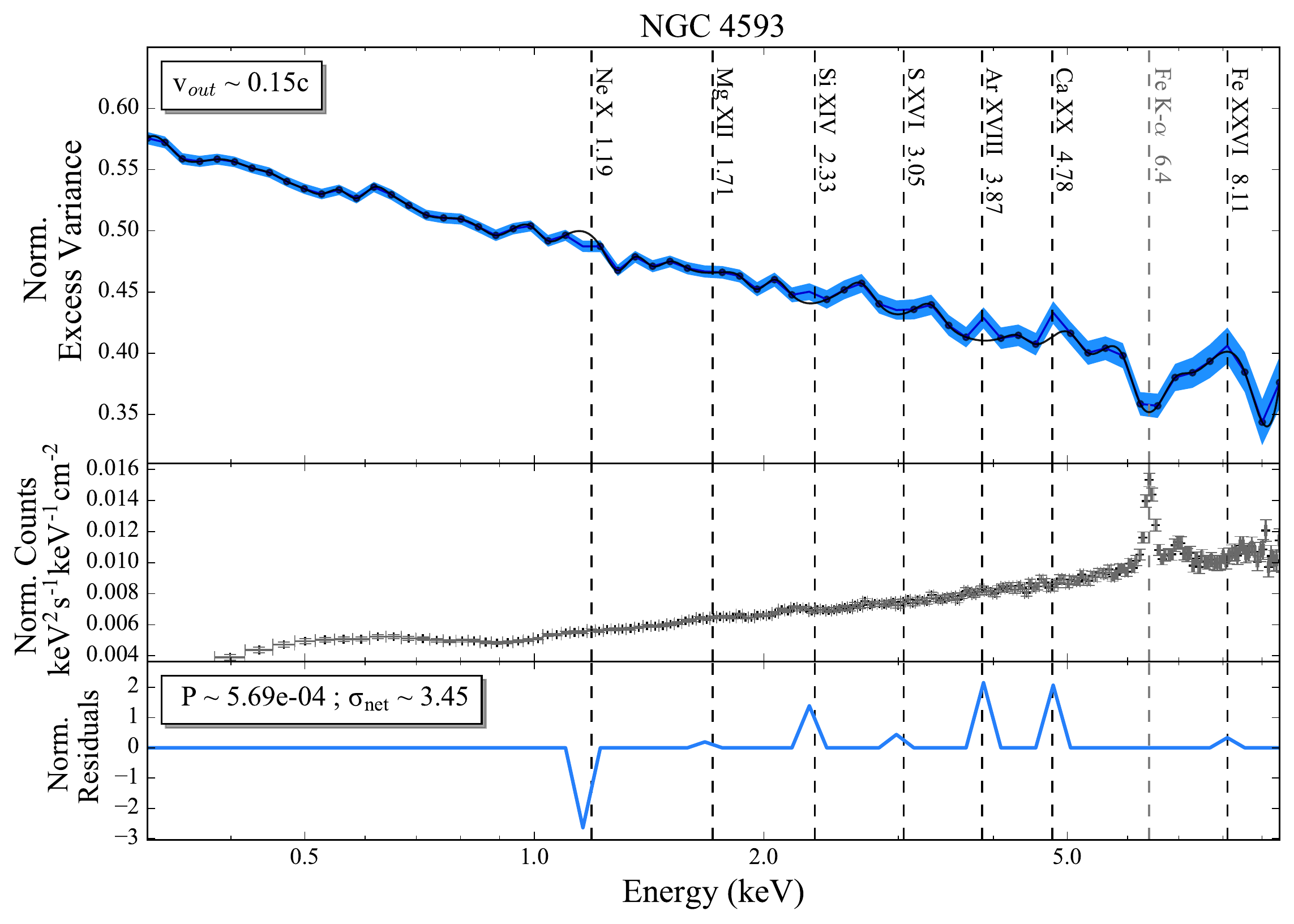}
    \includegraphics[width=85mm]{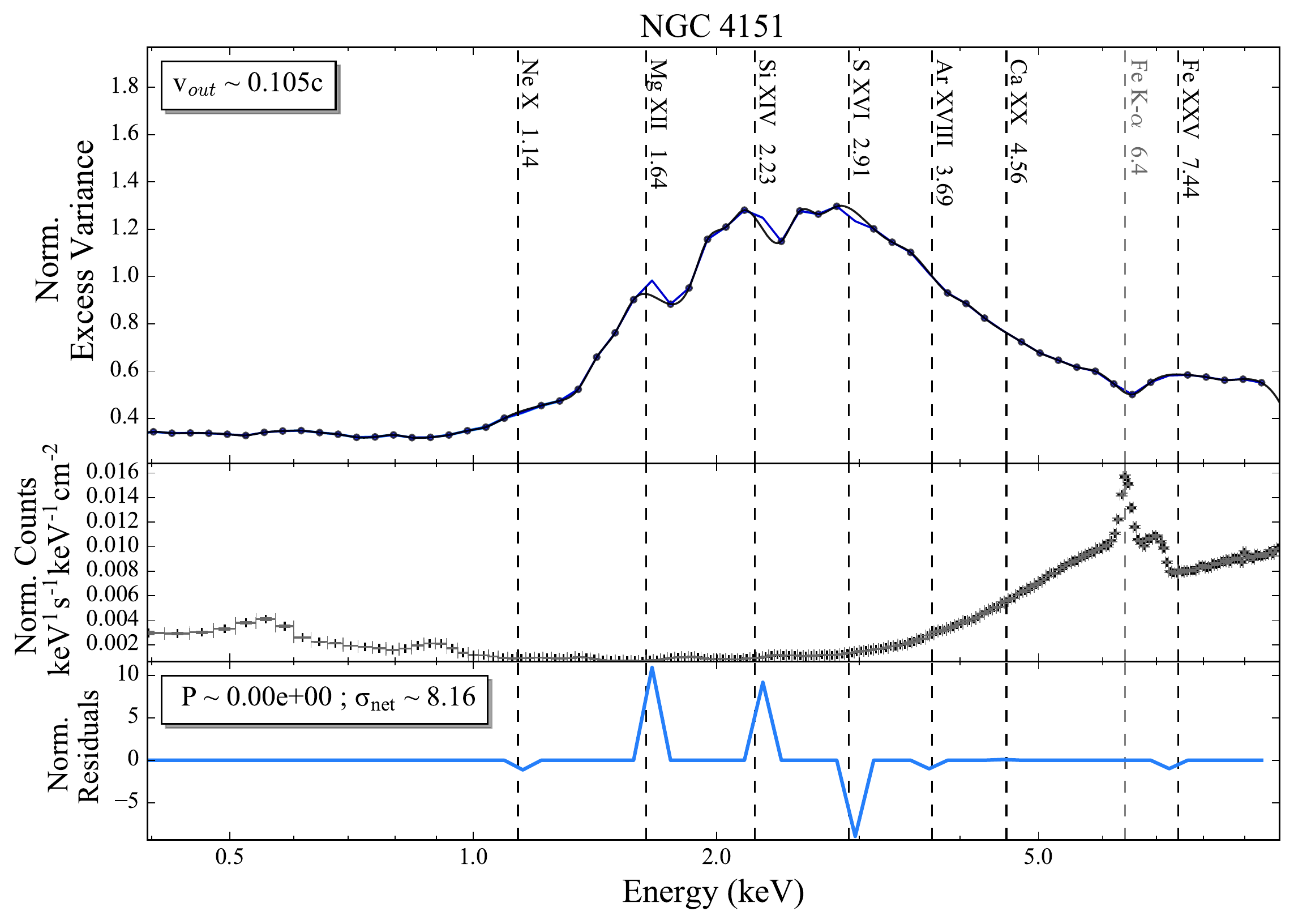}
    \caption{Mrk 509 (top left), NGC3227 (top right), NGC4593 (bottom left) and NGC 4151 (bottom right) are examples of outflows with weak evidence. The three panels in each graph have the same meaning as in Figure 3.}
\end{figure*}

\begin{figure*}
    \centering
    \includegraphics[width=85mm]{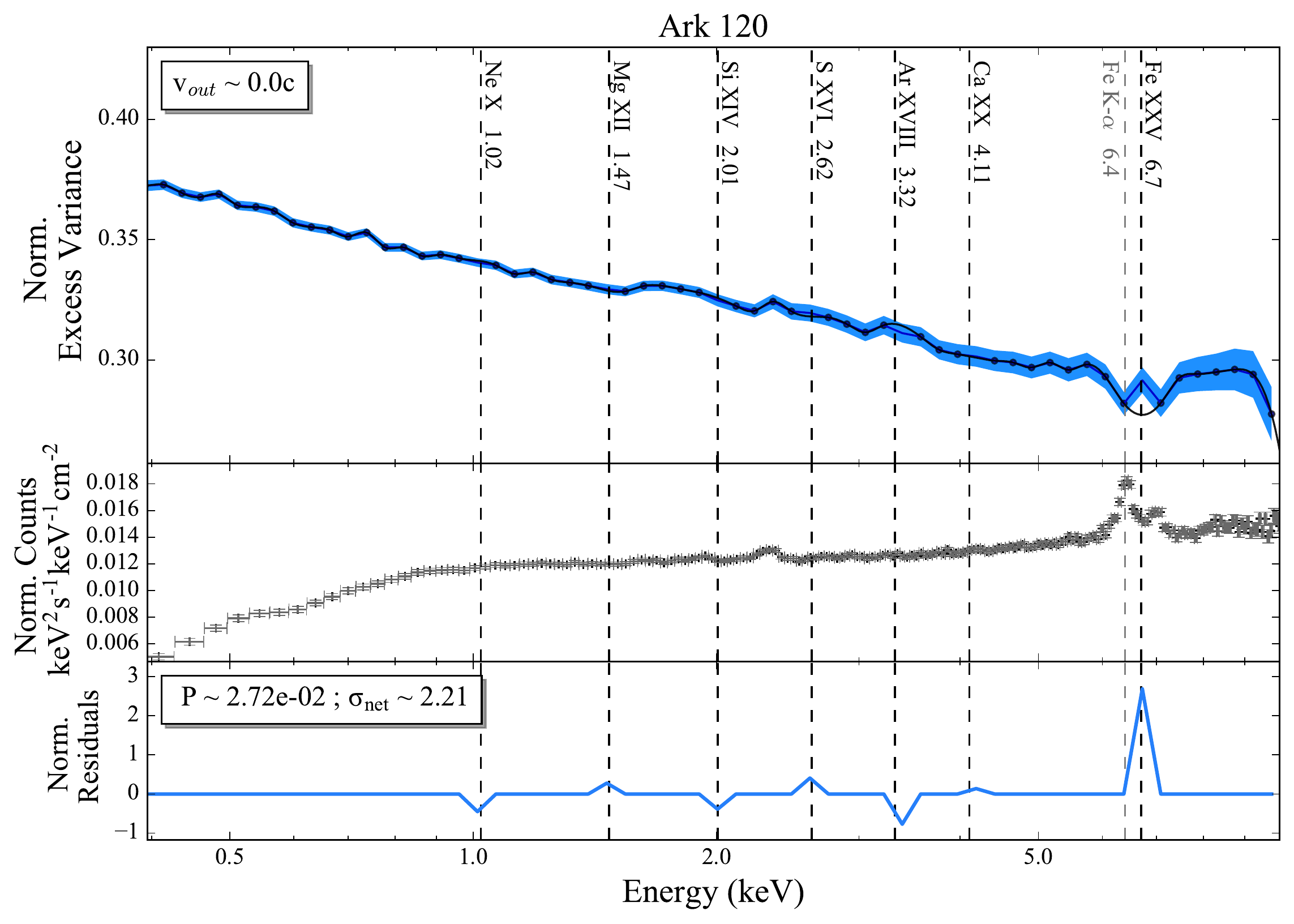}
    \includegraphics[width=85mm]{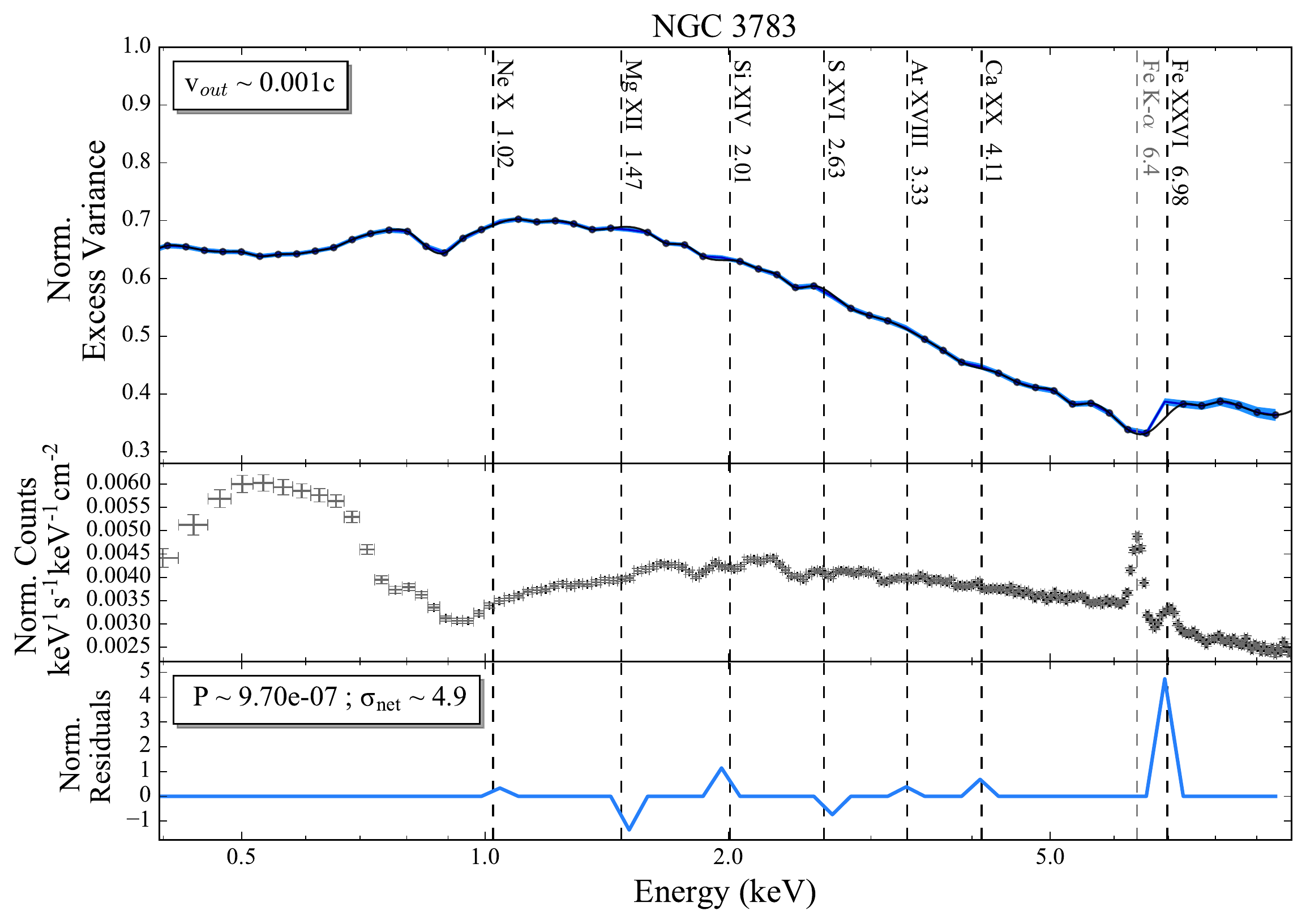}
    \includegraphics[width=85mm]{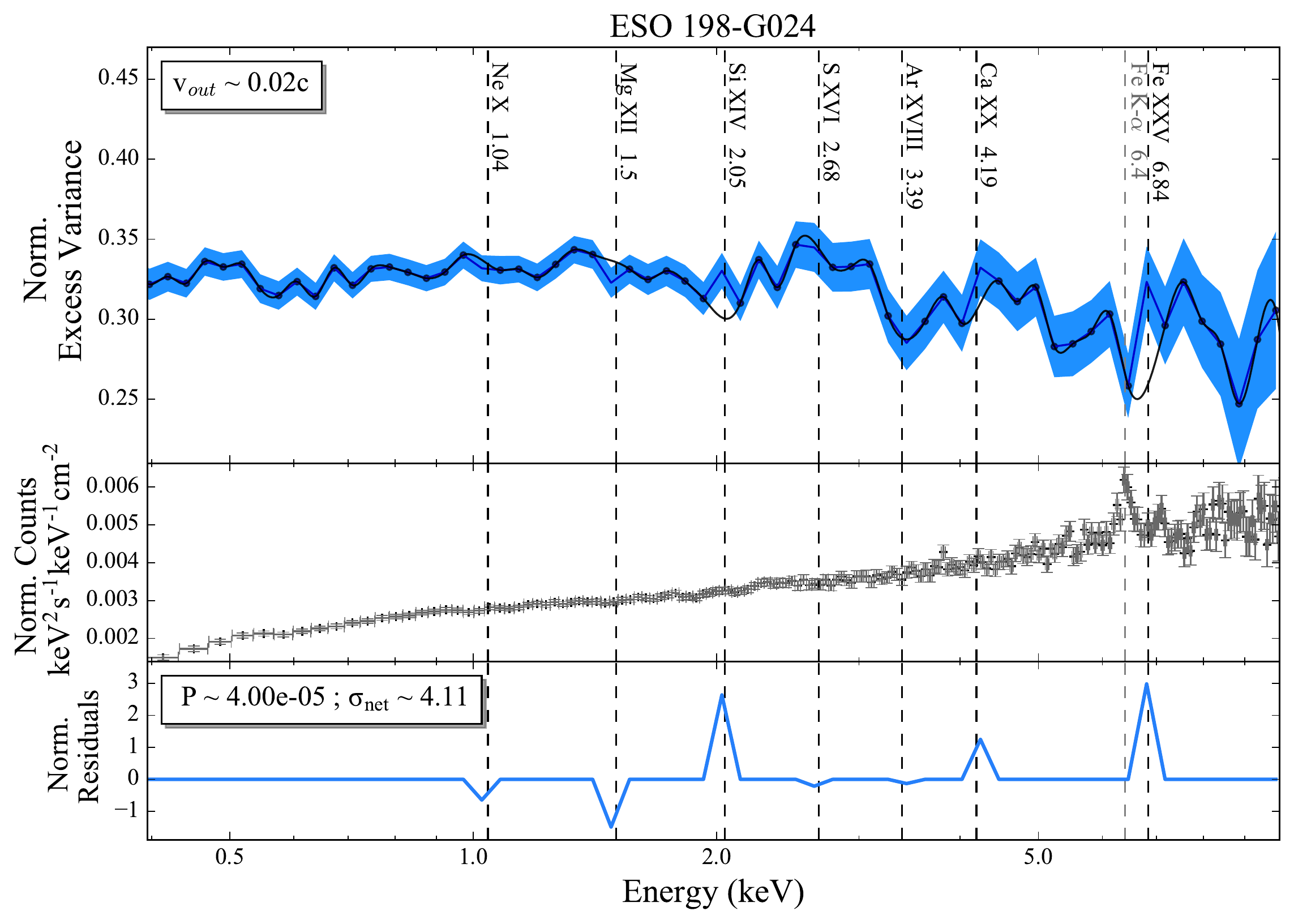}
    \includegraphics[width=85mm]{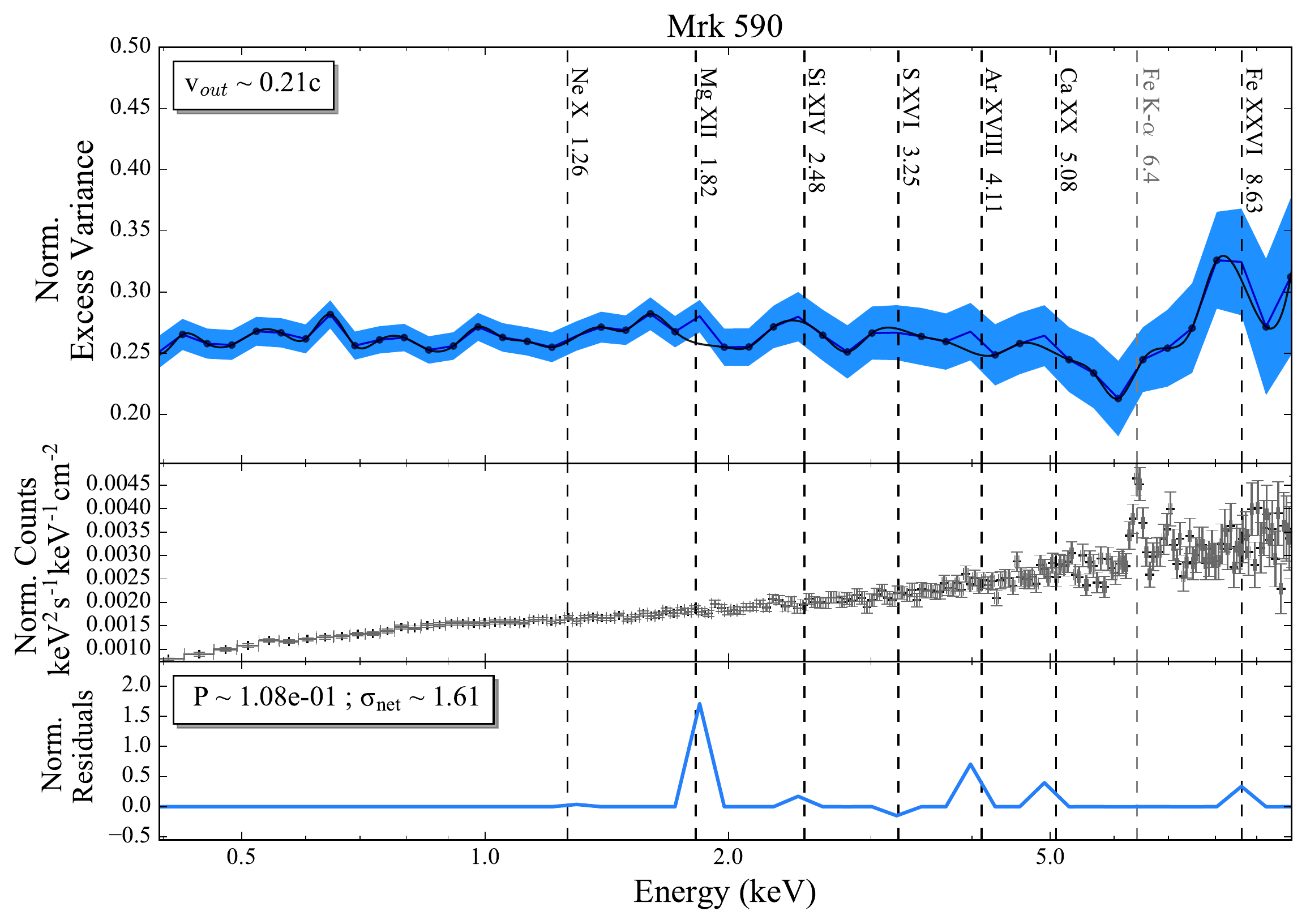}
    \caption{No evidence points towards the existence of outflows in this cases of Ark 120, NGC 3783, ESO 198-G024 and Mrk590. The three panels in each graph have the same meaning as in Figure 3.}
\end{figure*}

However, another plausible reason is the overestimated significances due to the spline fitting with $\chi^2_{\nu} << 1$, causing more sources to be classified as ``possible'' UFO sources, when in fact the matched peaks were not genuine features. As mentioned previously, more robust modeling methods, like the one presented in \citet{Parker19_RMS}, will be essential to quantifying the significance and verity of the UFO lines in future work to try to overcome random noise biases. 

Furthermore, it is important to note that our AGN sample consists only of bright targets at redshifts $\leq$ 0.2 as the data quality and source signal from \xmm\ is otherwise not high enough. Therefore, reporting on the total percentage of AGN which likely host UFOs is biased by our ability to actually detect all AGN. Even nearby AGN hosting outflows not intersecting our line of sight would pass undetected. It is out of the scope of this study to perform corrections based on data quantity and biases in our sample selection, but will be an essential metric for future studies aiming to quantify the global percentage of the whole AGN population to host UFOs. 

Additionally, \citet{Vaughan08} discuss the problem of false detections of narrow lines, filtered by publication bias which selects for the strongest outliers. This naturally leads to a strong correlation between the strength of the claimed feature and the size of the error on the feature strength.
This is why our study, along with the ones of \citet{Tombesi10} and \citet{Gofford13} are important, as they undertake ``uniform and systematic analysis'', without publication bias, and report a ratio of detections to non-detections, thus motivating deeper analysis into the sample sources. 
Figure 2 shows the derived outflow significance against the total excess variance and the total exposure (both proxies for the $F_\mathrm{var}$ spectrum quality). In both cases, there is a weak correlation between spectral quality proxy and the likely UFOs are predominantly found at higher variances and longer exposures. In the case where false detections dominate, as discussed by \citet{Vaughan08}, we would expect the significance to be independent of the spectral quality, so this indicates that we are seeing genuine evidence of outflows. The large spread in the points is likely due to the flaws in our simple significance test, as discussed above. We will revisit this in future work with more sophisticated tests.

Interestingly, as shown in Table 1, we do not find outflows in 8/19 of the \citet{Tombesi10} sources (namely 1H 0419-577, Ark 120, MCG-5-23-16, Mrk 279, Mrk 290, NGC 3516, NGC 3783 and NGC 7582), but we do find three UFOs supported by good evidence, which are classed as non-detections therein (namely, MCG-6-30-15, NGC 7314 and TON S180). Overall, for the common UFO detections ($\sim$~60\%), our outflow velocity results agree well within their uncertainties with the mean outflow velocity from \citet{Tombesi10}. This highlights the power of excess variance methods as they can produce comparable results to traditional blind Gaussian spectral line fitting using Monte Carlo methods, whilst being less dependent on the continuum and faster, albeit having less rigorous statistical significances at present time. Furthermore, the reproducibility of the results for $\sim$ 60\% of the sources further improves their respective veracity, especially as \citet{Tombesi10} found no correlation of outflow velocity to host galaxy redshift, meaning that the observed features do not merely originate from local intervening material. One may argue that $\sim$ 60\% agreement is not a particularly great reproducibility. However, upon examining each of the affected sources more closely, it is clear that the UFOs reported in \citet{Tombesi10}, but not reported here, are from sources in which the UFO feature was detected in only one individually analysed obsID and was virtually non-existent in all others. These include 1H 0419-577, Ark 120, MCG 5-23-16, Mrk 279, Mrk 290 and NGC 7582 (refer to Figure C.5 in the Appendix of \citet{Tombesi10}). Therefore, it makes sense that when combining all available observations on each source that these effects will be dominated by a case presenting no evidence for an outflow. This is further discussed below and in the individual explanations of the sources in Appendices A and B. 

A selection of four sources, which best describe the range of features seen in AGN likely to host UFOs are presented in Figure 3. Starting with IRAS 13224-3809, one can clearly see that the $F_\mathrm{var}$ peaks are much more prominent than the absorption features in the count spectrum, which has even been multiplied through by several factors of energy. This is a clear advantage of the method. The blue-shifted UFO feature at $0.238^{+ 0.003}_{- 0.049}$c shows to be an excellent match not only for the He-like \ion{Fe}{xxv} line emission at 8.54~keV, but also for the lower energy features, especially the H-like \ion{Ne}{x}, \ion{Mg}{ii}, \ion{Si}{xiv} and \ion{Ca}{xx}. The veracity of the lines is also underlined by the alignment of the peaks to the small absorption features in the count spectrum. Moreover, this UFO velocity agrees within its uncertainties with \citet{Parker17_nature} who find a best fit value at $0.236\pm0.006$c from spectral fitting. Additionally, \citet{Jiang18} find two UFOs at $0.267^{+0.04}_{-0.03}$c and $0.225\pm0.002$c, using stacked \xmm\ and \nustar\ spectra. We also note that the \ion{Ca}{xx} Ly-$\beta$ line aligns to the peak around 6.5~keV, however, it is not marked as it is difficult to detect and has a weak abundance. The prominent peaks correspond to even-even elements, as these have larger nucleonic stability arising from spin-coupling effects, among other phenomena. Moreover, IRAS 13224-3809 is an extreme NLSy1 galaxy, accreting near the Eddington limit and possessing high X-ray variability \citep{Parker17_irasvariability}, much like Mrk 766, 1H 0707-495, IRAS 13349+2438 and NGC 4051, to name a few. Around 54\% (7/13) of sources with good evidence for an outflow in this study are classed as NLSy1, as opposed to only around 13\% (3/24) of sources with no evidence. Although, this statistic may be due to detection bias, as NLSy1s are more X-ray variable in general. On the other hand, a possible correlation between the type of galaxy, along with the accretion power of the AGN, and the likeliness of hosting a UFO could be present. This relates back to the aforementioned feedback processes and the interplay between the X-ray emission from the inner accretion disk and the ionizing winds at a much larger scales \citep{Parker17_nature}.

ESO 323-G77 is another example of a strong UFO candidate, found to host a $0.085^{+ 0.006}_{- 0.048}$c velocity wind. Our result disagrees with the average velocity of 0.007c found by \citet{Tombesi10} as this would match to the small $F_\mathrm{var}$ peak around 6.7~keV but overall shows worse alignment in the lower energy features. It is a highly absorbed source for which the variability seems to be dominated by changes in absorption, rather than source flux. Small $F_\mathrm{var}$ values at low energies are observed due to soft X-ray photons being absorbed by the edges of the intervening clumpy torus and the broad line region, when viewed from an inclination angle of $\sim$ 45$^{\circ}$ \citep{MiniuttiESO323}. Another source of absorption could be from warm absorbers (WAs) with outflow velocities around 1000--4000~km~s$^{-1}$ \citep{MiniuttiESO323}; interestingly, \citet{Tombesi13} find strong correlations between the ionization, column densities and velocities of UFOs compared to WAs. They report that in their sample of 35 Seyfert Type 1 galaxies, $>$34\% host UFOs, of which 67\% also show the presence of WAs and their properties lie at opposite ends of parameter space distributions, which could possibly imply that both of these processes form part of a ``single large-scale stratified outflow observed at different locations from the black hole'' \citep{Tombesi13}.  

MCG-6-30-15 is one of the most-studied X-ray bright Seyfert galaxies, possessing a broad asymmetric and gravitationally redshifted Fe K-$\alpha$ emission line \citep{Tanaka95}. This study finds a relatively fast $0.08^{+0.02}_{ - 0.05}$c velocity outflow, compared to the value of 0.007 $\pm$ 0.002c reported in \citet{Gofford15}, to align well over the \xmm\ bandpass. However, MCG-6-30-15 is a complex source that could possibly possess multiple outflow velocity components as well as spectral features, thus making the reproducibility of results difficult. For example, our study also shows a 4.12~$\sigma_{net}$ agreement of a zero velocity outflow component.

An extreme example of a UFO found at $0.35^{+ 0.05}_{- 0.02}$c is TON S180. Such outflows create significant amounts of controversy due to the relative unreliability and lower SNR of the EPIC-pn at high energies making it difficult to locate absorption features in the count spectrum (see Figure 3). Future investigation of this source (and for example PKS 0558-504), possibly with \nustar\ which has a larger energy range of 3-79~keV albeit lower resolution, is solicited to confirm or deny the existence of a UFO.

Figure 4, featuring Mrk 509, NGC 3227 and NGC 4593 are classified as sources with weak evidence for outflows, mainly due to their low significances (which we recall are already largely overestimated) and lack of clear narrow $F_\mathrm{var}$ peaks for the \ion{Fe}{xxvi} and \ion{Fe}{xxv} lines. NGC 3227 with v$_\mathrm{out} = 0.038^{+ 0.020}_{- 0.031}$c agrees well with the result of \citet{Gofford13} of 0.005 $\pm$ 0.004c, whilst Mrk 509 is comparable to \citet{Tombesi10} and no literature values exist for NGC 4593. Cases like Mrk 509, where the results are comparable but not consistent, highlight the issue of outflows being transient in nature and changing locations over time. For example, \citet{Tombesi10} locate the outflow in 3/5 of the obsIDs of Mrk 509 from 2000--2006, whereas this study combines data from 16 independent observations of the source over 9 years (see supplementary data for a table containing detailed obsID information). \citet{King_2015} also argue that UFOs are more likely a series of expanding shells than a continuous wind and that current X-ray observations of AGN are too sparse to detect impacts of the UFO on the variability/count spectra within days of the UFO launch, so a large fraction of such winds remain undetected. Furthermore, a common trend seen in these sources is that the best alignment is achieved when matching the \ion{Fe}{xxvi} line instead of \ion{Fe}{xxv}. Thus, as a consequence of higher ionization levels, it is reasonable to not see great alignment for lines at lower energies as these would be fully ionized by the source flux, resulting in no absorption lines. Therefore, future improvements of the method of calculating significances should take the above into account. 

Additionally, NGC 4151, a commonly studied source due to its complex spectral and variability features, shows weak but intriguing evidence of a UFO of $0.105^{+ 0.032}_{- 0.032}$c. This is in agreement with the result of \citet{Tombesi10}. It was only classified as a possible detection, regardless of the $>7\sigma_{net}$ significance, as the spline, weighted by the inverse square of the errors, is largely overestimating the peaks and the $F_\mathrm{var}$ spectrum does not show a clear peak at high energies. 

The timescales obtained from such variability studies also show potential in determining the distance of these winds from the central super-massive black hole, which is a crucial piece of information to understand more about their formation and launch mechanisms. This study, however, presents a very diverse set of timescales, ranging from the minimum time-bin size of 1000~s to the maximum time between different observations of several years. Therefore, it makes it difficult to find what frequencies the material is varying at. This will be more thoroughly addressed in future work that will focus on only a small subset of the sources presented in this paper.

As an example, NGC 4151 was found to have a large variation of the $F_\mathrm{var}$ spectra over different observation periods (timescales of years). Causes of this change could be due to cumulative random errors in the lightcurves as opposed to intrinsic variability effects \citep{Vaughan03_variability}. However, another reason could be the transient and variable nature of this source. This is highlighted in \citet{Tombesi10} who only detect the aforementioned outflow in 1/6 obsIDs of NGC 4151. Furthermore, \citet{Gofford15} find an outflow velocity of 0.055$\pm$0.023c for this same source, except from \suzaku\ spectra taken less than a month after the findings of \citet{Tombesi10} (obsID: 0402660201). All of these effects, plus possible multi-component UFOs, make understanding and modelling of this phenomenon extremely tasking, but highlight important correlations between the varying source flux and/or column density of the absorber and the properties of the UFOs \citep{Parker17_nature}.

Subsequently, Figure 5 depicts the spectrum of Ark 120 (top left) to be an example of a non-UFO source, i.e. a false detection where v$_\mathrm{out} = 0$c is fully feasible with both the peak and trough observed in the $F_\mathrm{var}$ and count spectra, respectively. This is in stark disagreement with \citet{Tombesi10} who report a very fast outflow of v$_\mathrm{out}=0.269$c.  Possible reasons for this may stem from Ark 120 being a very unobscured, bare Seyfert 1 type AGN being viewed face-on such that our line of sight does not intersect with any UFO wind \citep[e.g.][Matzeu et al., in prep.]{Giustini19}. Alternatively, the wind could be completely ionized so no absorption features would be visible \citep[e.g.][]{Pinto18}. Both of these factors could induce biases in finding the global fraction of AGN hosting UFOs, especially because  Ark 120, for example, has an estimated black hole mass of around $1.5 \times 10^{8}$ \ms\ and is accreting at $L_\mathrm{bol}/L_\mathrm{Edd}$ = 0.2, which should be sufficient to observe small broad absorption lines \citep{Porquet_2019}. \cite{Reeves16ARK120} actually point out that soft X-ray emission lines are seen in the high SNR \xmm\ RGS data of Ark 120, meaning that substantial amounts of circumnuclear gas is present, but it is out of our line of sight, in line with the hypotheses outlined above.

Additionally, Figure 5 shows NGC 3783 to be a non-UFO candidate because, much like MCG-5-23-16, NGC 7469, NGC 3516 and Ark 120, the veracity of the UFO line is questionable. This is because the alignment coincides either with the neutral iron K-$\beta$ emission line/neutral K-edge or the trough in between the K-$\alpha$ and K-$\beta$ emission lines. ESO 198-G024 is also a common example of ``no detection'' case as the $F_\mathrm{var}$ spectrum is very noisy and the observed peaks are highly dependent on the choice of energy binning. This means that when the number of energy bins combined increased to $\Delta E/E = 0.035$ (instead of the optimal $\Delta E/E = 0.026$ as explained in Section \S 3), the \ion{Fe}{xxv}/\ion{Fe}{xxvi} peaks disappeared or shifted in location (NGC 4748 also presented this problem for example). Lastly, Mrk 590 portrays the often recurring example of low significances, lack of any clear features and large error margins, and so we were unable to attain strong evidence for any UFOs. The concept of publication bias is brought to light here as many non-detections were as a result of less numerous observing campaigns and shorter exposure times of the sources. More specifically, the average (good) exposure time for a UFO-detected source in this study was around 580~ks, compared to 280~ks for non-detection sources, sometimes barely reaching a total of 100~ks of observation time (see supplementary data for more information). Therefore, this study aims to motivate further observation on these sources in order to gain a more comprehensive view of the AGN population as a whole.

As mentioned in the discussion of IRAS 13224-3809, a myriad of avenues remain open for future investigation on the correlations regarding ultra-fast outflows. This study in particular strongly supports the anti-correlation between UFO line and continuum flux, previously noted by \citet{Parker17_irasvariability, Parker18_pds456} in IRAS 13224-3809 and PDS 456. This is because the peaked variability spectra presented in this paper \textit{require} that they be anti-correlated, not simply arbitrarily correlated. For example, if absorption lines would get stronger as flux increased, there would be less variability in the given energy band, resulting in a series of dips, rather than the clear peaks we observe in our sources. This relates back to one of the previously mentioned drawbacks of the method about ruling out the presence of UFOs if the UFO lines are not responding to the continuum. An example could be NGC 1365 (see Figure B7 in Appendix B), which is an extremely variable source with clear Fe absorption lines (although at sub-UFO velocities) \citep{Risaliti2007, Risaliti13}, yet it is classed as presenting no evidence for a UFO in its variability spectrum.

Further correlations include \citet{Pinto18} tentatively reporting that higher luminosity sources produce faster velocity outflows. This investigation has shown that NLSy1 galaxies may have a higher tendency to host outflows, due to them generally accreting at high rates and hosting smaller mass black holes, thus having variability on shorter timescales. \citet{Tombesi14} also find that UFOs are present in $>27$\% of radio-loud sources, not just radio-quiet ones, so it could be that these outflows form a crucial part of unifying AGN. Furthermore, \citet{Klindt19} suggest that accretion disk winds could play a major role in the driving out of dust and gas from red, dust obscured quasars as they evolve into unobscured blue quasars. In particular, seeing as these winds are observed in a large fraction of the AGN population, their wide angle geometry coupled with the high velocities would be more effective at impacting the host galaxy compared to strongly collimated axial jets which are more likely to deposit material in the ISM, far from the central black hole. Additionally, as mentioned in the introduction, studies have found that UFOs can have a significant impact on their host galaxies if their mechanical energy is larger than the threshold around 0.5--5\% of $L_\mathrm{bol}$ \citep{Hopkins2010}. This could be investigated in future work by modelling the count spectra to find a measurement of the respective column densities for different sources to find the global percentage of sources which are above this threshold and see if this correlates with the detection/non-detection of UFOs.

It is important to note that our sample is strongly biased. We have preferentially selected bright, variable AGN, both of which correlate with accretion rate \citep{Koerding07}. Since powerful disk winds are thought to be launched at high accretion rates \citep{King03}, the sources in our sample are presumably more likely to host UFOs than AGN in general. This is unavoidable, given the limitations of current instrumentation and data, but means that our estimate of 30--60\% of sources having UFOs should be regarded as an upper limit on the UFO fraction.

At present time the origins and launch mechanisms, and therefore, precise feedback effects, of UFOs are largely unknown. Several authors have investigated the feasibility of thermally, radiatively and/or magneto-hydrodynamically driven winds, yet it seems that a common consensus is yet to be reached in terms of identifying how different AGN drive their disk winds or if all of these are in fact part of the same driving mechanism at different stages of its evolution \citep[e.g.][]{Fukumura17, Matzeu17, Kraemer18}. Overall, future studies on the statistical prevalence and origins of ultra-fast outflows, along with the properties of galaxies hosting them will be crucial to understanding their impacts on the environment and evolution of AGN through feedback mechanisms.

Future work will focus on modelling the fractional variance spectra to accurately assess the significance of the peaks and to break the degeneracy between matching the \ion{Fe}{xxv}/\ion{Fe}{xxvi} lines. This will hopefully enable a more precise estimate of the global fraction of AGN which host UFOs. Additionally, upcoming X-ray missions such as \textit{Athena}, \textit{XRISM}, and \textit{eROSITA} will prove very useful in ascertaining the presence of ultra-fast outflows in not only the nearby AGN studied in this paper but also ones at higher redshifts, due to the improved spectral resolution, effective area, and sky coverage of the instruments. Until then, more investigations using \nustar\ for looking at higher velocity outflows, for example, will be vital to gaining a more profound understanding of these extreme phenomena.

\vspace{+2em}
\section{Conclusions}

Fractional excess variance methods are used for the first time to search for ultra-fast outflows in the samples of \citet{Tombesi10} and \citet{Kara16}. We find that 28\% of this radio-quiet AGN sample strongly support the presence of UFOs and an additional 31\% may also have UFOs albeit with weaker evidence, totalling to $\sim$ 30--60\% overall. The mean and median velocities are $\sim$ 0.14c and 0.12c, respectively, and UFOs range from 0.038 to 0.35 times the speed of light.

Our results agree with past literature, finding that UFOs are a relatively widely observed phenomena in AGN, thereby asserting their importance in the study of AGN feedback and evolution. We have shown that a large number of sources have variable UFO lines, and that the relation between continuum flux and UFO line strength, previously found for IRAS 13224-3809 and PDS 456 \citep{Parker17_irasvariability, Parker18_pds456}, holds in general. This method also opens up an exciting avenue to explore the frequency of variability and hence determine the scales at which these winds are launched. Overall, searching for UFOs in variability spectra has shown to be a powerful and model-independent technique for a wide range of AGN types and future studies aiming to model these spectra will only provide more insight into the properties of ultra-fast outflows. 

\vspace{+1em}

\section*{Acknowledgements}

ZI is supported by the European Space Agency (ESA) trainee program, with additional funding from the Collingwood College 1972 Club Student Opportunities Fund.
MLP, GM and NAC are supported by ESA Research Fellowships.
This work is based on observations with \xmm, an ESA science mission with instruments and contributions directly funded by ESA Member States and NASA.
This research has made use of the NASA/IPAC Extragalactic Database (NED) which is operated by the Jet Propulsion Laboratory, California Institute of Technology.



\newpage
\bibliographystyle{mnras}
\bibliography{bibliography} 



\clearpage

\appendix
\section{Brief Notes on single AGN sources}

\subsection{Comments on UFOs with good evidence}

\textit{1H 0707-495} has a similar spectral shape to IRAS 13324-3809 as they are both NLSy1 galaxies \citep{Hagino16}. Clear variability peaks which match count spectra absorption features with a high significance are present. The outflow velocity agrees, within its uncertainties, with past studies including \citet{Kosec19, Dauser12}. The pattern of peaks is less clear than in IRAS~13224-3809, which is likely due to the strong change in velocity of the absorption observed by \citet{Dauser12}. 

\textit{IC 4329A} shows much lower amplitudes of variability but strong overall alignment through the whole energy range. It is also in agreement with the results of \citet{Tombesi10} and \citet{Markowitz06}. A relatively low $\sigma_{net}$ compared to other sources with good evidence for UFOs stems from the shorter total exposure time of 113~ks, leading to larger statistical uncertainties. 

\textit{IRAS 13349+2438} shows large variability amplitudes at higher energies and our result agrees with \citet{ParkerIRAS13349} who find v$_\mathrm{out} = 0.13 \pm 0.01$c. This time the relatively low sigma is due to fitting the higher-ionization-state emission line \ion{Fe}{xxvi}, meaning the spline is inaccurately fitting lower energy lines that are not expected to be present.

\textit{Mrk 205} shows good overall alignment and agreement with the \citet{Tombesi10} average value of $\sim$ 0.1c.

\textit{Mrk 766} is a popular source of spectral study \citep{Miller07, Miller06_mrk766, Turner07, Pounds03}, which report findings of a blue-shifted \ion{Fe}{xxvi} Ly-$\alpha$ absorption line around 7.3~keV. This agrees with our findings and the average outflow velocity of \citet{Tombesi10} of $\sim$~0.067c. However, Mrk 766 displays complex variable/multi-component features as later \citet{Tombesi11, Gofford13} find much lower outflow velocities. 

\textit{NGC 4051} is found to have similar outflow velocity as in \citet{Tombesi10}, although it is a difficult source to study as there are multiple absorption components likely to be present. \citet{Pounds04_NGC4051} also finds an absorption feature around 7.1~keV, in agreement with our findings.

\textit{NGC 4395} is an example where we find good evidence for a UFO while \citet{Tombesi10} does not. This is because the strength of the \ion{Fe}{xxv} line is comparatively weaker than the distinct peaks from 3~keV to 5~keV, which align very well with the overall blue-shifted emission lines. The variability here is more likely to be caused by changes in absorption rather than a response to changes in ionization. 

\textit{NGC 7314} is an example of a relatively slow UFO, for which \cite{Tombesi10} report no detection. However, the approximate alignments of the peaks throughout the energy range and the clearly visible high energy iron peak mean it was classed as a UFO with good supporting evidence. 

\textit{PDS456} is a widely studied variable source reporting outflows over a range of velocities: $\sim$ 0.19-0.31c \citep{Reeves03, Reeves09, Matzeu17, Gofford13}, in agreement with the results of this paper. This is likely due to the strong response of the UFO lines to the changes in source flux and the multi-component nature of the outflow.

\subsection{Comments on UFOs with weak evidence}

\textit{Ark 564} is a luminous, soft X-ray spectrum Seyfert 1 galaxy \citep{Edelson02} but our study is the first report of a possible UFO located in this source. The variability spectrum shows to be largely energy independent over the whole \xmm\ band but some clear peaks emerge and align well with a blue-shifted velocity of $0.16^{+ 0.02}_{- 0.05}$c, even though the statistical significance is largely overestimated.  

\textit{IRAS 18325-5926} is studied by \citet{Iwasawa16}, reporting a UFO at $\sim$ 0.2c, also obtained from \xmm\ EPIC-pn, which is similar to the result obtained here of $0.15^{+ 0.02}_{- 0.05}$c. This UFO is likely to be responding to the changes in ionization levels and flares of the source. 

\textit{IZW1}, alternatively \textit{UGC00545}, is a luminous NLSy1 galaxy, accreting at near-Eddington rates with literature outflow values $>$ 0.25c from \citep{Reeves19}, which is in slight disagreement to our results. However, the energy binning had to be increased from optimal values to obtain physical fractional variance results. As a result,the less precisely binned variability spectra may be the reason for the discrepancy. 

\textit{MCG-02-14-009} was classed as a source with weak evidence as the \ion{Fe}{xxvi} feature looks significant but the errors are rather large and the overall significance is quite low, due to only \ion{Ca}{xx} showing good alignment at lower energies. No literature outflow velocities exist for this source. 

\textit{Mrk 1040} could be a promising source in terms of the high alignment of variability peaks throughout the energy spectrum and corresponding absorption in the count spectra. However, the errors remain large and the significance too low to claim an outflow. \cite{Reeves_2017MRK1040} find a wind velocity profile that is consistent with the systematic velocity of the AGN, alluding to an interesting ``stalled'' outflow at large scales.

\textit{Mrk 335} is a NLSy1 galaxy with a detected outflow at $0.12^{+ 0.08}_{- 0.04}$c, in agreement with that found in our study of $0.051^{+ 0.053}_{- 0.020}$c \citep{Gallo19}. However, only weak evidence supports this outflow as only \ion{S}{xvi} and \ion{Ar}{viii} align comparatively well at lower energies.

\textit{Mrk 79} is classed as a possible detection due to lack of strong alignment over the whole energy range but visible \ion{Fe}{xxv} peak, even though it agrees very well with the results of \citet{Tombesi10, Tombesi11}. Lack of alignment at low energy may be a result of a multi-component outflow or one that is variable in location. 

\textit{Mrk 841} is similar to the case above, where clear variability peaks are present but for example \ion{Ar}{xviii} and \ion{Ca}{xx} do not correspond well to the outflow velocity of $0.051^{+ 0.022}_{ - 0.050}$c, which is meanwhile in excellent agreement with \citet{Tombesi10, Tombesi11}. 

\textit{NGC 4507} is a common example of a source where the large neutral iron line absorption feature in the $F_\mathrm{var}$ spectrum makes the subsequent peak have questionable veracity. However, in this case the count spectrum highlights the feasibility of this feature being a UFO line. Nevertheless, the higher binning required to produce the variability spectrum and low total exposure time leading to large errors makes this UFO detection uncertain. 

\textit{NGC 5506} shows high absorption of low energy X-ray photons, causing low amplitudes of variability in the soft X-ray part of the spectrum. It is classed as a possible detection as the peaks are not in themselves highly significant nor prominent with regards to the continuum, even though they show good overall alignment and agree with \citet{Gofford13}. This is another example case of the spline over-fitting the spectrum. 

\textit{NGC 6860} is very interesting as there are clear peaks matching very well to the different H- and He-like ion lines, except these are not supported by corresponding absorption troughs in the count spectrum. This could be an example of a case where $F_\mathrm{var}$ statistics are better at identifying UFO lines than traditional spectral modelling; however, this remains to be verified once the statistical significance of variability peaks can be better parameterised and greater exposure time is obtained. 

\textit{NGC 7213} has a best-matching velocity of $0.18^{+ 0.02}_{- 0.07}$c in terms of statistical significance, but an outflow at $\sim$ 0.09c also has a moderate 3$\sigma_{net}$ and aligns better with the visible absorption feature around 7~keV. Further observations of this source and independent analysis will be needed to claim or reject the presence of a UFO. 

\textit{PG 1211+143} is a very bright NLSy1 galaxy at soft X-ray energies and has varied results about the presence of UFO. For example, \citet{Zoghbi15} argue for no UFO, albeit the \nustar\ observations were acquired at a time when the line at $\sim$7.1~keV was weaker in strength, whilst \citet{Tombesi10} find an outflow at $\sim$ 0.128 and \citet{Pounds03} at 0.095$\pm$0.005c. This study only finds weak evidence for a UFO as no prominent peaks feature in the low-energy part of the spectrum, nor is there alignment to a high enough significance (bearing in mind that the significances are overestimated). Nevertheless, the \ion{Fe}{xxv} is clearly visible as a peak and trough in the variability and count spectra, respectively.

\textit{PG 1244+026} clearly shows a prominent variability peak and absorption line at 7.56~keV. However, it cannot be considered as good enough evidence to ascertain a UFO because one can see the spline over-fits the spectra at lower energies, causing a significant overestimation in the overall significance. This remains to be fixed in future studies. 

\textit{PKS 0558-504} is found to possess an ultra-fast outflow at $0.30^{+ 0.02}_{- 0.05}$c which is bordering on the high energy limit of the \xmm\ EPIC-pn detector. Intriguing peaks scatter the whole variability spectrum, yet this time they appear broader, thus having a lower significance as calculated by the spline fitting method. 

\textit{RE J1034+398} is classed as a possible outflow as the variability spectrum was highly dependent on the energy binning used. For example, when it was increased to $\Delta E/E= 0.044$, the peak at 8.72~keV disappeared completely, whereas decreasing down to the optimal binning created two separate peaks (none of which matched better than the final figure plotted). However, considering that \ion{Fe}{xxvi} is being matched and one does not expect to see lower energy absorption features, there is good alignment with both \ion{Ca}{xx} and \ion{Ar}{xviii}. 

\textit{SWIFT J2127.4+5654} shows good alignment in general and a prominent relativistically blue-shifted iron feature at 7.73~keV, matching with the absorption dip in the count spectrum. However, this value disagrees with the 0.231$\pm$0.006c outflow velocity value of \citet{Gofford13} as when using this result to match the \ion{Fe}{XXV} to the peak around 8.5~keV, the significance overall is more than two times worse. 

\subsection{Comments on UFOs with no evidence}

\textit{1H0419-577} has literature outflows of $\sim$ 0.037c and 0.079 $\pm$ 0.007c found by \citet{Tombesi10, Tombesi11} but $F_\mathrm{var}$ methods fail to align corresponding absorption lines to peaks, which are not just noise in the spectrum. 

\textit{ESO 113-G010} shows no detection as the $\sigma_{net}$ is low, errors are large and peaks are not significant enough over the noise. This target would solicit more observing time to improve the $F_\mathrm{var}$ spectral quality. 

\textit{ESO 362-G18} is classed as a non-detection as only the \ion{Ar}{xviii} feature is a prominent variability feature and the relatively high overall significance is merely due to inaccurate spline fitting. \citet{Humire2018} does report on 100--250~km~s$^{-1}$ outflowing gas within the AGN ionization cone, however, these are much below the 0.033c velocity limit to be considered UFOs.  

\textit{Fairall 9} produces a clear, low-noise variability spectrum; however, there are no significant enough peaks over the whole energy range to claim an outflow, in agreement with \citet{Tombesi10}. Additionally, as with many other non-detections, the \ion{Fe}{xxv} feature aligns in between the iron K-$\alpha$ and K-$\beta$ emission lines so it is not a true feature. 

\textit{H 0557-385} has large errors associated with the variability spectra, not only making it difficult to see any prominent features but also the features are highly dependent on the choice of energy binning, as explained above. \cite{Tombesi10} also do not find any narrow Fe K absorption line. 

\textit{IRAS 17020+4544} does not produce adequate alignment and the \ion{Fe}{xxvi} feature in the variability spectra aligns with an emission feature as opposed to an absorption feature in the count spectrum. 

\textit{MCG-5-23-16} is an interesting case where the normalized residuals show considerable alignment with the variability features. However, when examining by eye, one observes that they are not clear peaks, merely manifestations of over-fitting. Moreover, the outflow measurements of $\sim$0.118c and 0.116$\pm$0.004c, found by \citet{Tombesi10, Tombesi11} do not match with any higher significance compared to the one depicted on Figure B6. 

\textit{MCG+8-11-11} does not show any evidence for a UFO as the variability spectrum is largely governed by noise and sensitivity to energy binning (most likely as a result of shorter exposure time). Neither \citet{Matt06} nor \citet{Tombesi10} report evidence for ultra-fast outflows.

\textit{Mrk 279} is classed as a non-detection due to low statistical significance and lack of agreement between the variability and count spectra features. \citet{Tombesi10} report an inflow of $\sim$ -0.001c and \citet{Gofford13} find an outflow of 0.220 $\pm$ 0.006c. 

\textit{Mrk 290} shows a very noisy $F_\mathrm{var}$ spectrum which is highly dependent on energy binning, hindering the detection of any strong features. \citet{Tombesi10} reports an outflow at $\sim$ 0.141c, which would not agree with the variability spectra presented here. 

\textit{Mrk 704} is consistent with \citet{Tombesi10} claiming no outflow. 

\textit{MS22549-3712} is classed as a non-detection for the usual reasons. Interestingly, alignment with variability troughs is common in this source, possibly due to photo-ionised emission lines in the outflow being less variable than the continuum ``due to smearing out of the variability caused by the wide ranges of light travel time'' \citep{Parker17_irasvariability}. 

\textit{NGC 1365} is a highly absorbed source, with a rapidly spinning black hole, that has been found to show strong relativistic disk reflection \citep{Risaliti13} and extreme variability on timescales of days and weeks, where it changed from being Compton-thick to Compton-thin \citep{Risaliti2007}. However, not many UFO studies have been conducted, other than \citet{Gofford13} who found an upper limit of $<$~0.014c. We find no strong evidence for an outflow due to the lack of significant features and alignment (see text for more detail).

\textit{NGC 3516} shows the variability spectra aligned with the best fit velocity value of \citet{Tombesi10} of $\sim$~0.008c, which clearly does not produce strong evidence for this to be considered a UFO. \citet{Gofford13} also report a very slow outflow of 0.004$\pm$ 0.002c (below the limit to be considered a UFO), meanwhile detailed investigation by \citet{Markowitz08} finds that a ``dip near 6.9~kev, at the rest-frame energy for \ion{Fe}{xxvi}'', suggesting no relativistically blue-shifted outflow.

\textit{NGC 4748} analysis produce a best fit velocity of $\sim$ 0.3c outflow. However, it is very sensitive to energy binning and has a noisy $F_\mathrm{var}$ spectrum, which is why we are skeptical about the reality of a very fast outflow in this source.

\textit{NGC 526A} is again very noisy with high levels of over-fitting. Future investigations focusing on the modelling of the variability spectra may accentuate the \ion{Fe}{xxvi} peak which seems to present a corresponding absorption dip in the count spectrum.

\textit{NGC 5548} is probably one of the few examples where variability methods fail to highlight the absorption features better than the count spectrum. Given the large 1~Ms exposure time, the errors in the lightcurves are small, resulting in the significance values being overestimated.

\textit{NGC 7172} is consistent with a zero velocity outflow, owing to the additional alignment with a corresponding absorption feature in the count spectrum. However, there are no prominent H- nor He-like Fe lines to align to at high energies as the variability plateaus. 

\textit{NGC 7469} shows the recurring feature of aligning to the drop after the iron K-$\beta$ line/neutral K-edge and low significance, thus reducing the veracity of a UFO line being present. This source is interesting though, as the neutral iron line is not present as a dip in the variability spectrum, which may result from inter-observation variability where distant reflection has adequate time to respond and so does not leave a negative feature. 

\textit{NGC 7582} is classed as a non-detection due to similar issues as above. One can also observe that the $\sim$ 0.255c result obtained by \citet{Tombesi10} is not consistent with the variability spectra, as there is no peak at $\sim$ 9~keV. However, future investigation with telescopes of higher energy ranges could provide more accurate data at these high energies where the EPIC-pn detector reliability dwindles.

\newpage
\onecolumn
\section{Variability spectra analysis for remaining sources in AGN sample.}

\begin{figure*}
    \centering
    \includegraphics[width=85mm]{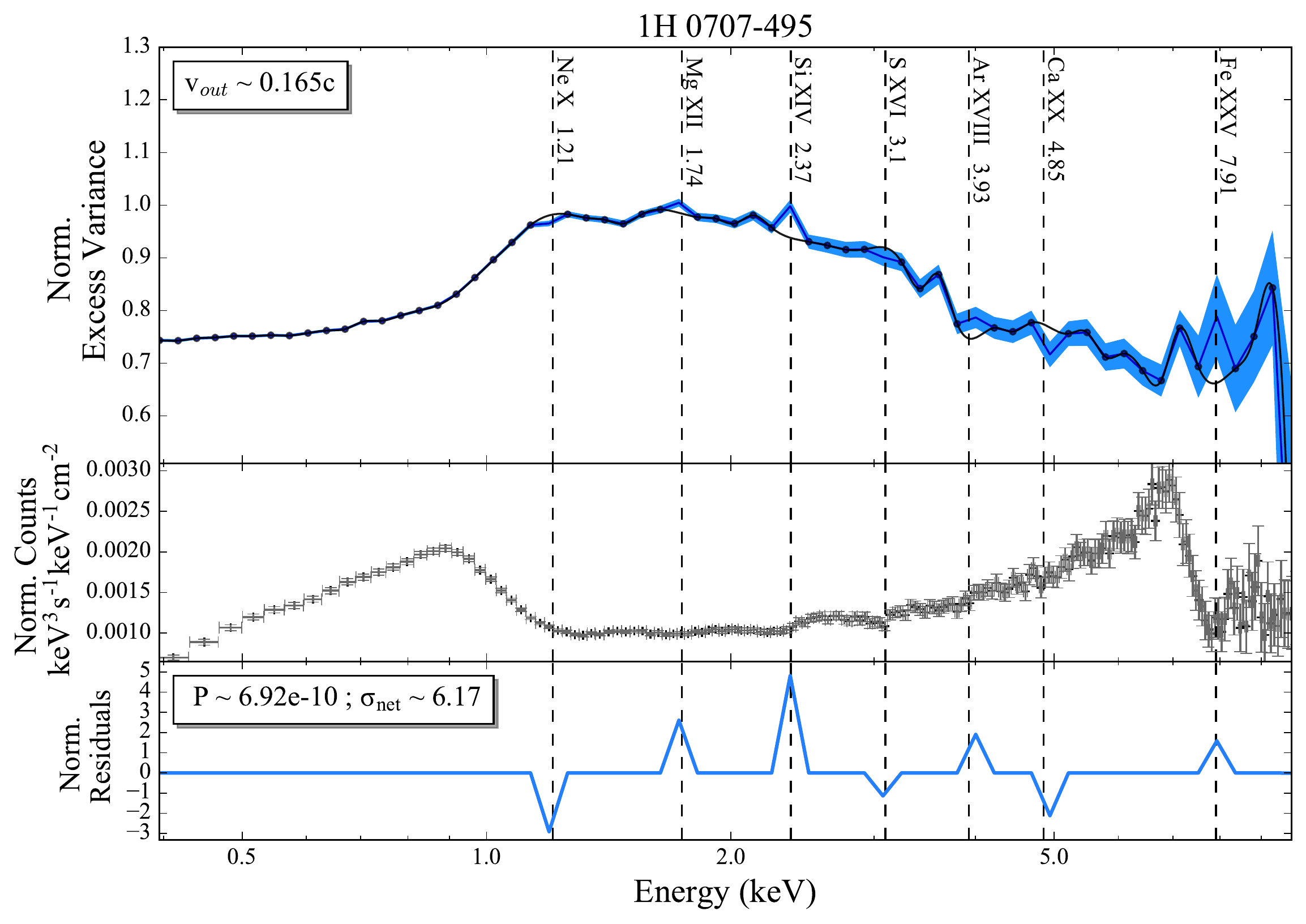}
    \includegraphics[width=85mm]{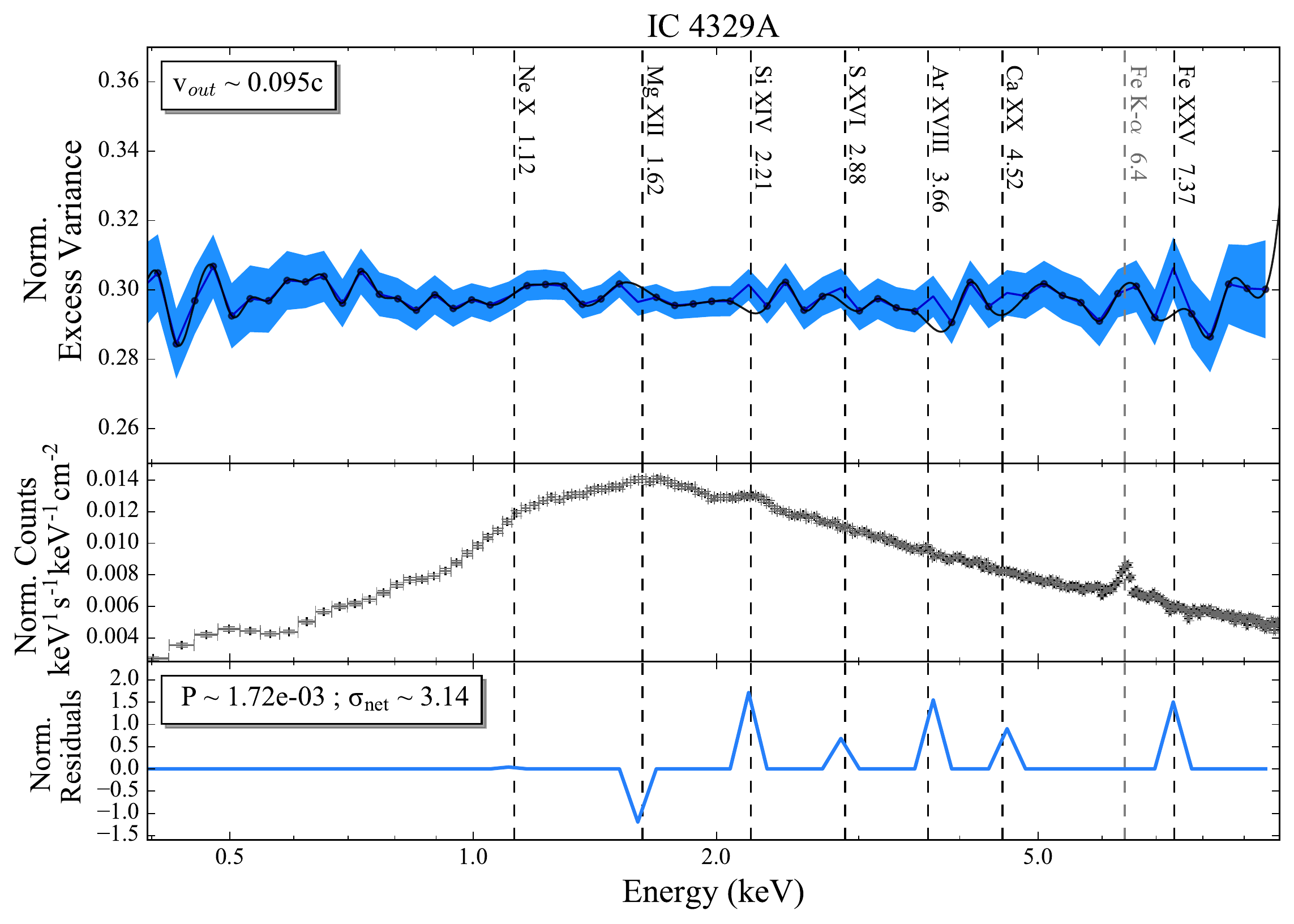}
    \includegraphics[width=85mm]{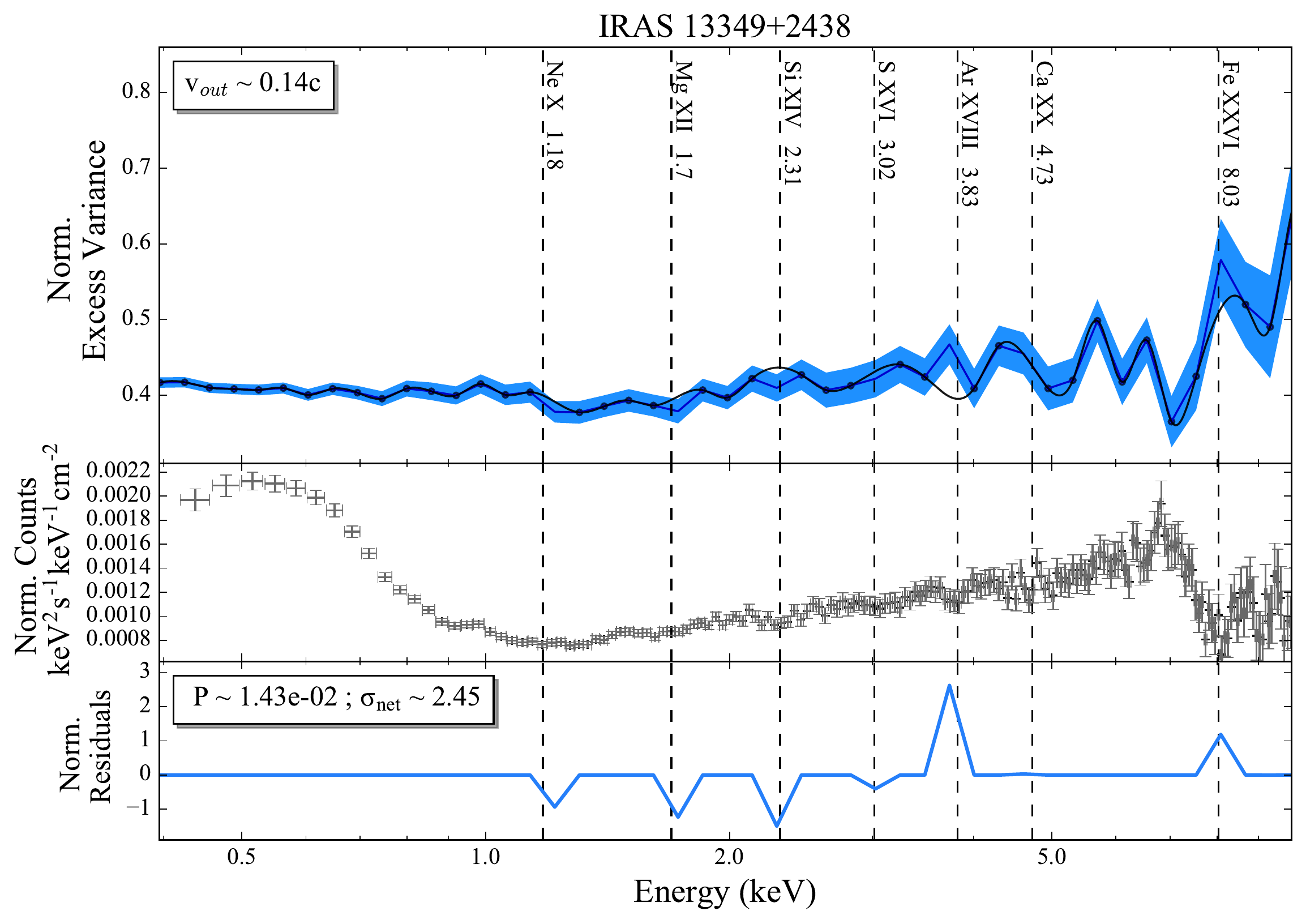}
    \includegraphics[width=85mm]{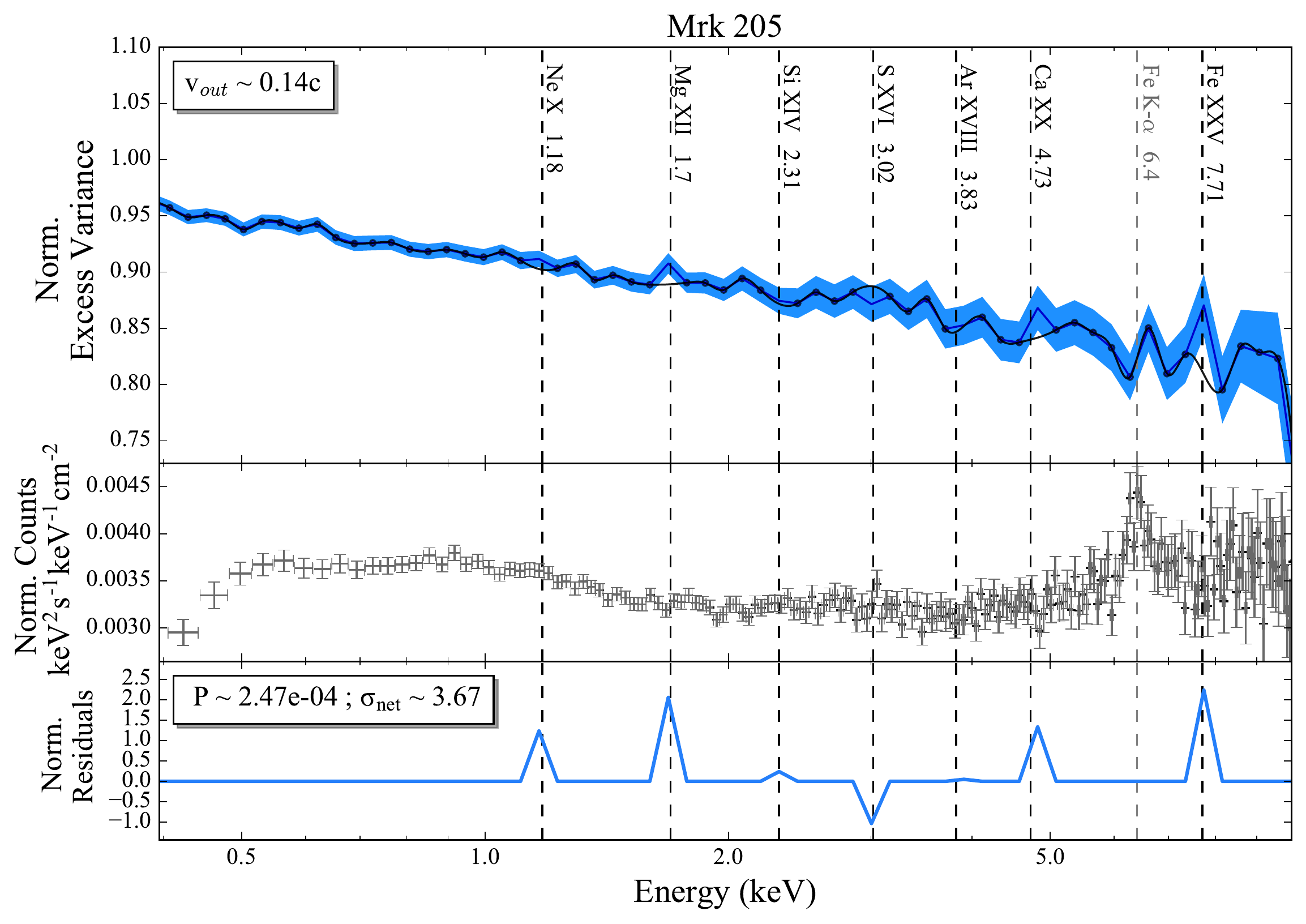}
    \includegraphics[width=85mm]{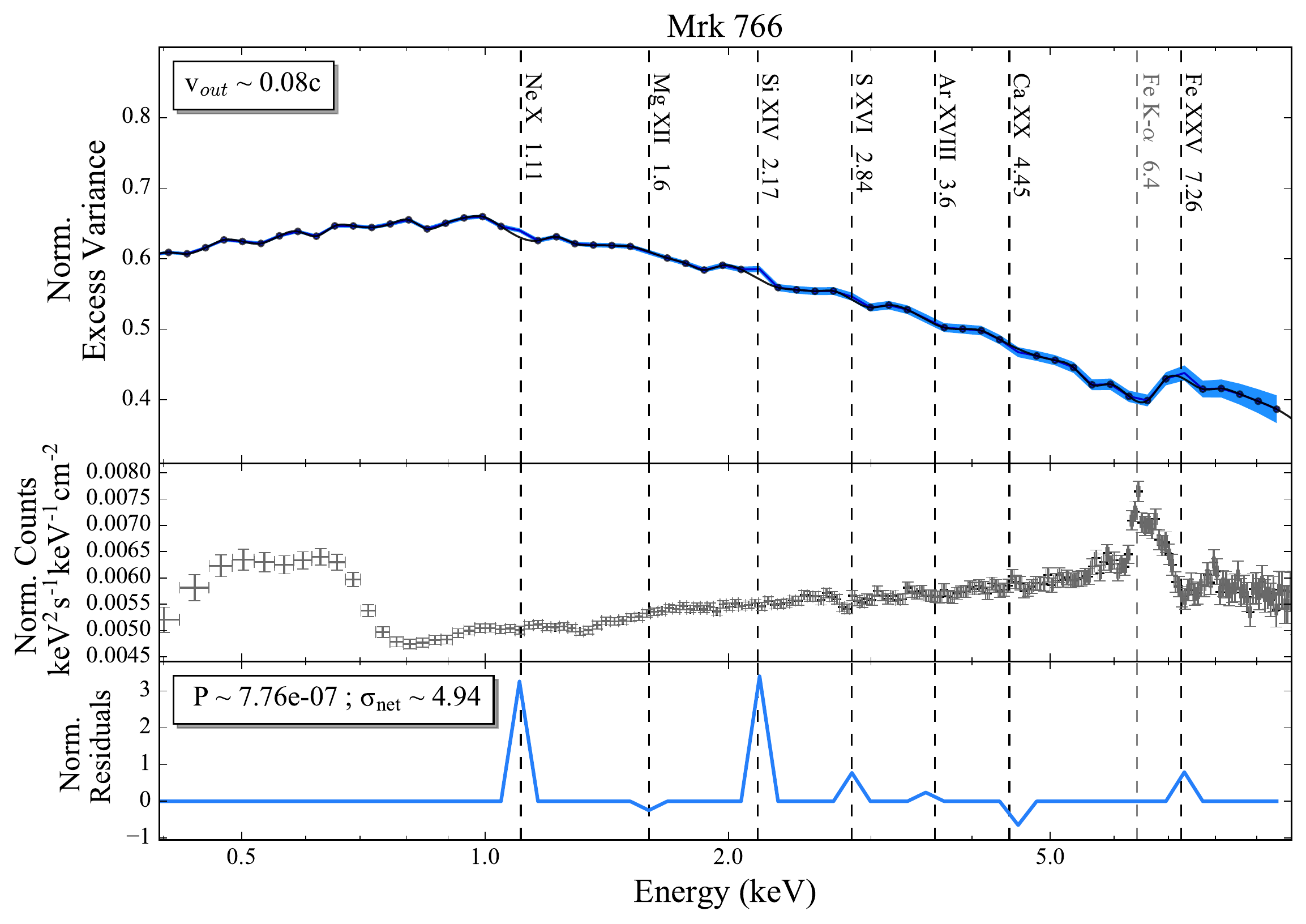}
    \includegraphics[width=85mm]{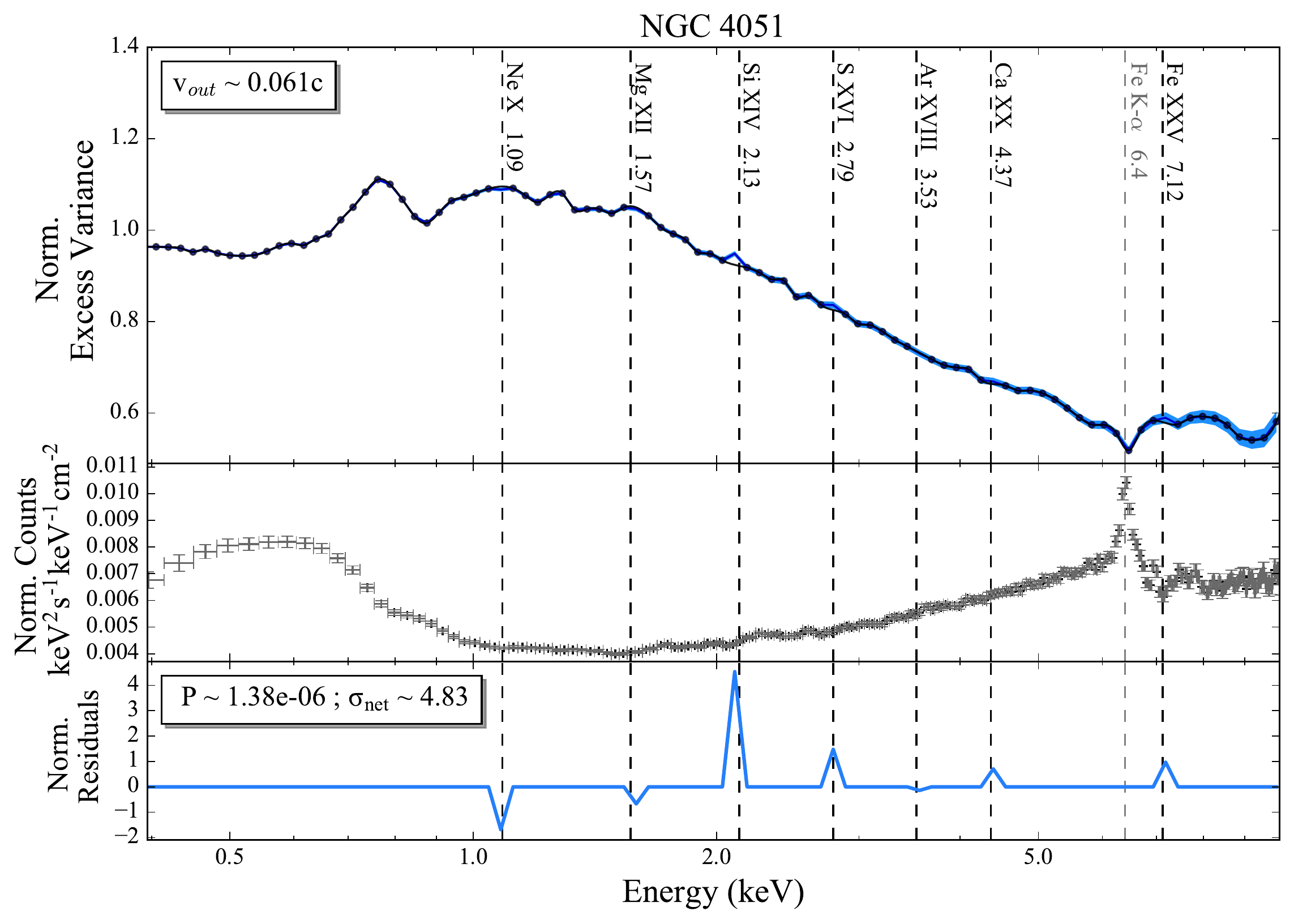}
    \caption{Examples of sources with strong evidence for a UFO.}
\end{figure*}

\begin{figure*}
    \centering
    \includegraphics[width=85mm]{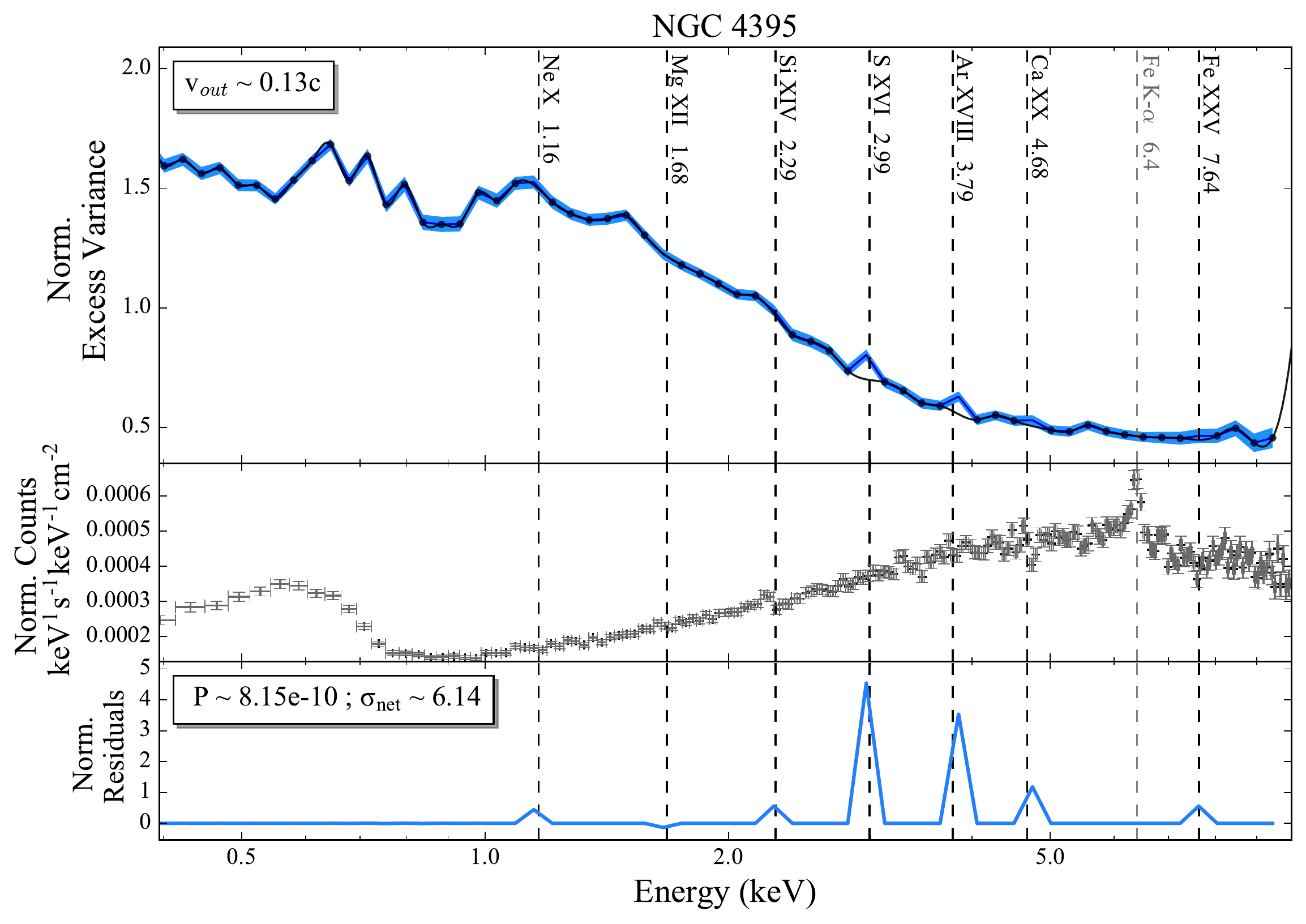}
    \includegraphics[width=85mm]{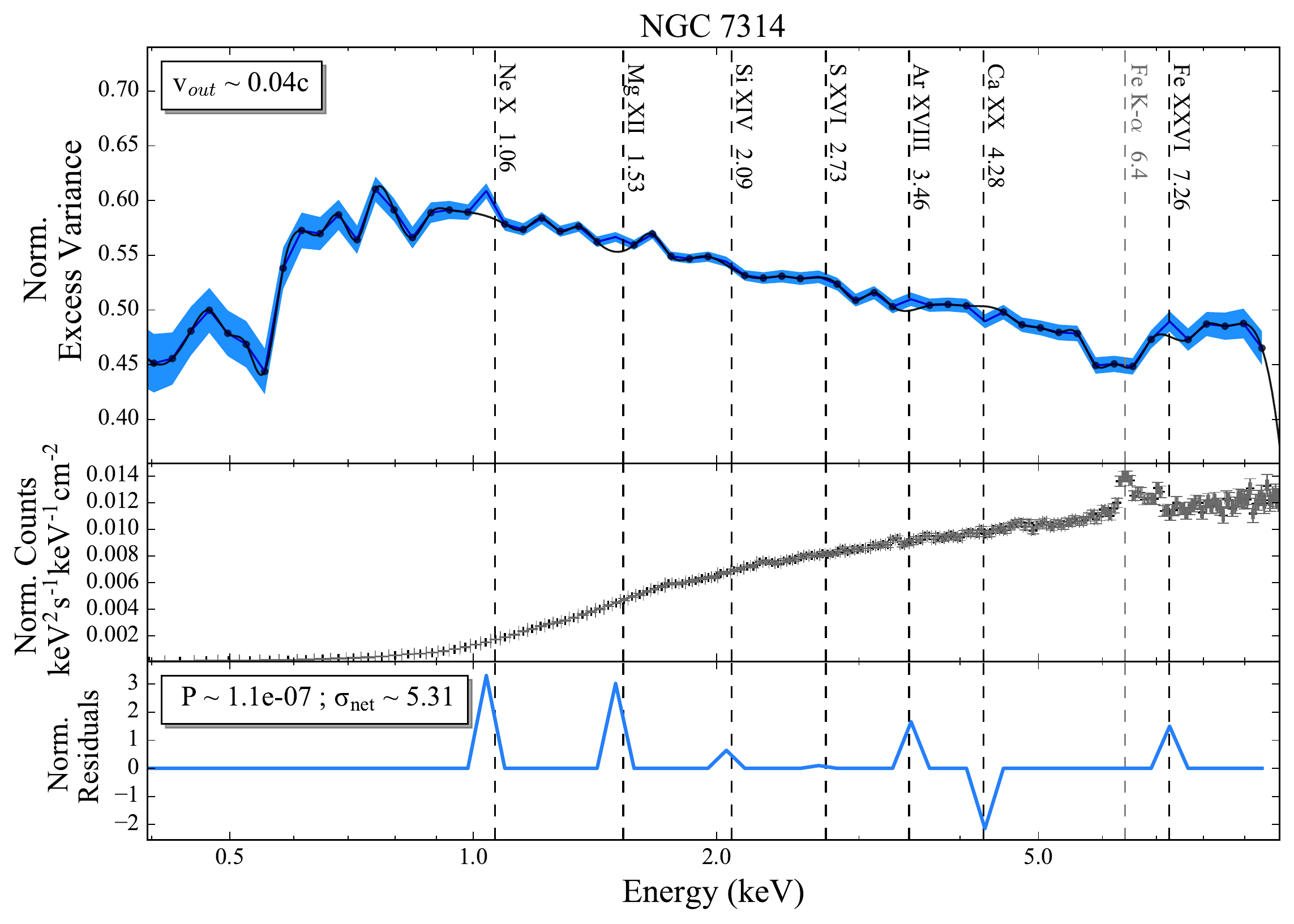}
    \includegraphics[width=85mm]{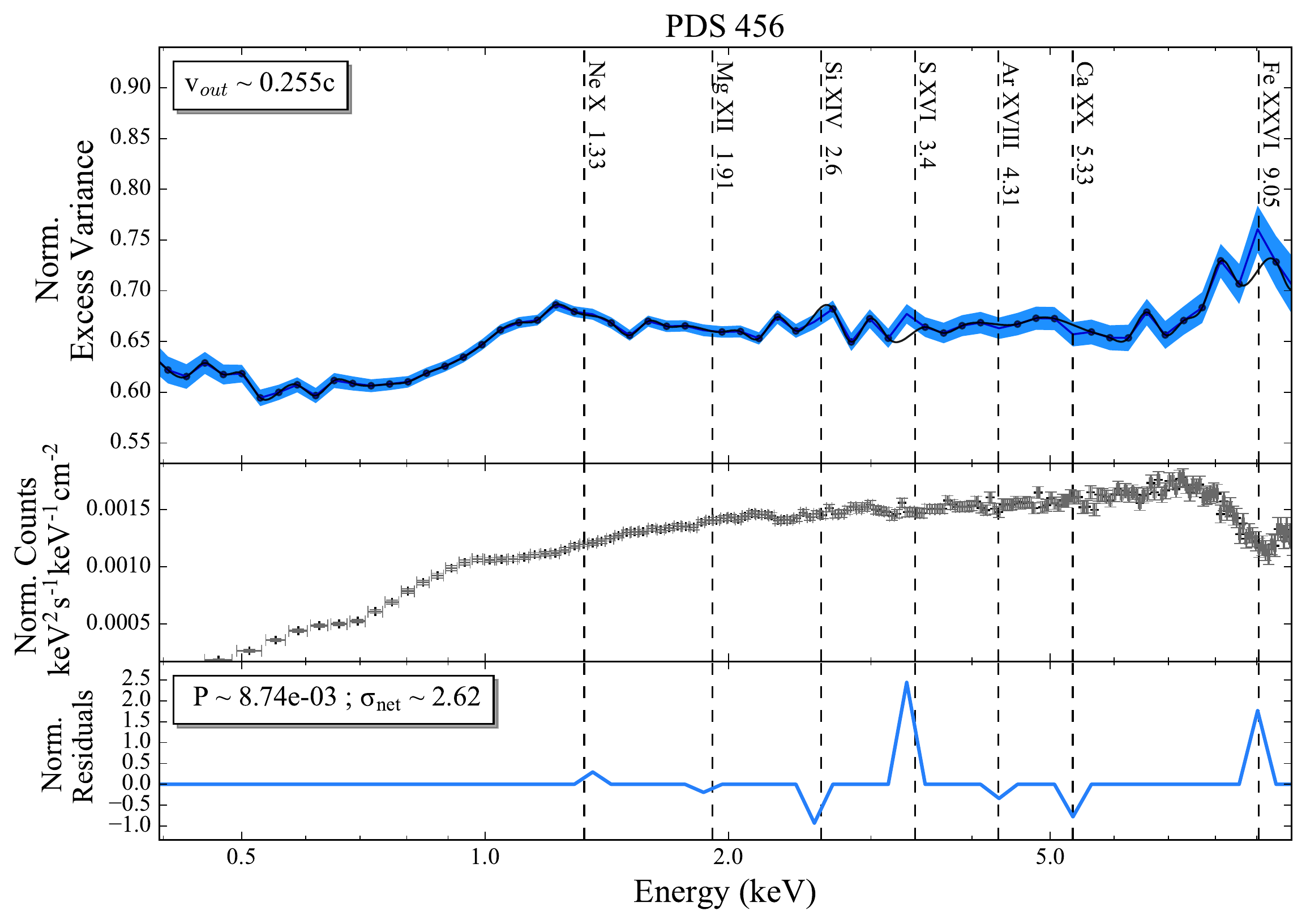}
    \includegraphics[width=85mm]{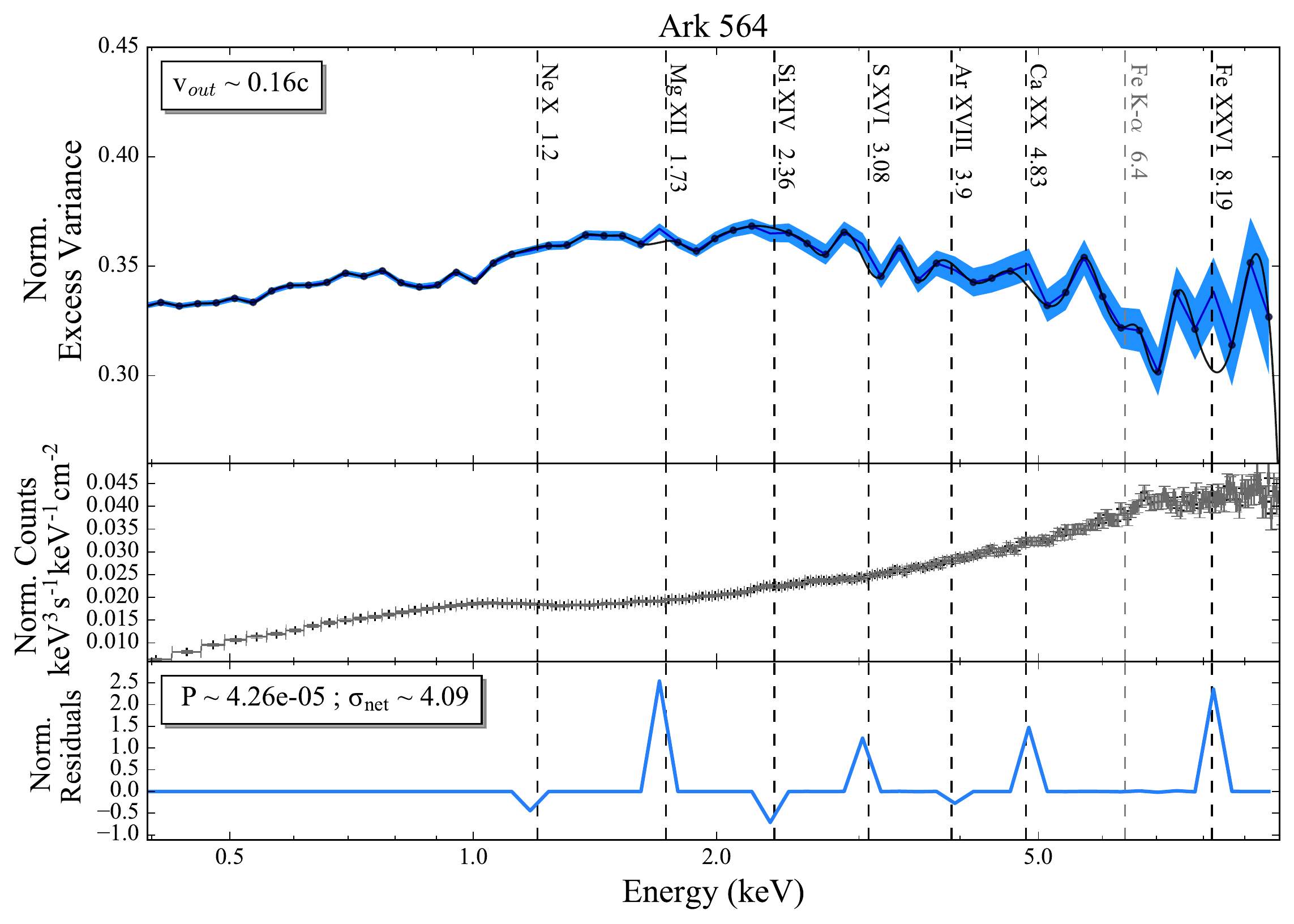}
    \includegraphics[width=85mm]{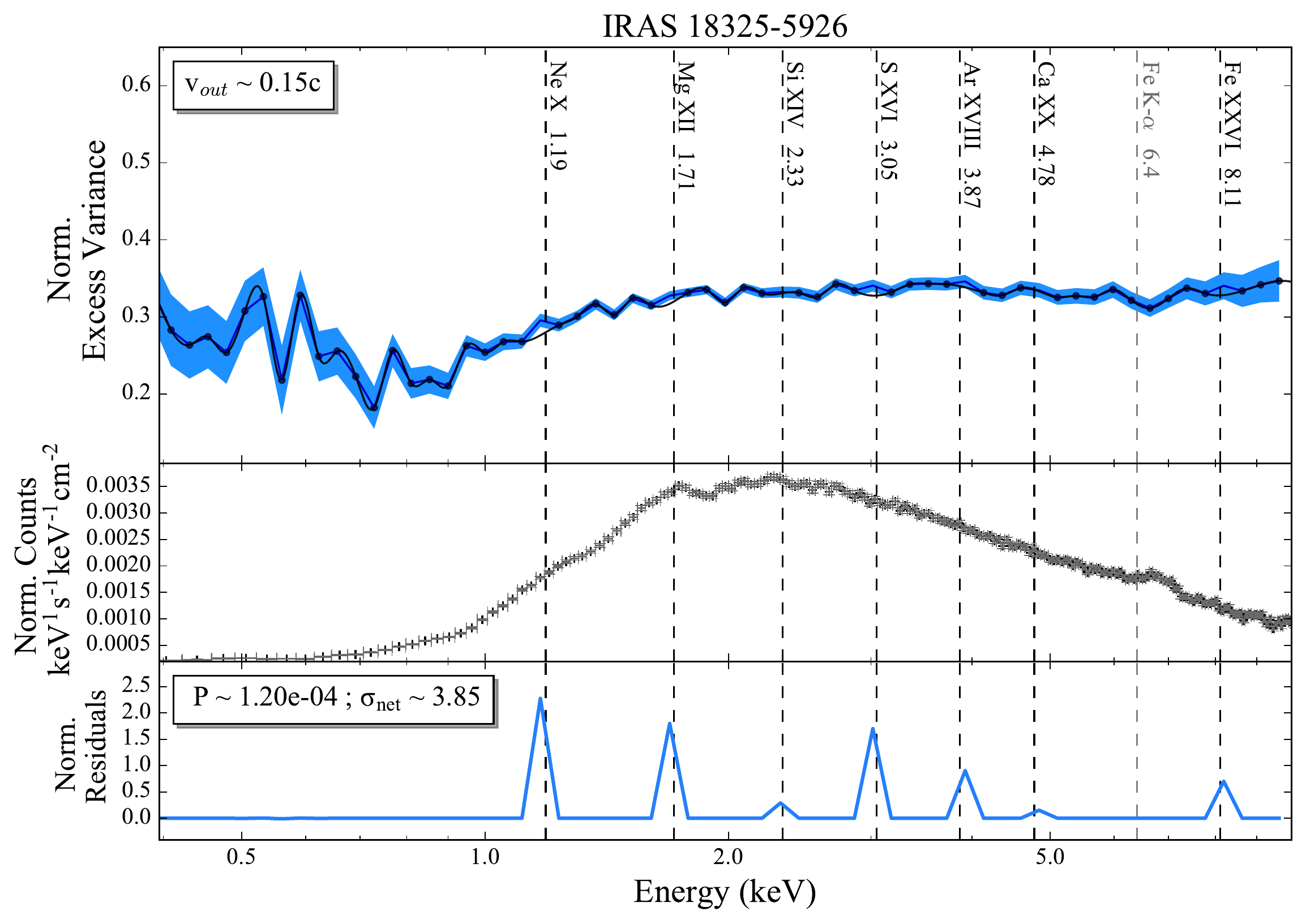}
    \includegraphics[width=85mm]{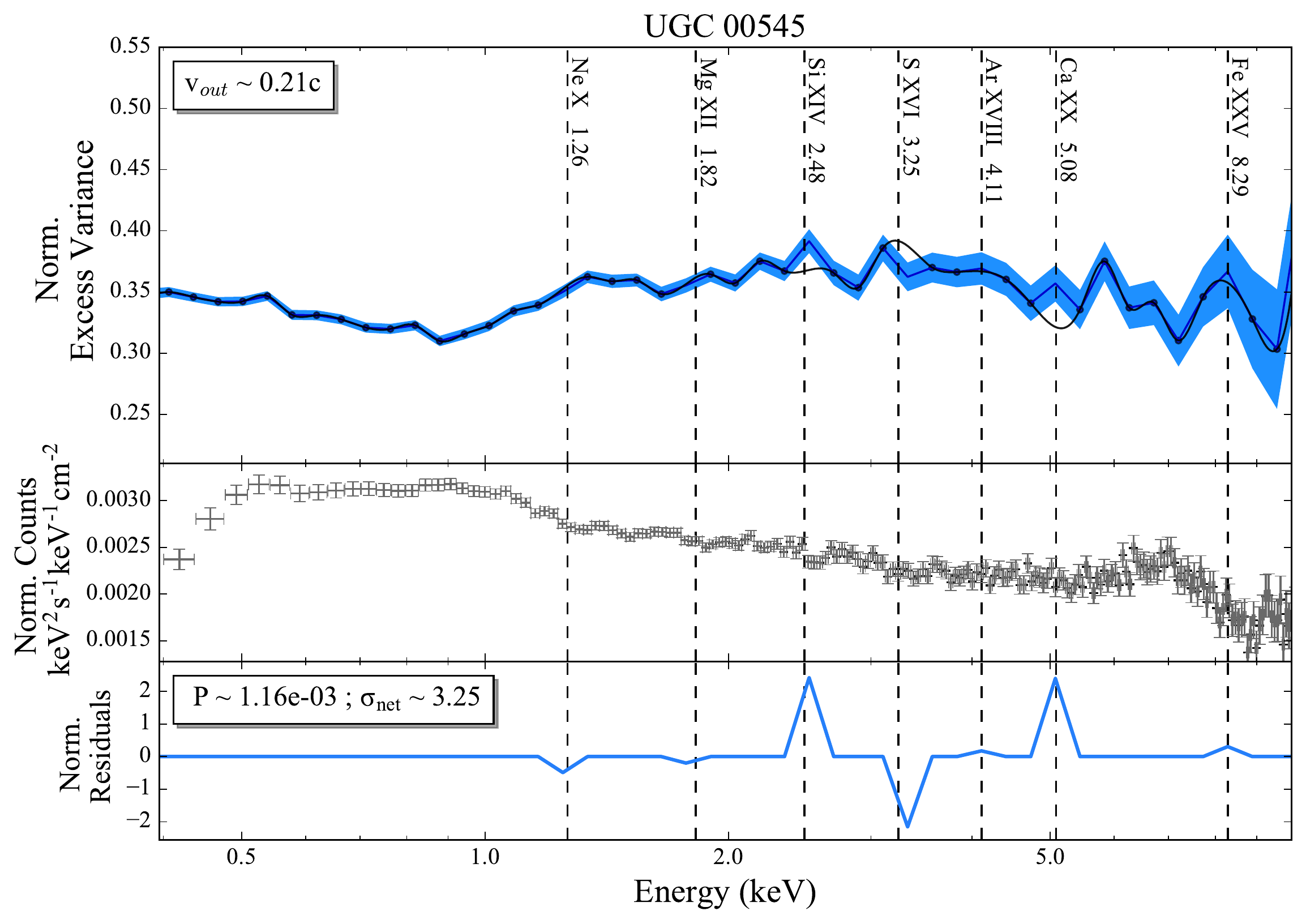}
    \caption{NGC 4395, NGC7314 and PDS 456 are sources with good evidence for UFOs, while Ark 564, IRAS 18325-5926 and UGC00545 have only weak evidence.}
\end{figure*}

\begin{figure*}
    \centering
    \includegraphics[width=85mm]{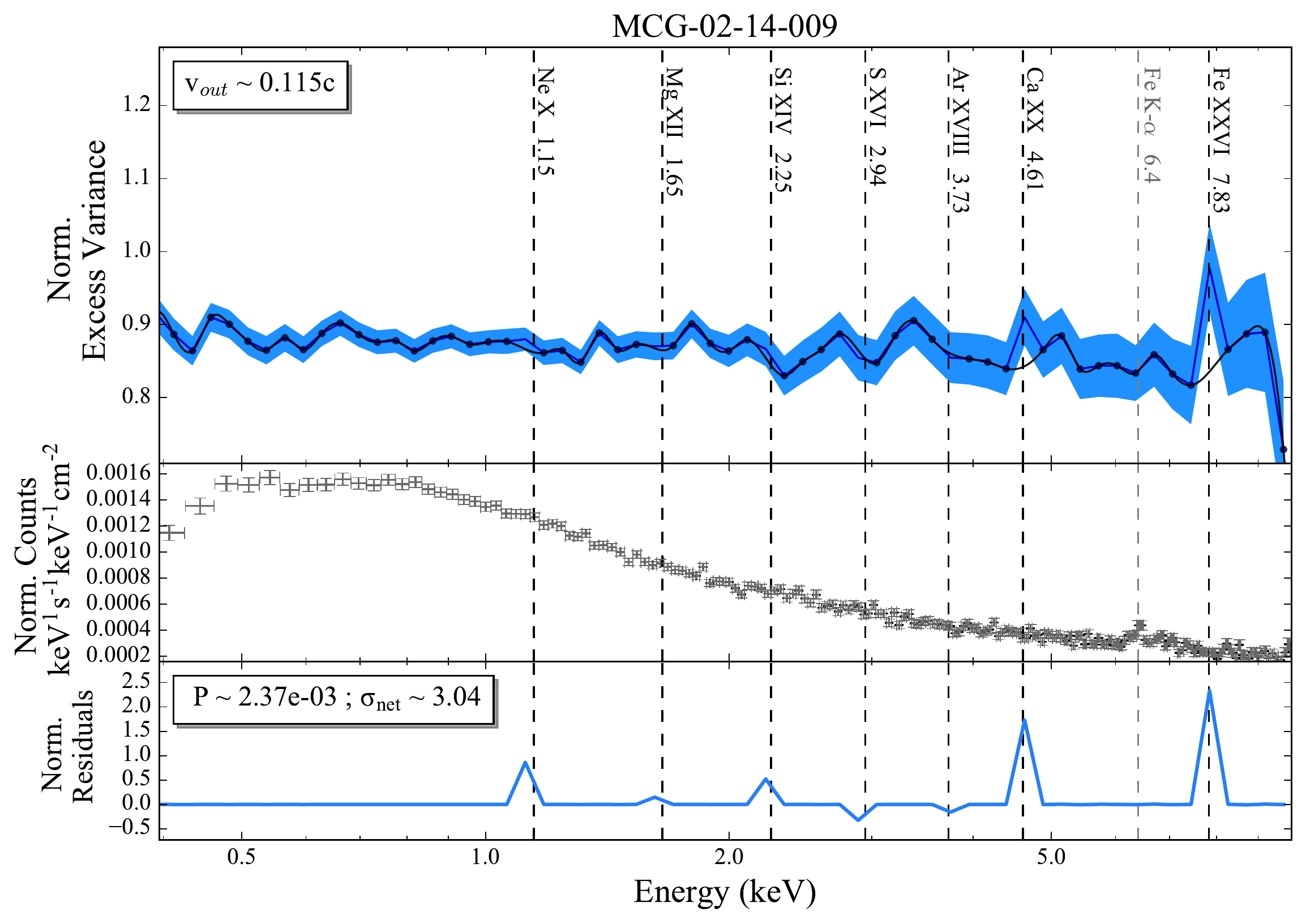}
    \includegraphics[width=85mm]{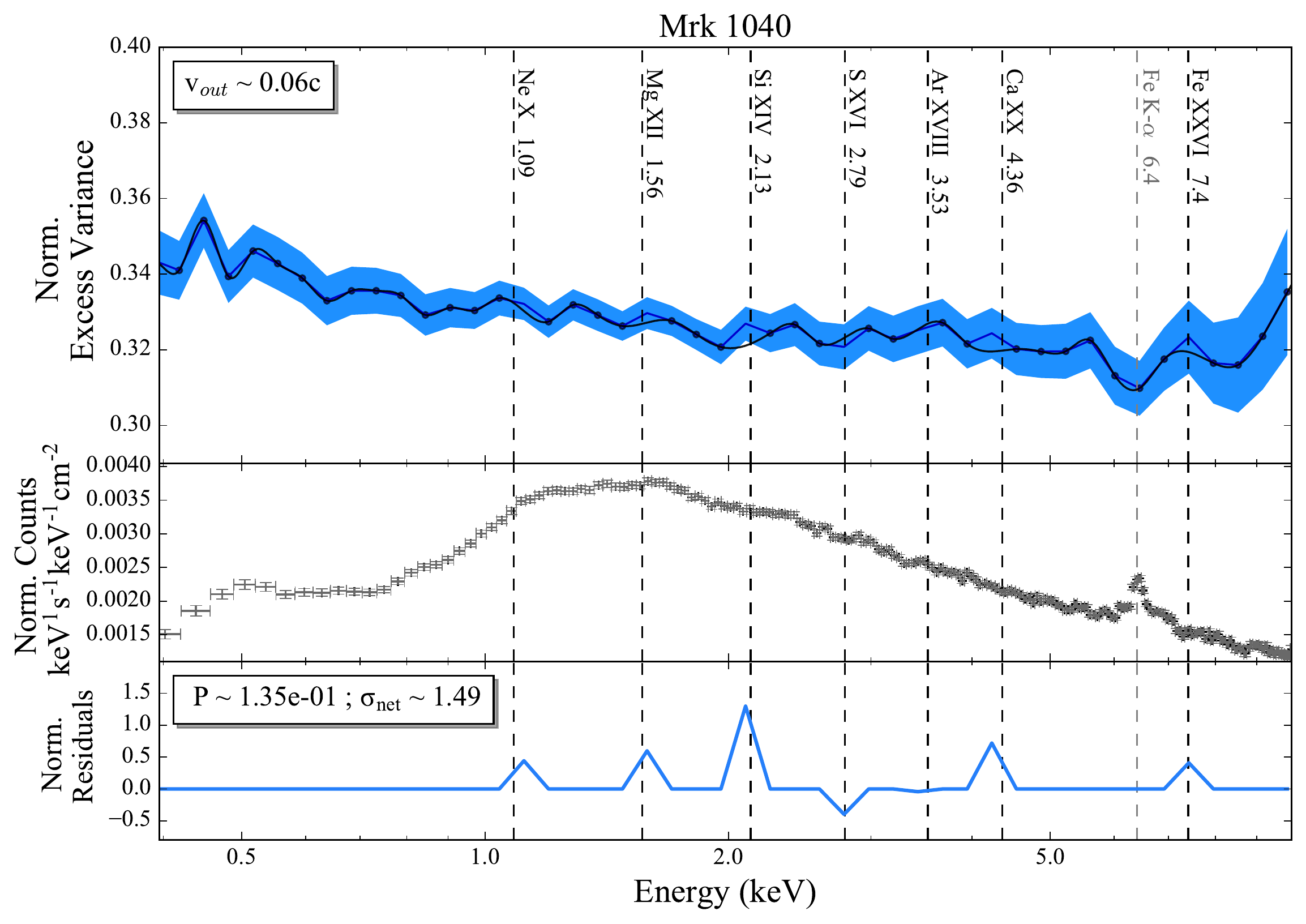}
    \includegraphics[width=85mm]{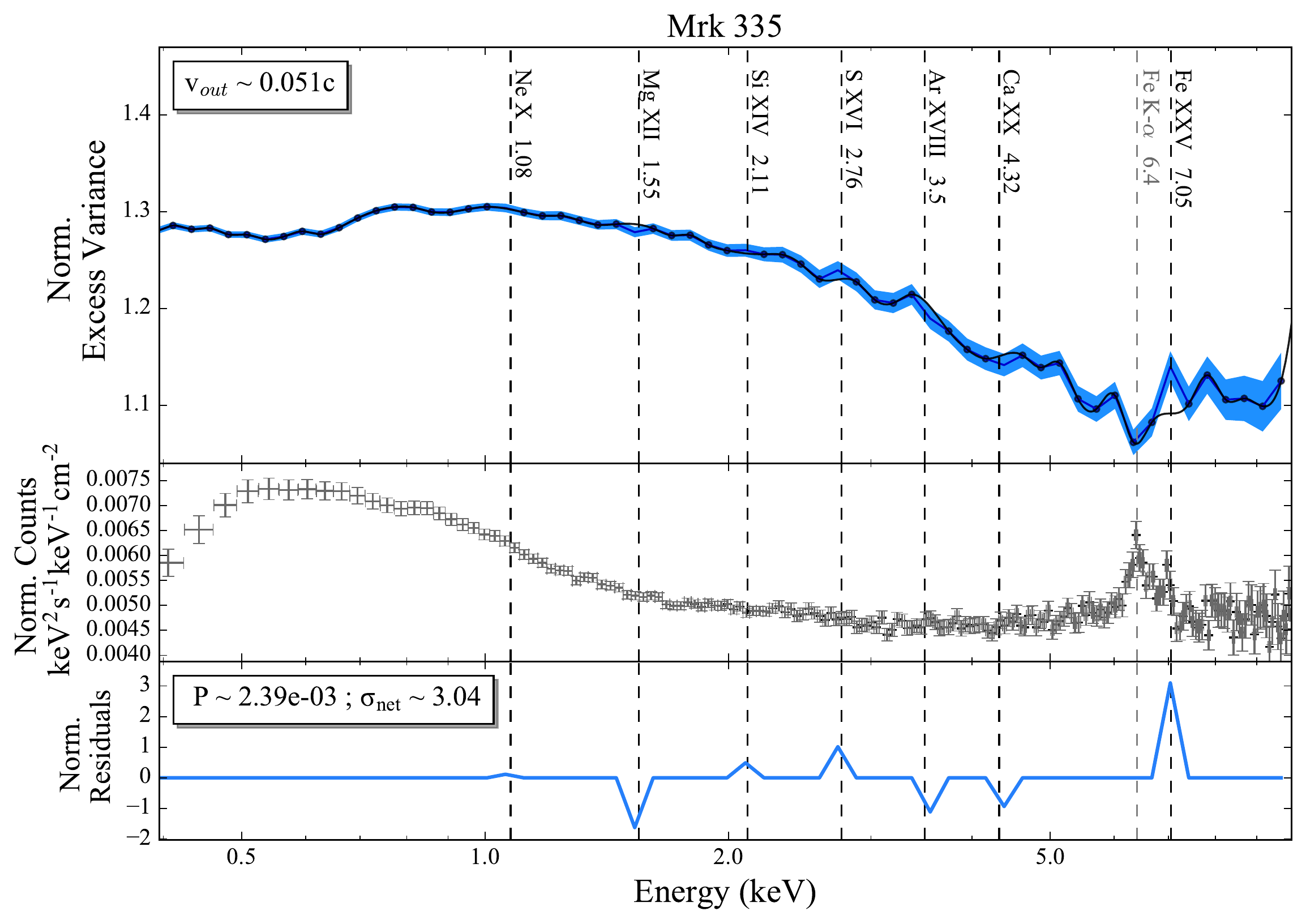}
    \includegraphics[width=85mm]{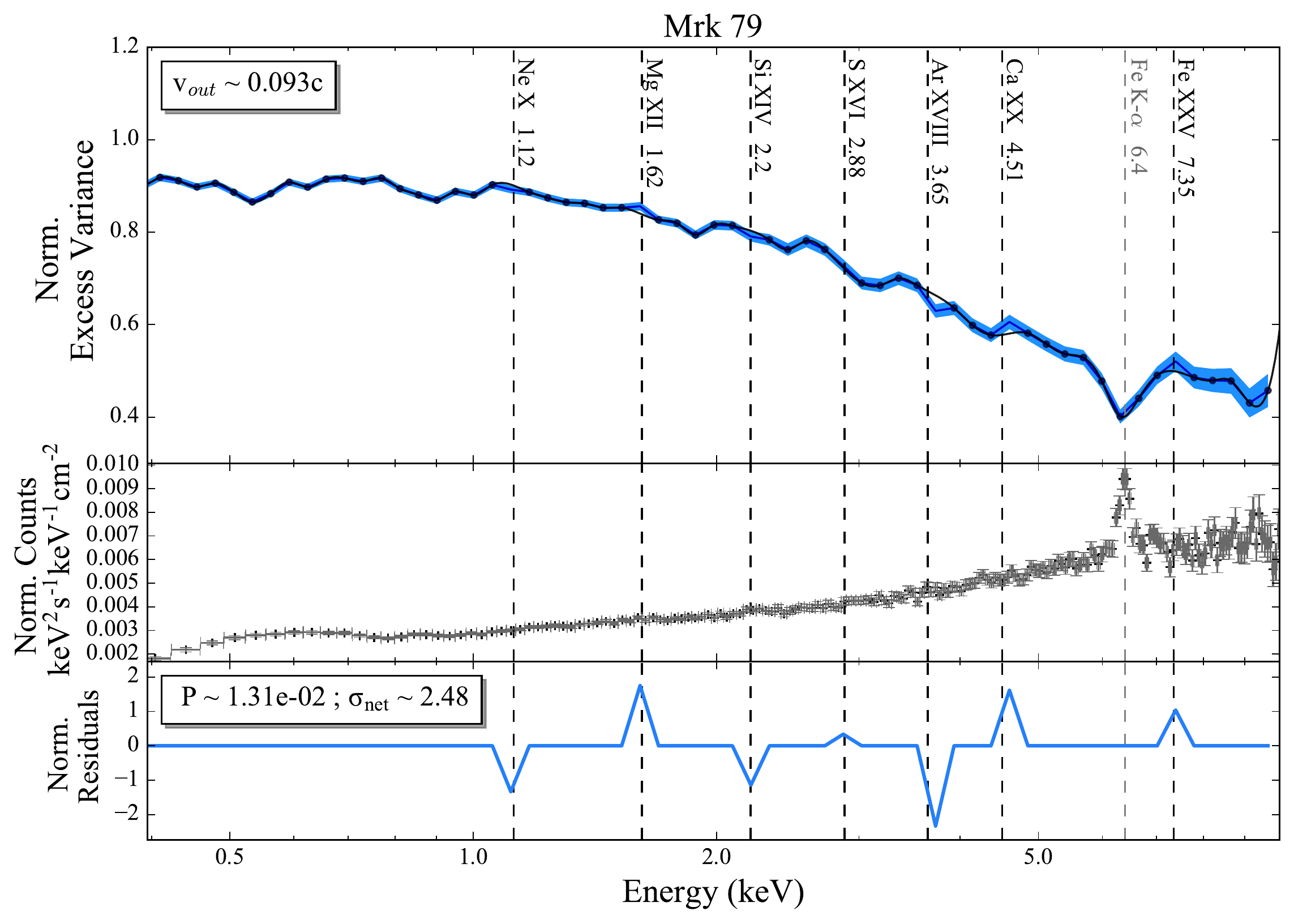}
    \includegraphics[width=85mm]{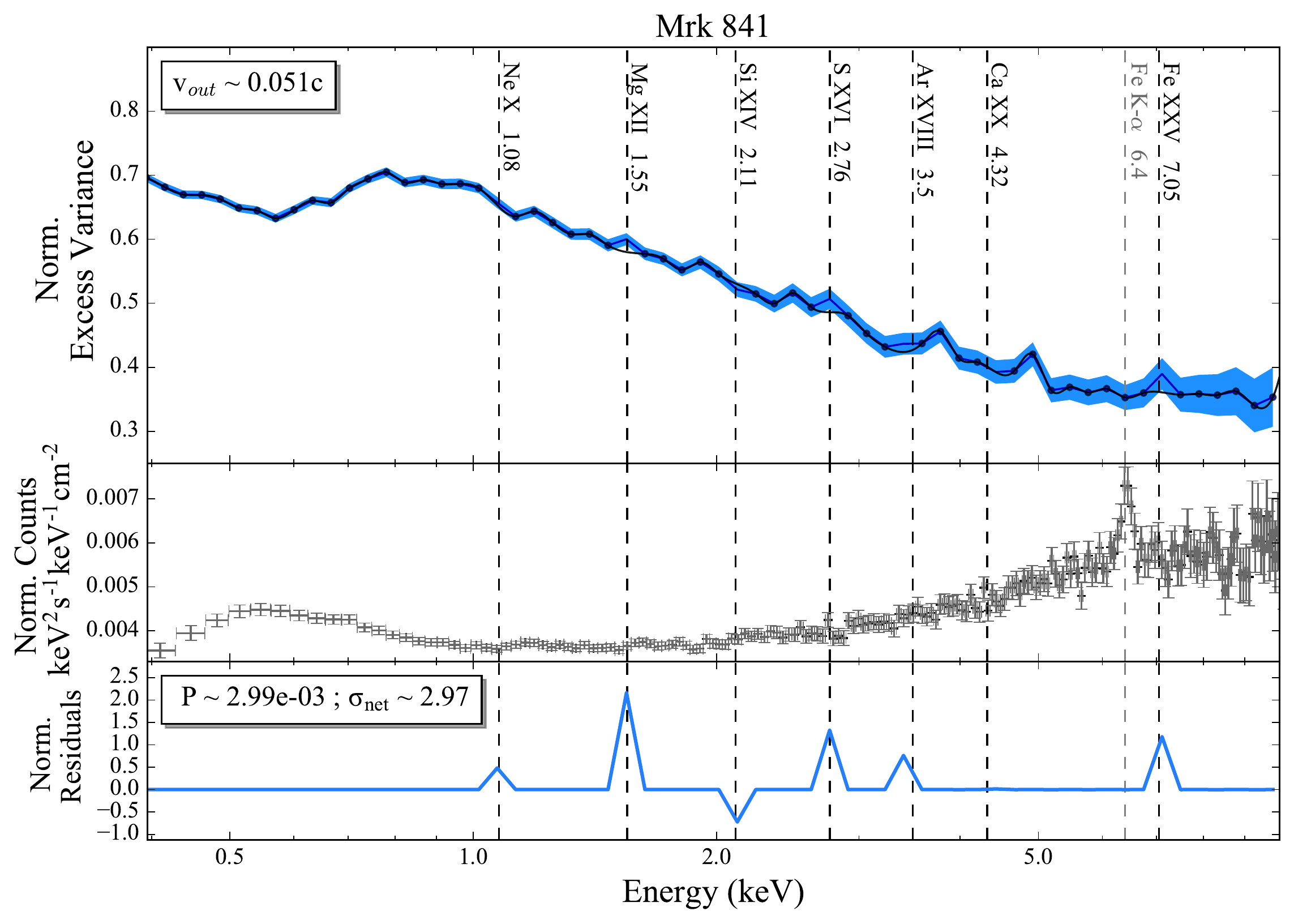}
    \includegraphics[width=85mm]{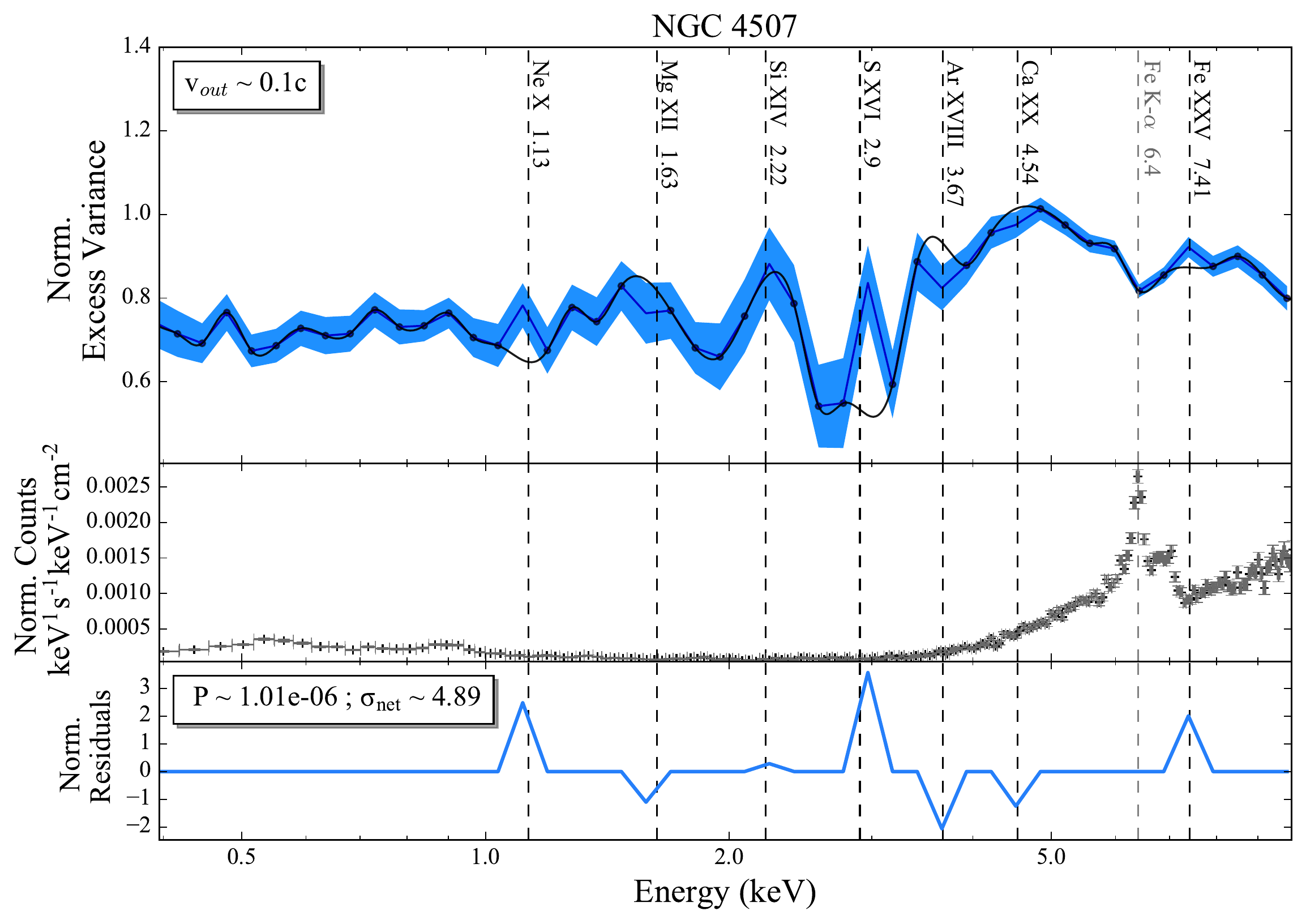}
    \caption{Examples of sources with weak evidence for a UFO.}
\end{figure*}

\begin{figure*}
    \centering
    \includegraphics[width=85mm]{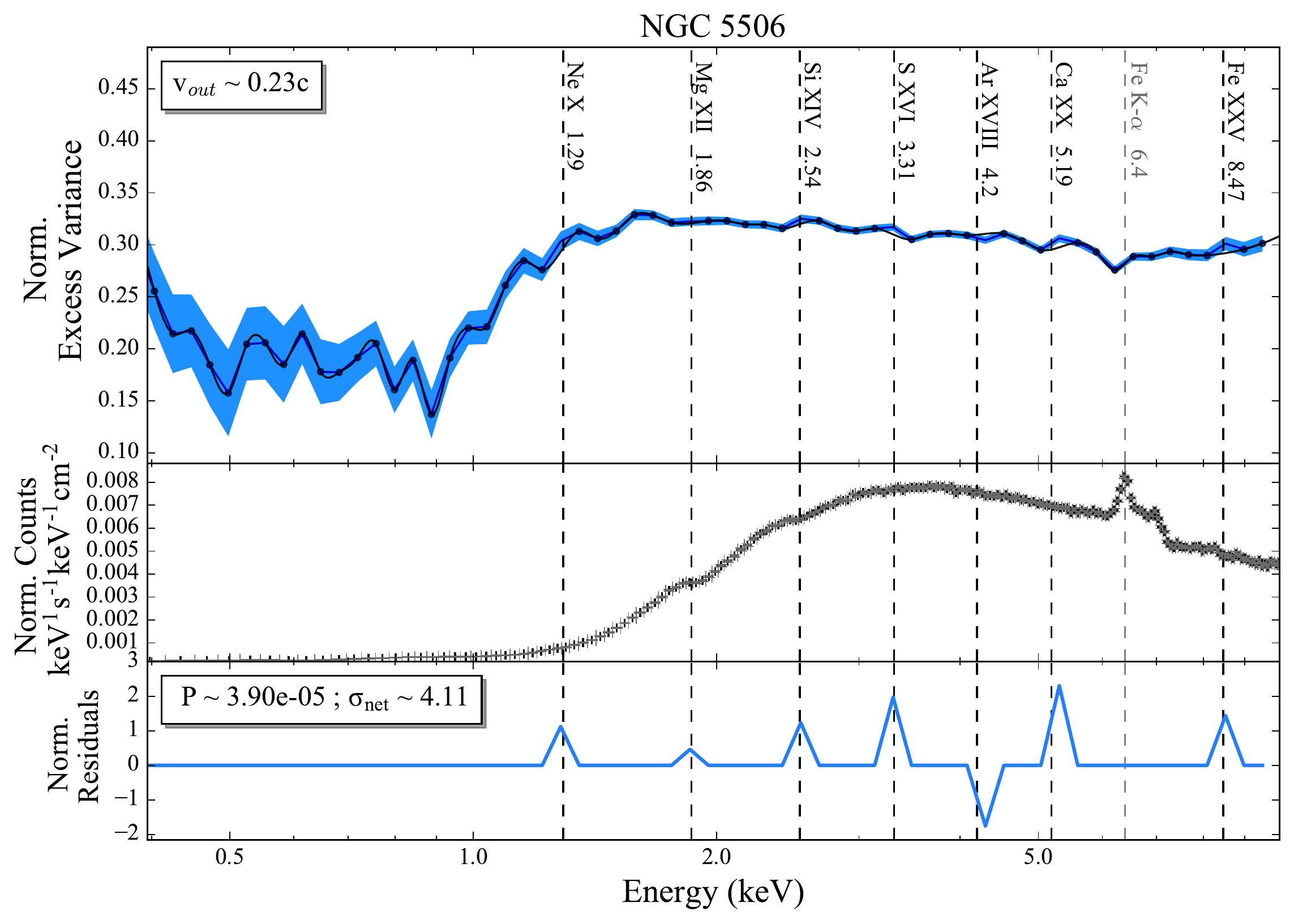}
    \includegraphics[width=85mm]{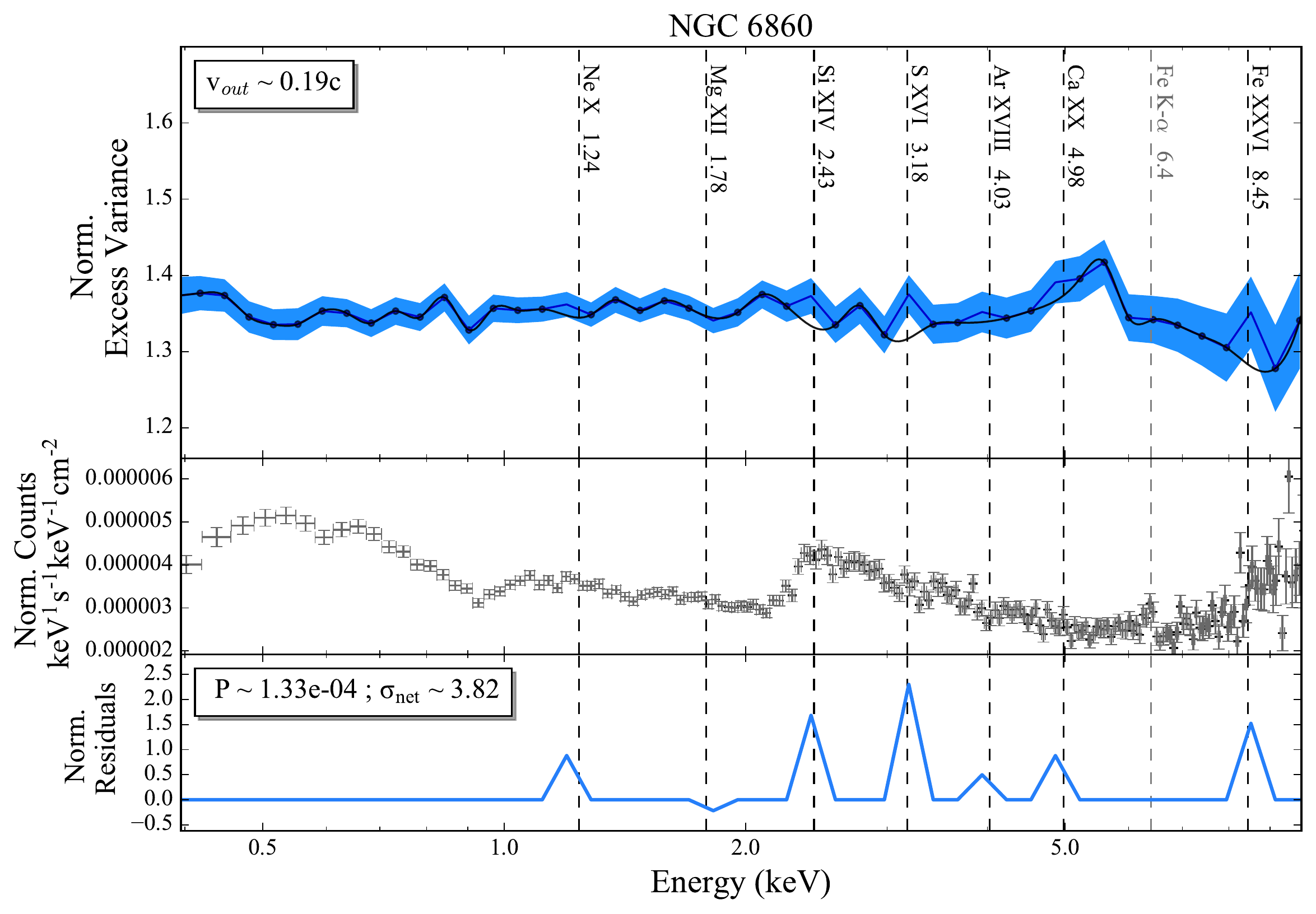}
    \includegraphics[width=85mm]{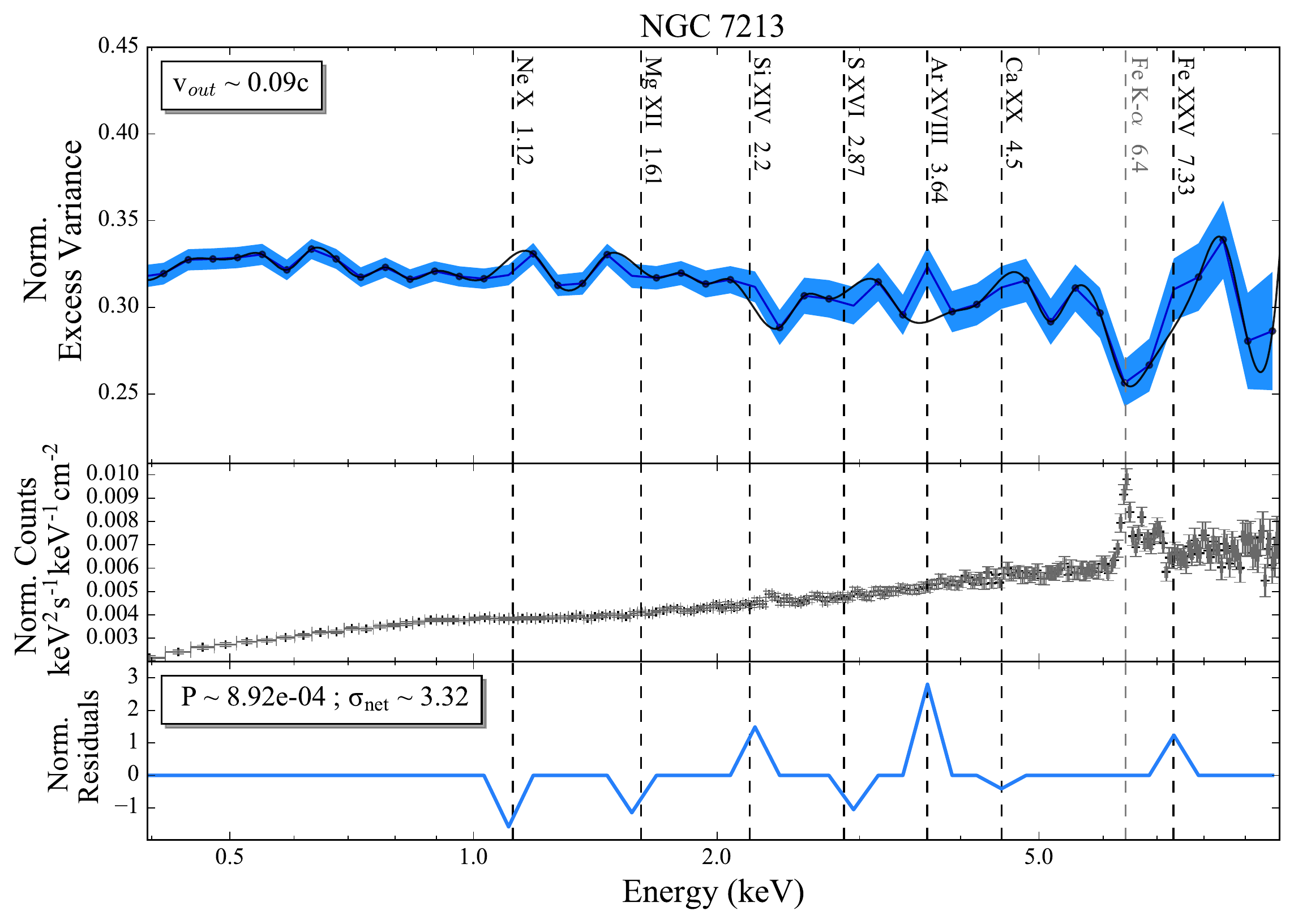}
    \includegraphics[width=85mm]{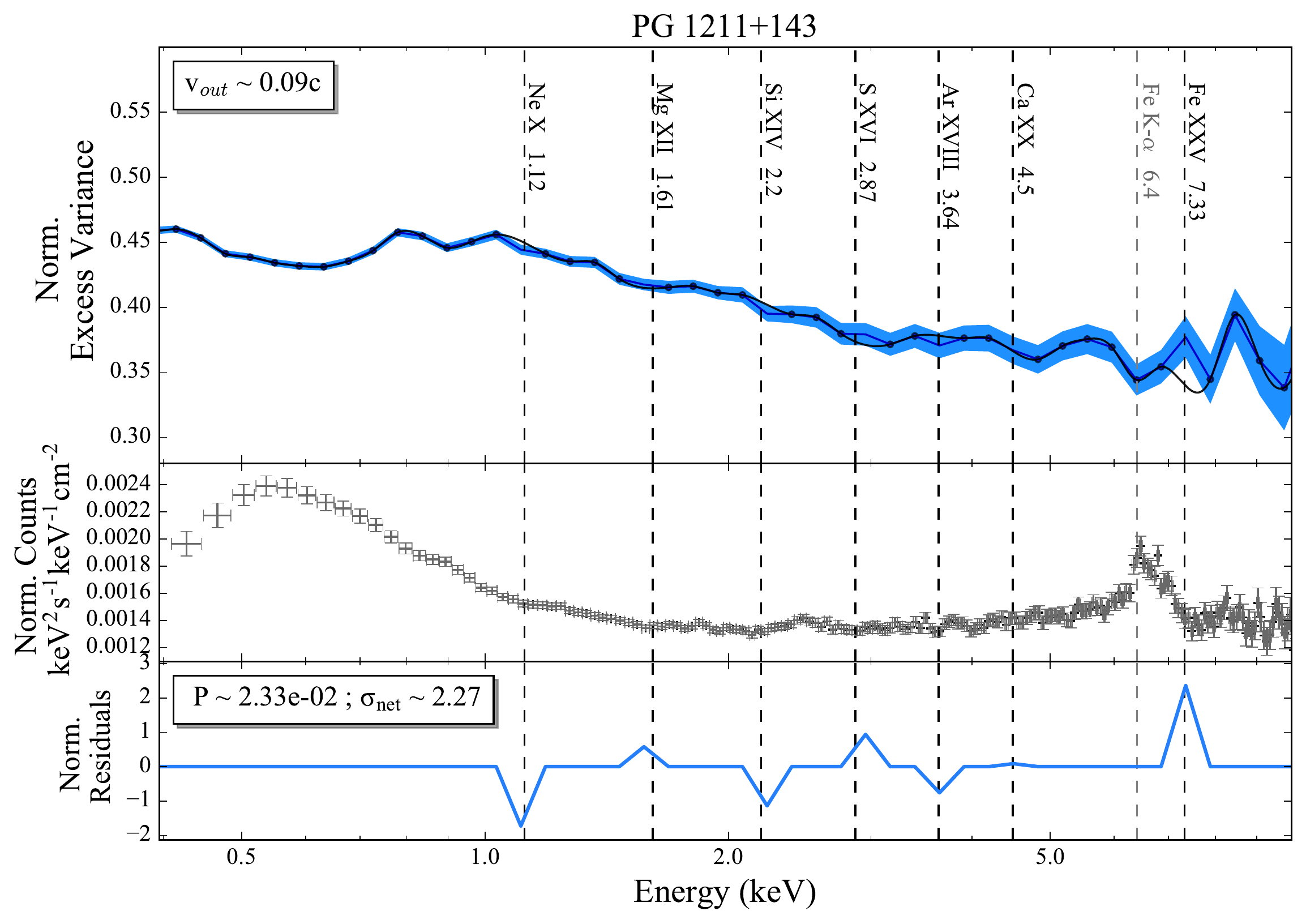}
    \includegraphics[width=85mm]{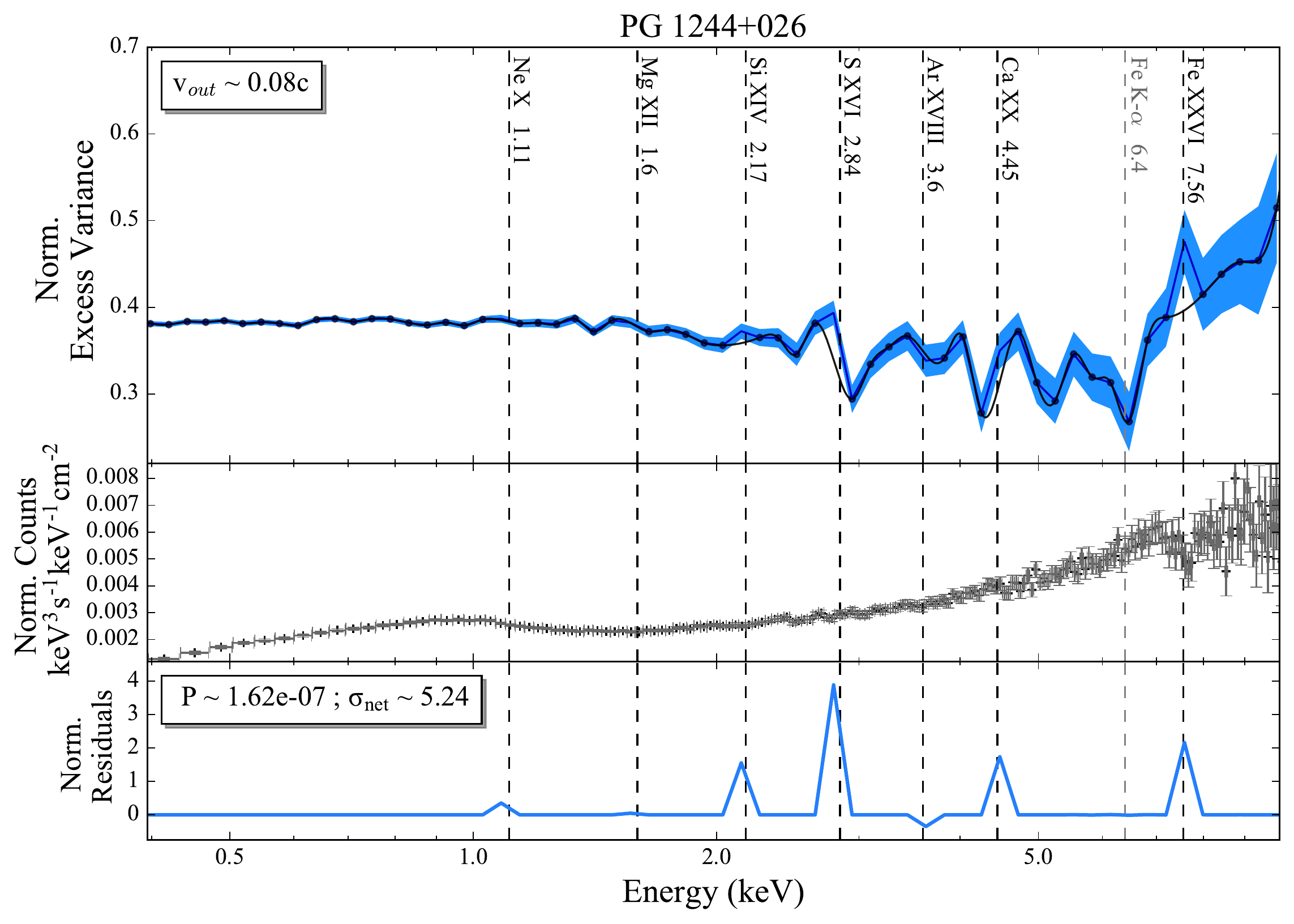}
    \includegraphics[width=85mm]{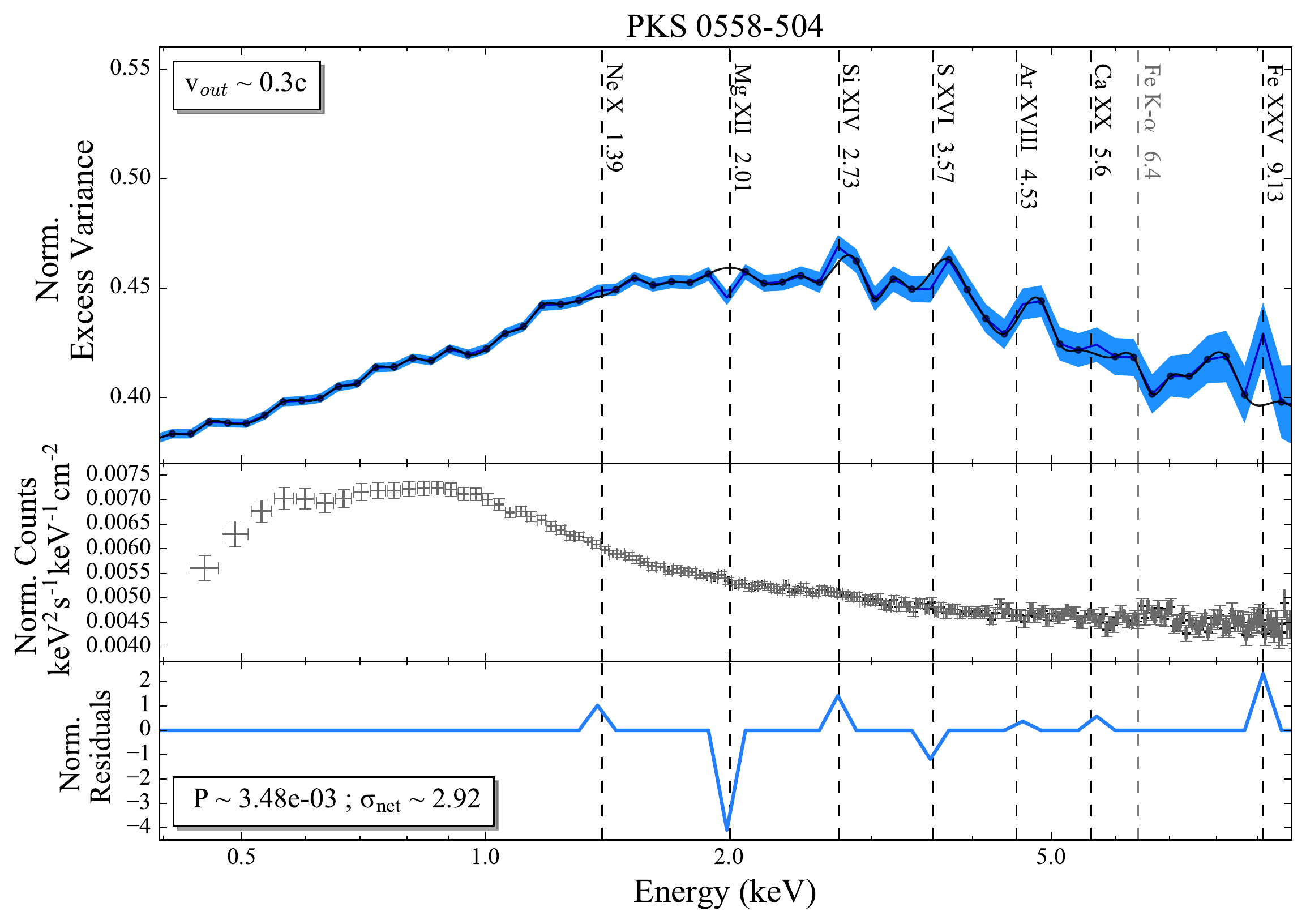}
    \caption{Examples of sources with weak evidence for a UFO.}
\end{figure*}

\begin{figure*}
    \centering
    \includegraphics[width=85mm]{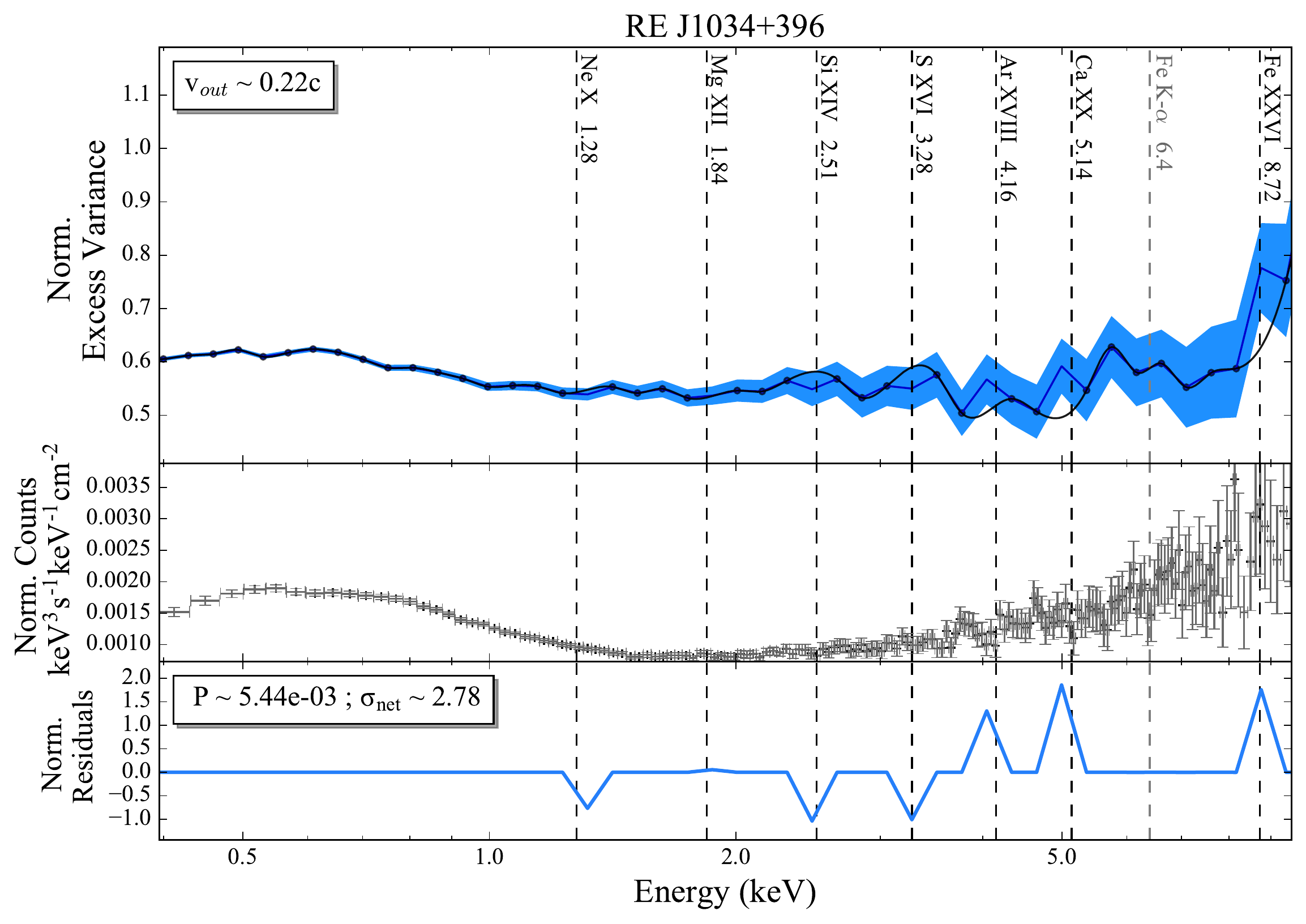}
    \includegraphics[width=85mm]{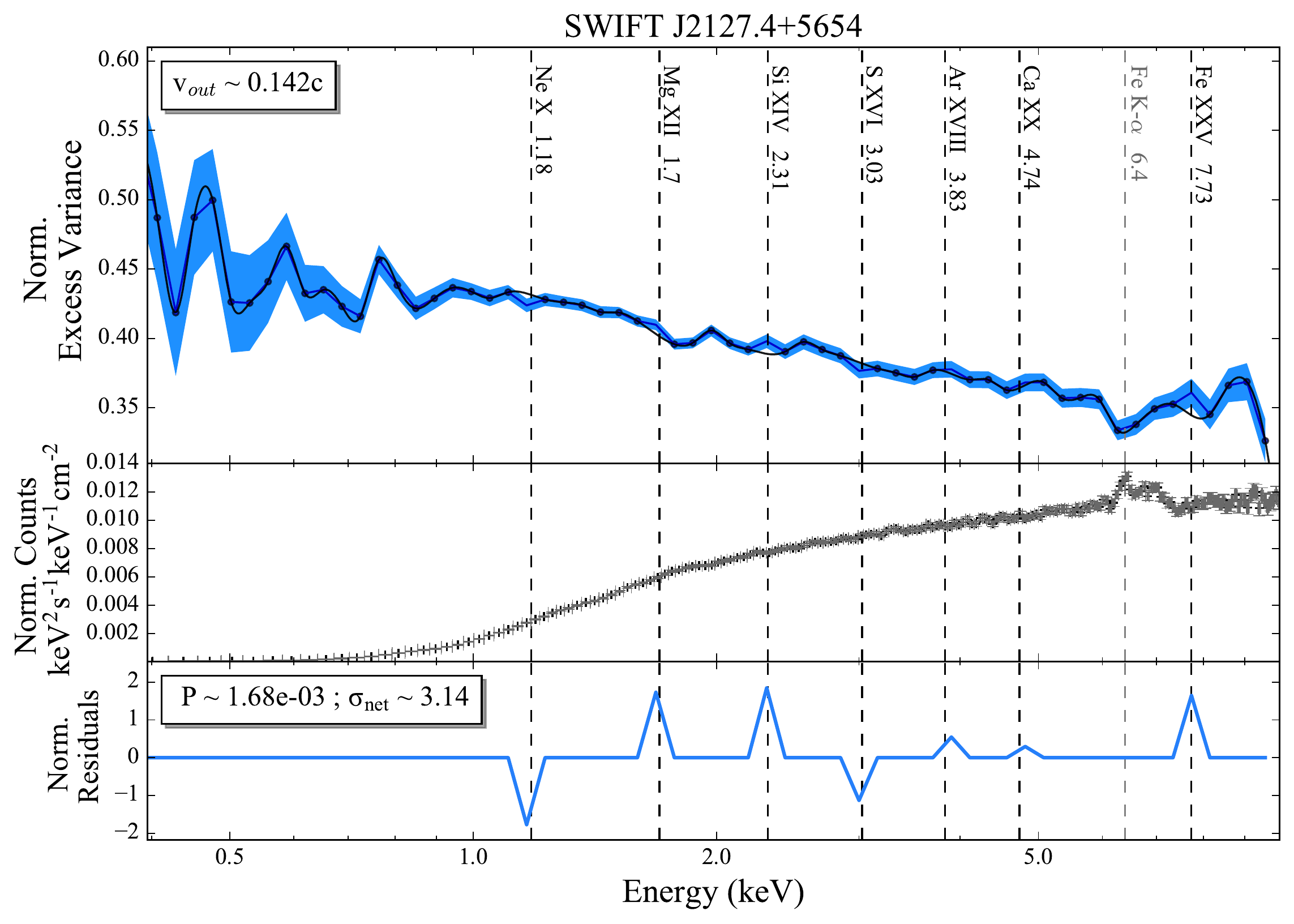}
    \includegraphics[width=85mm]{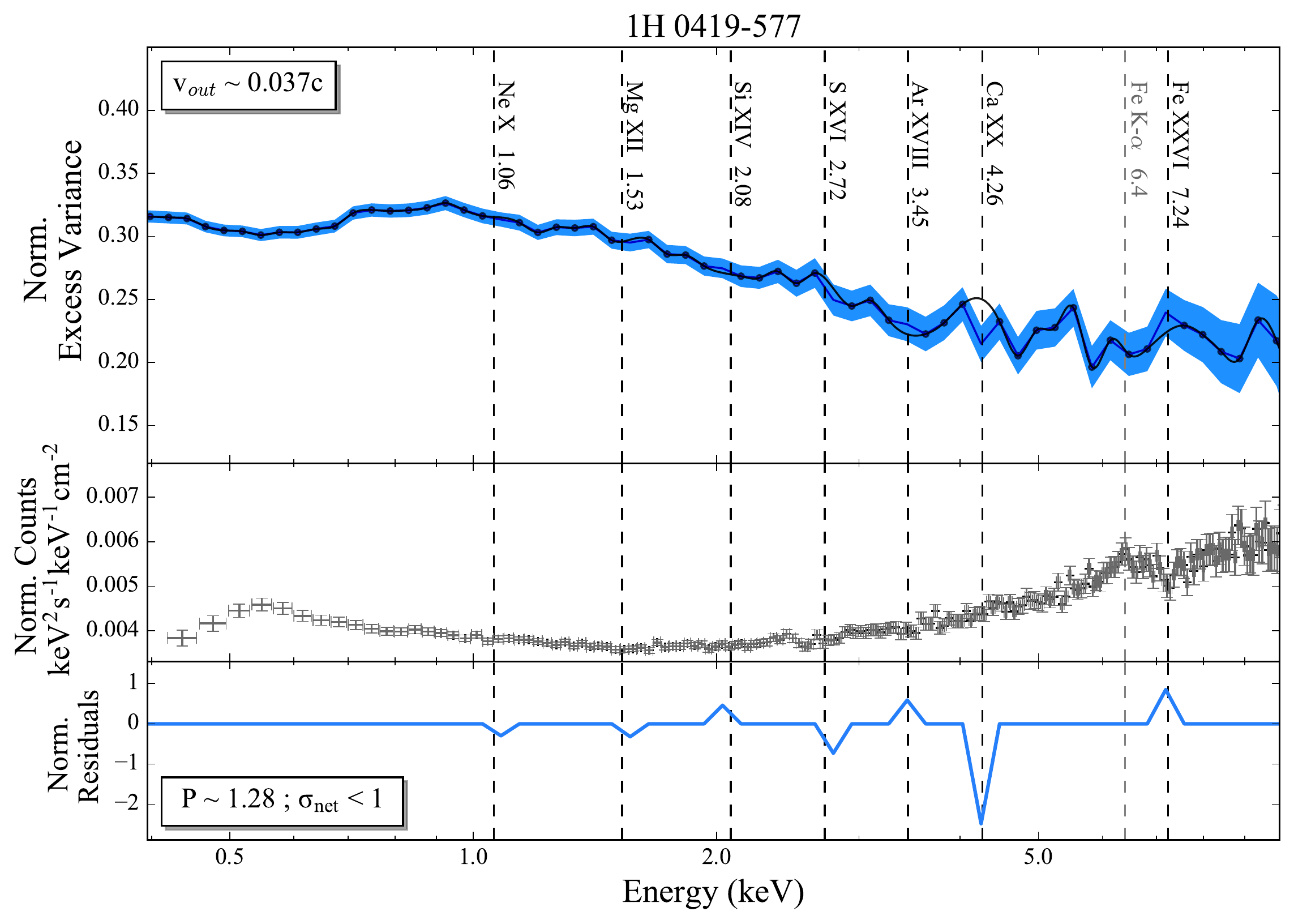}
    \includegraphics[width=85mm]{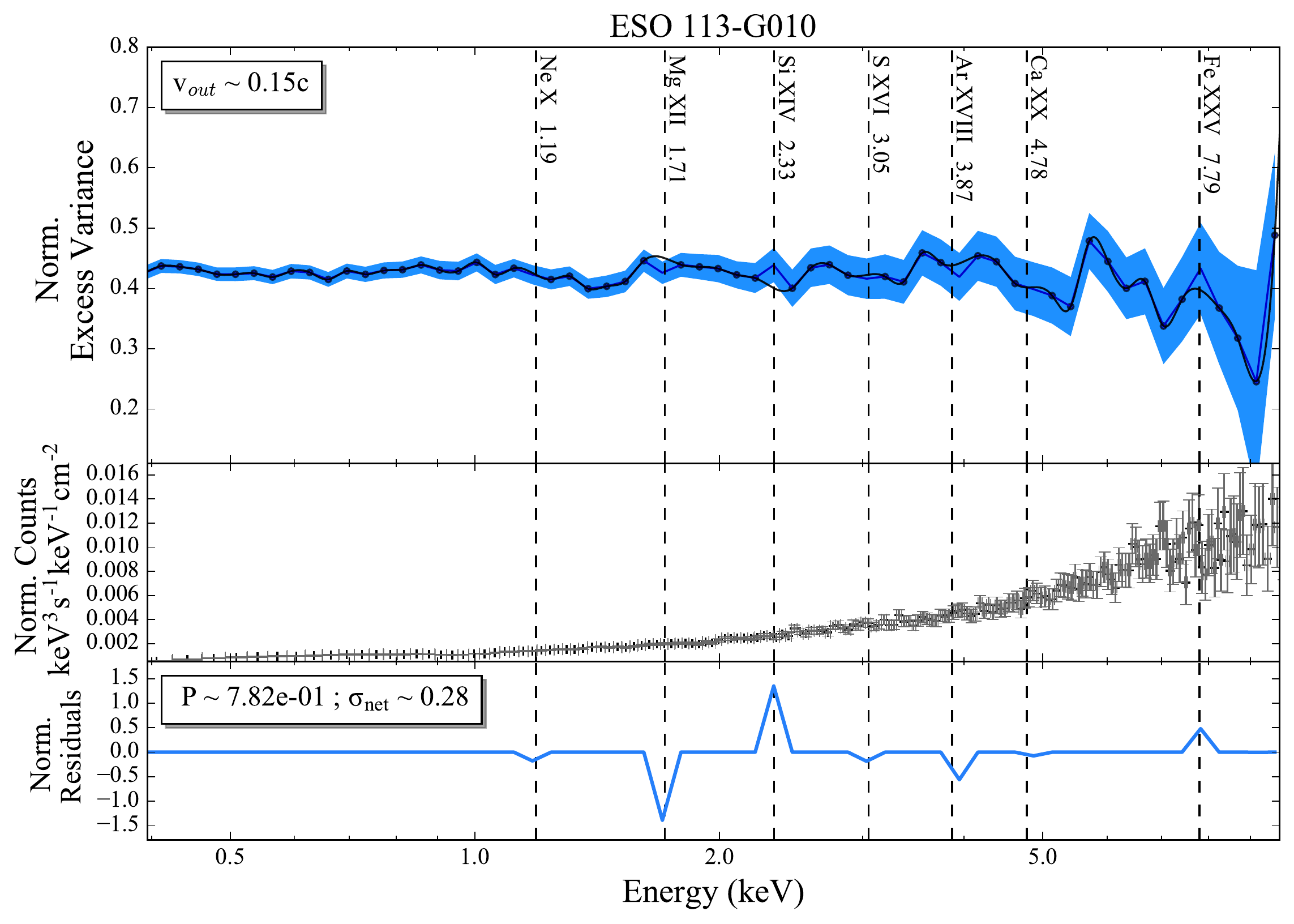}
    \includegraphics[width=85mm]{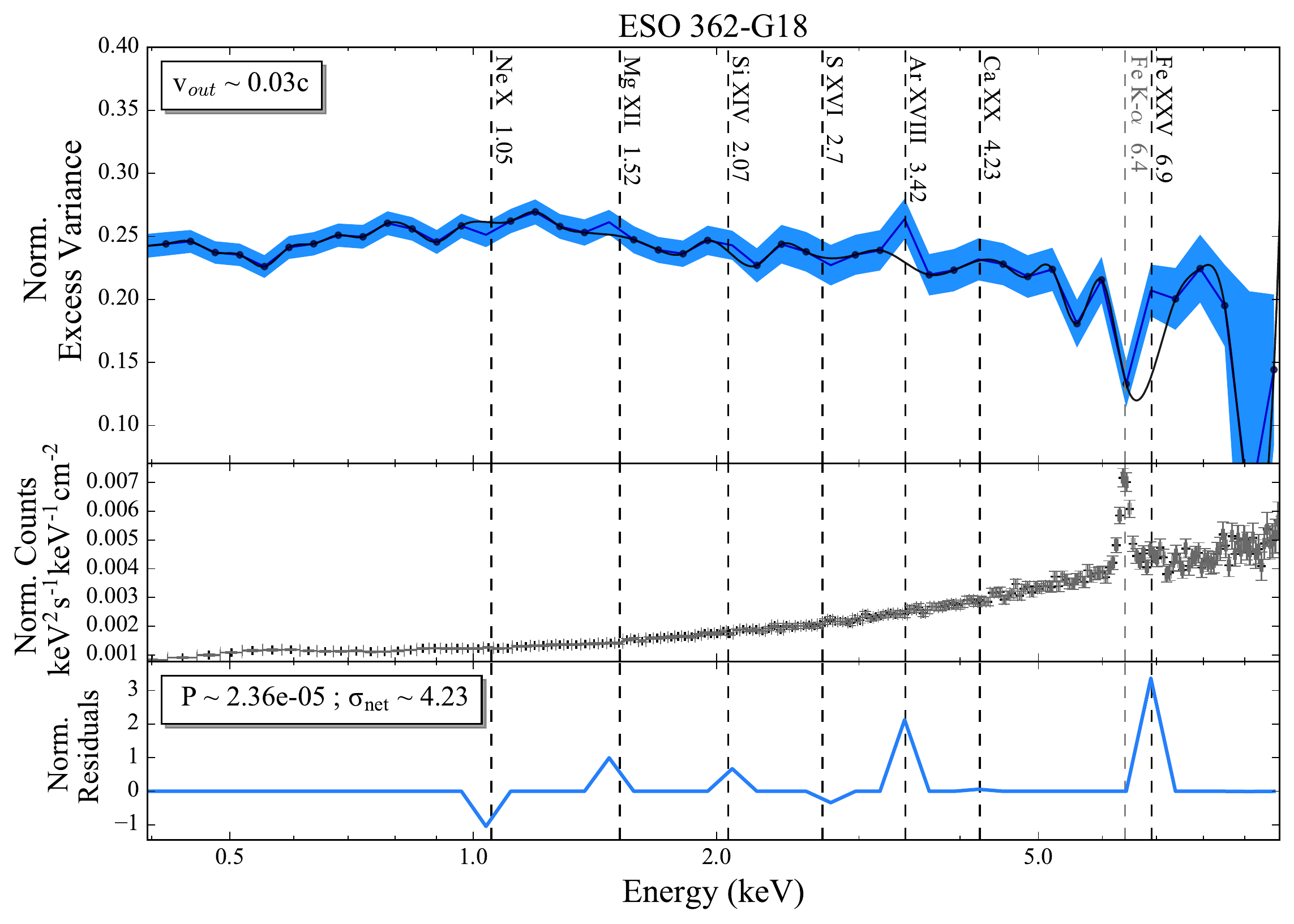}
    \includegraphics[width=85mm]{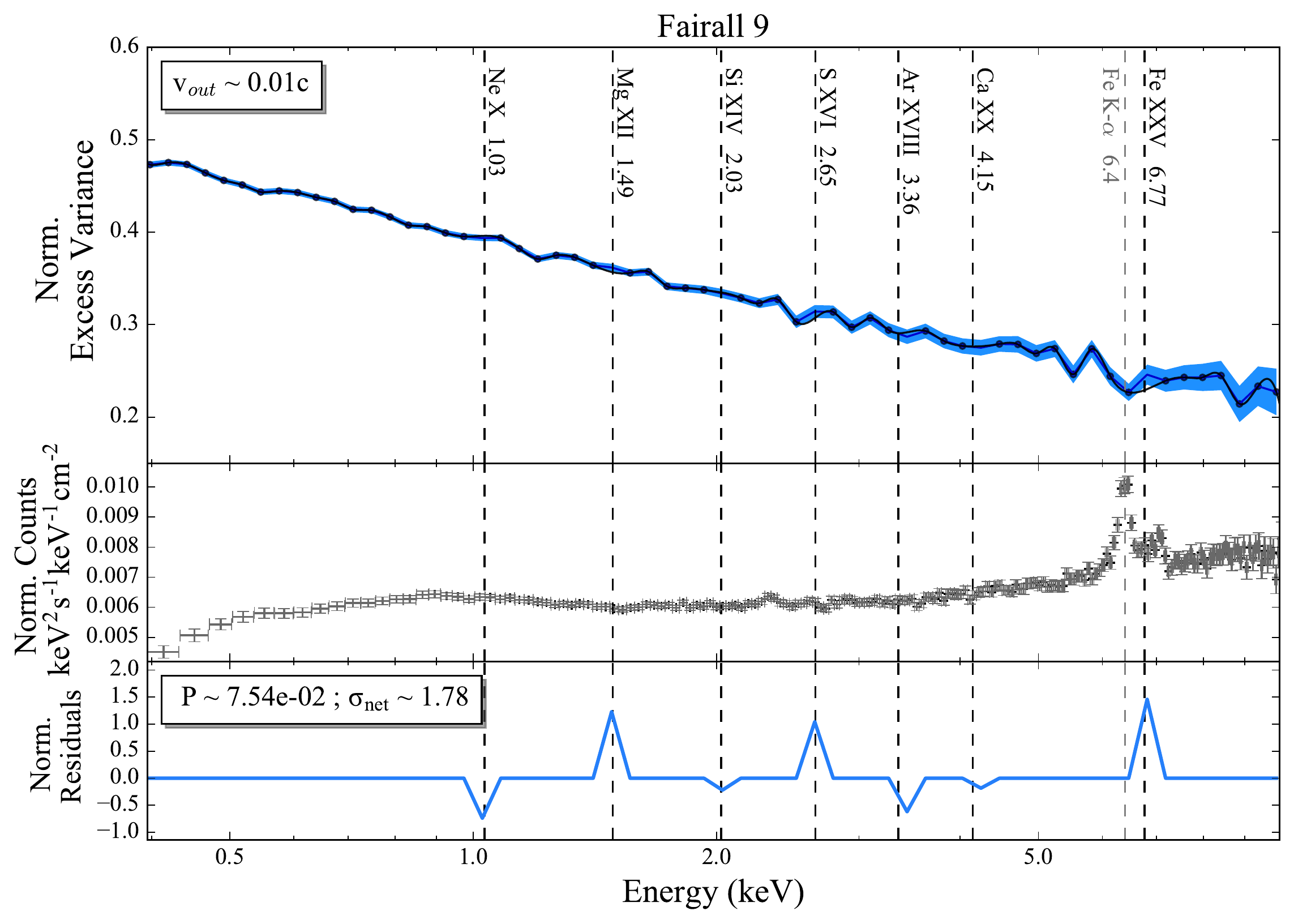}
    \caption{RE J1034+396 and SWIFT J2127.4+5654 are examples of UFOs with weak evidence for UFOs, whereas 1H 0419-577, ESO 113-G010, ESO362-G18 and Fairall 9 have no evidence.}
\end{figure*}

\begin{figure*}
    \centering
    \includegraphics[width=85mm]{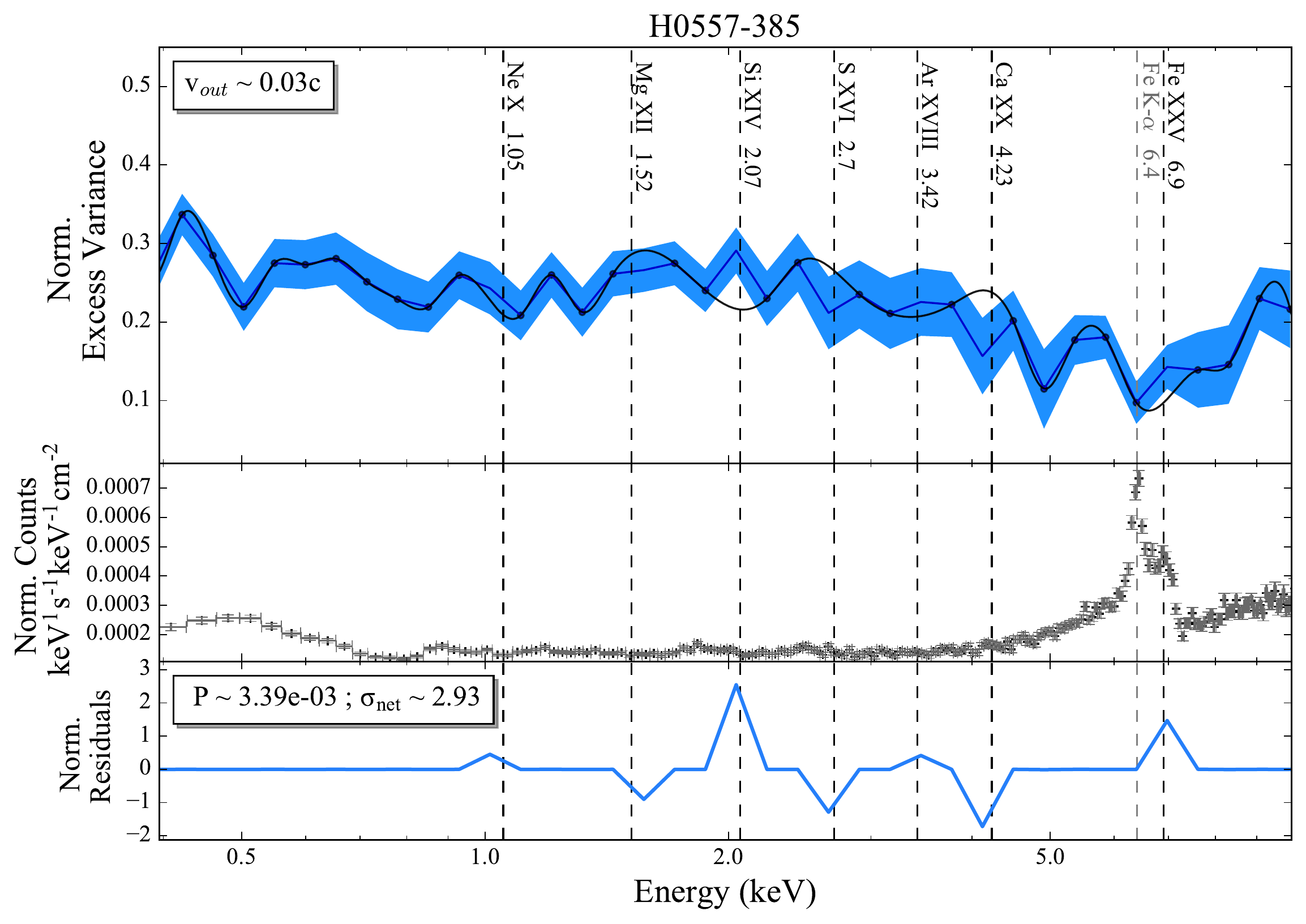}
    \includegraphics[width=85mm]{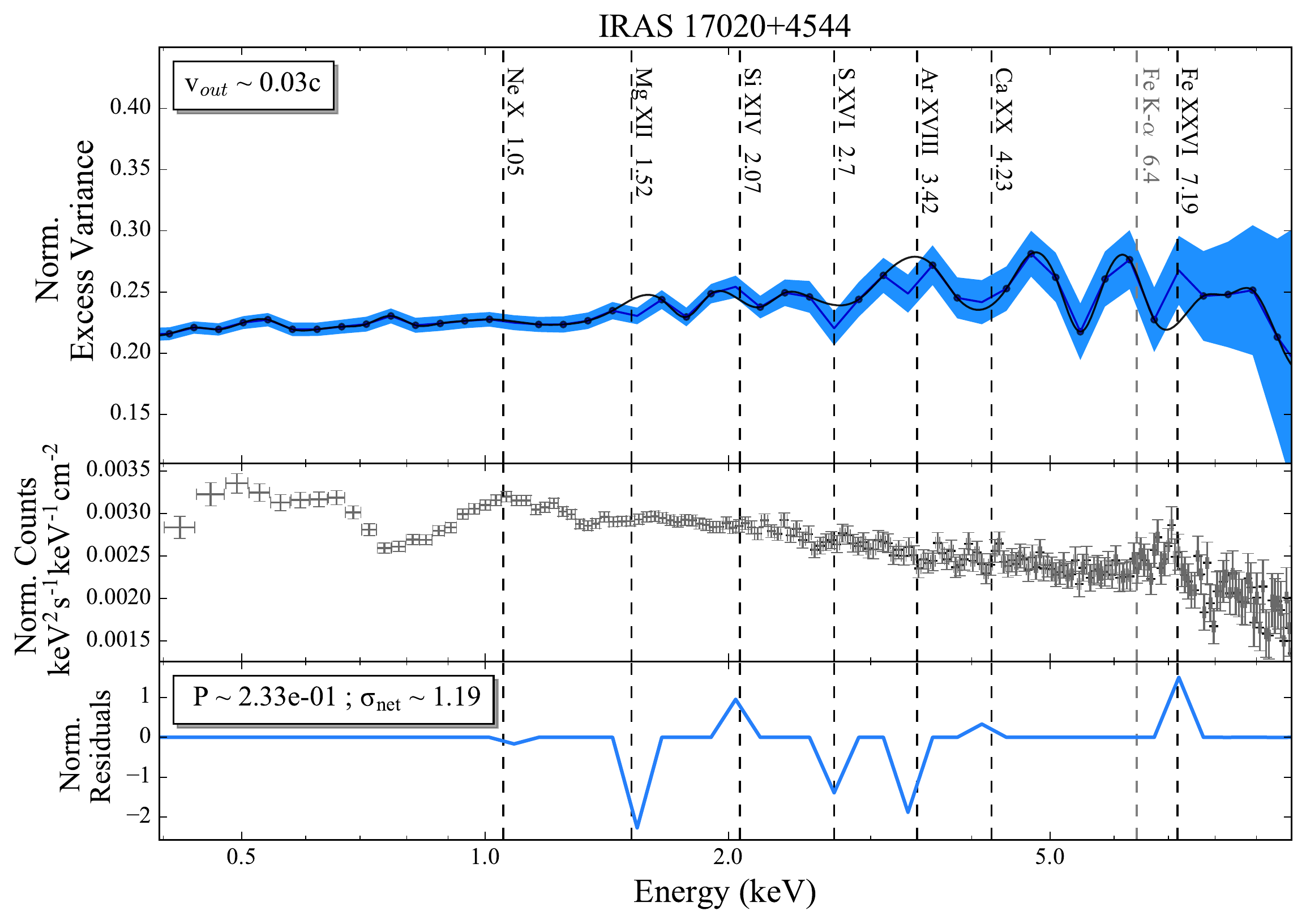}
    \includegraphics[width=85mm]{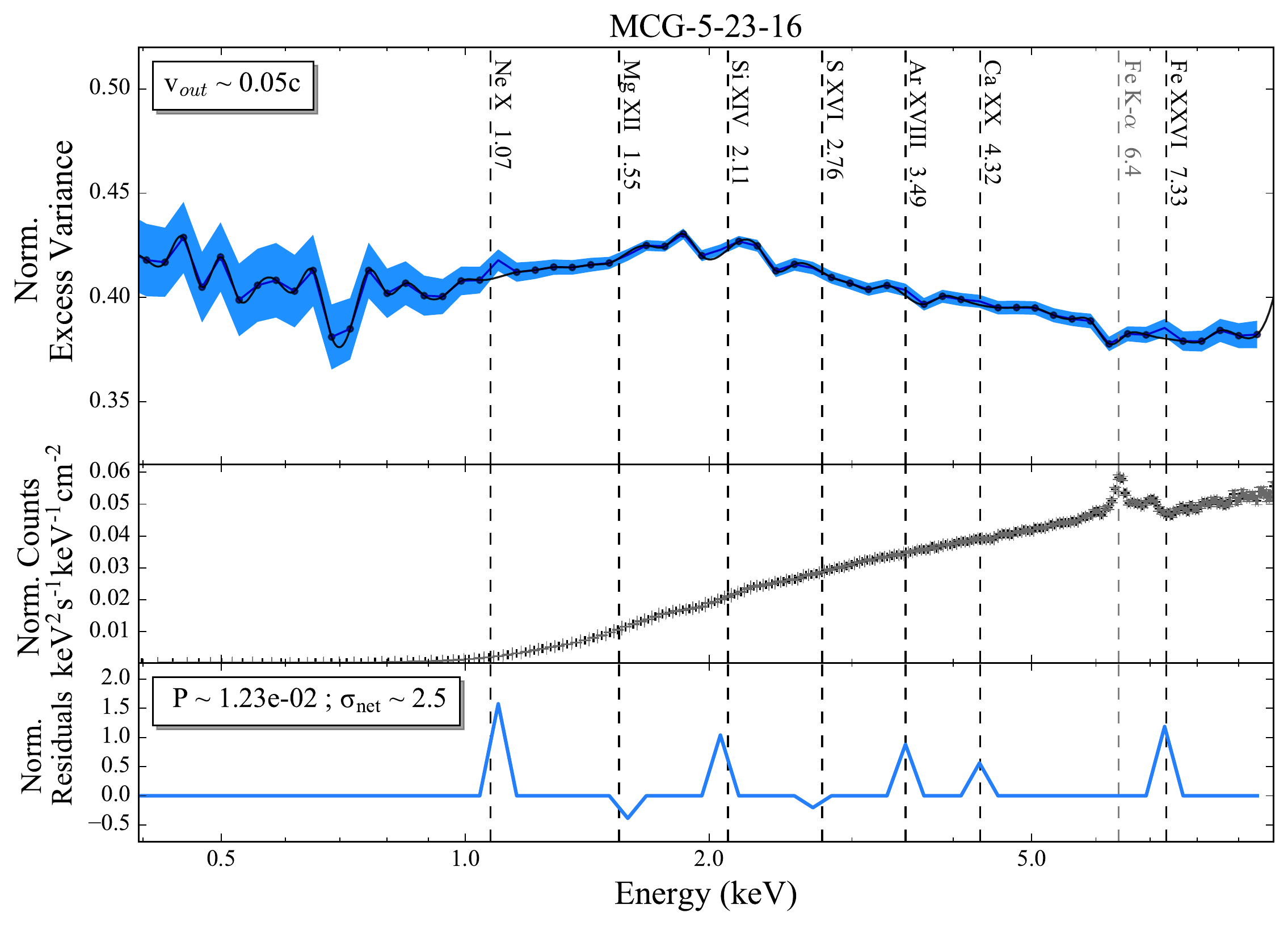}
    \includegraphics[width=85mm]{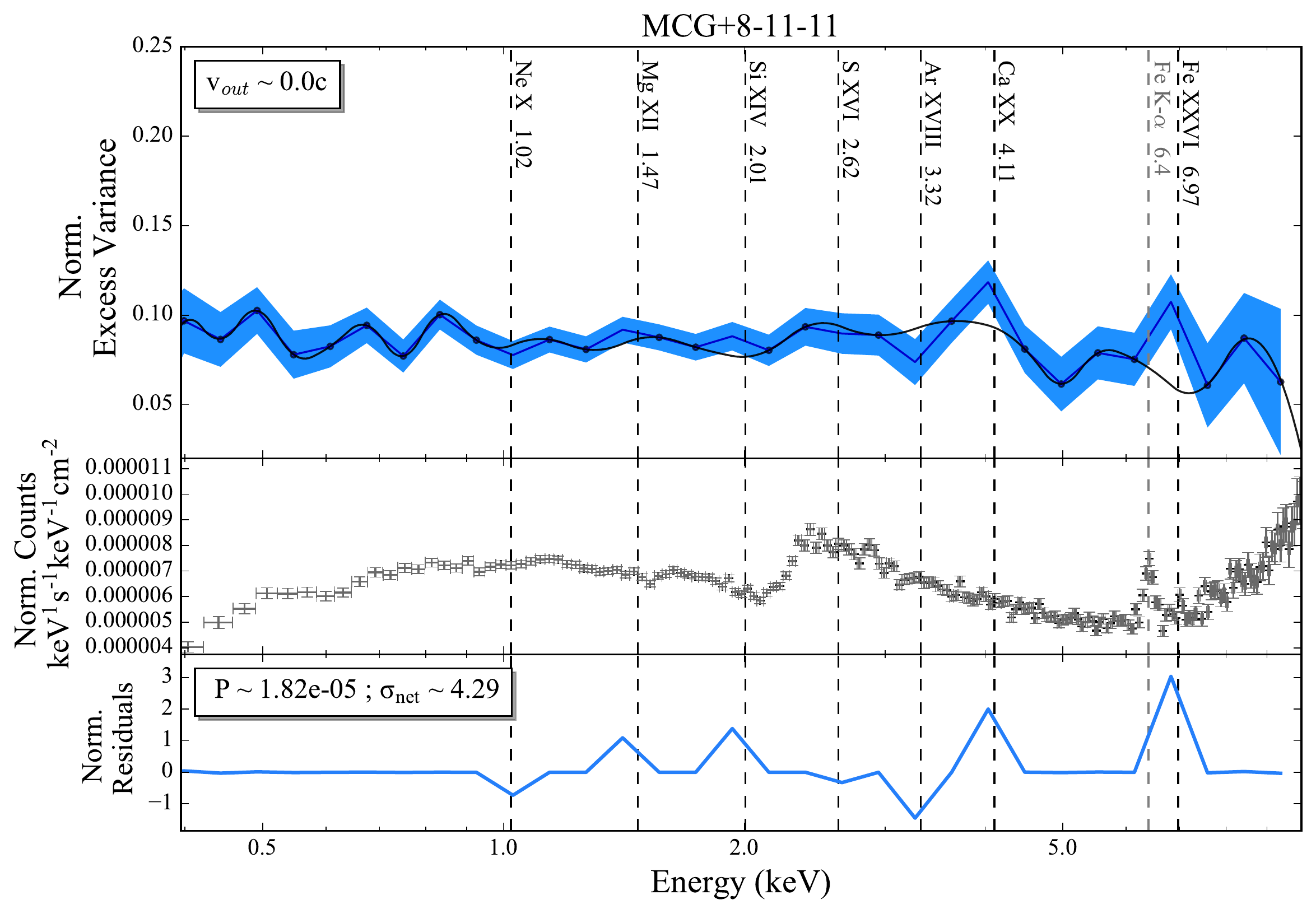}
    \includegraphics[width=85mm]{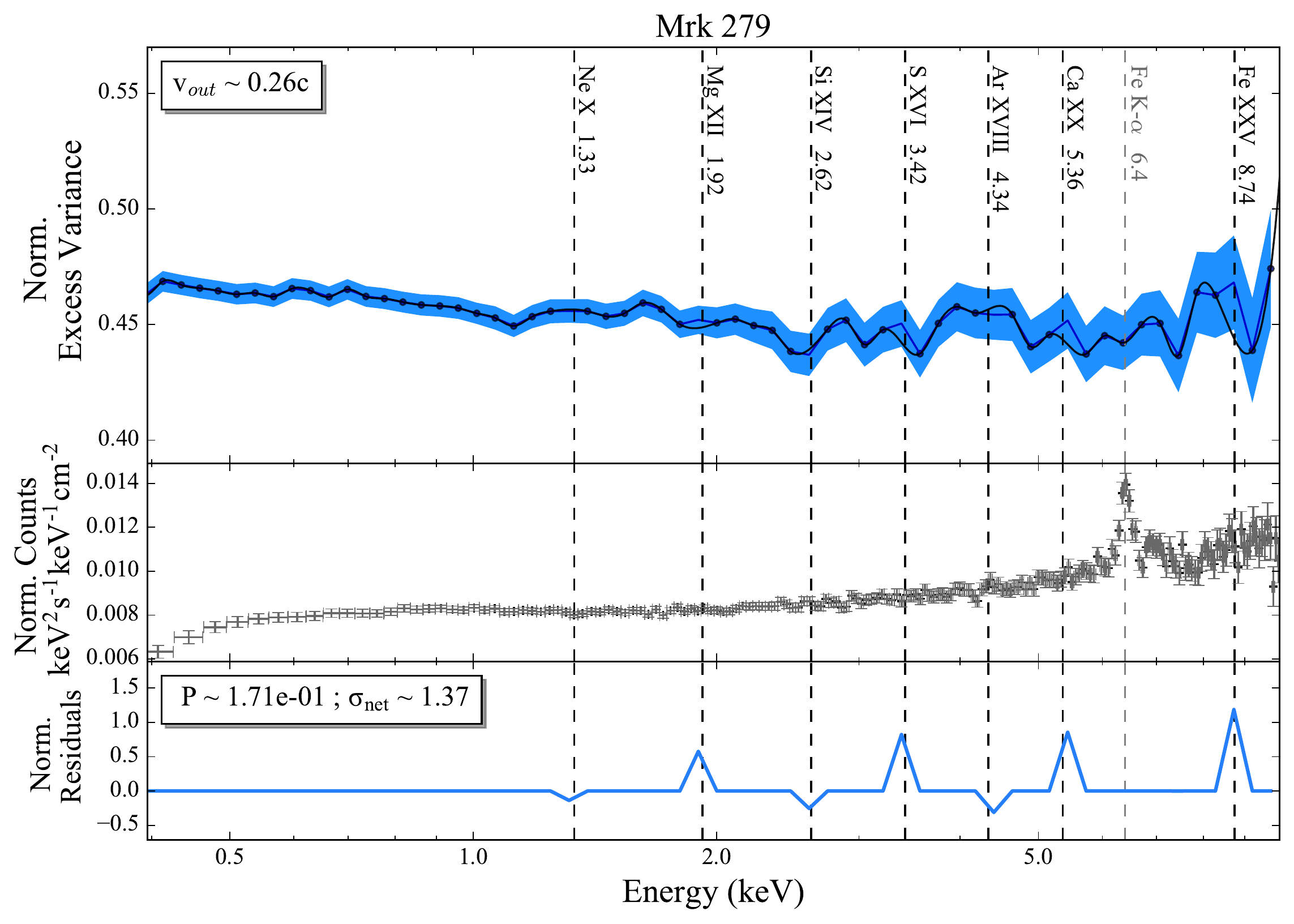}
    \includegraphics[width=85mm]{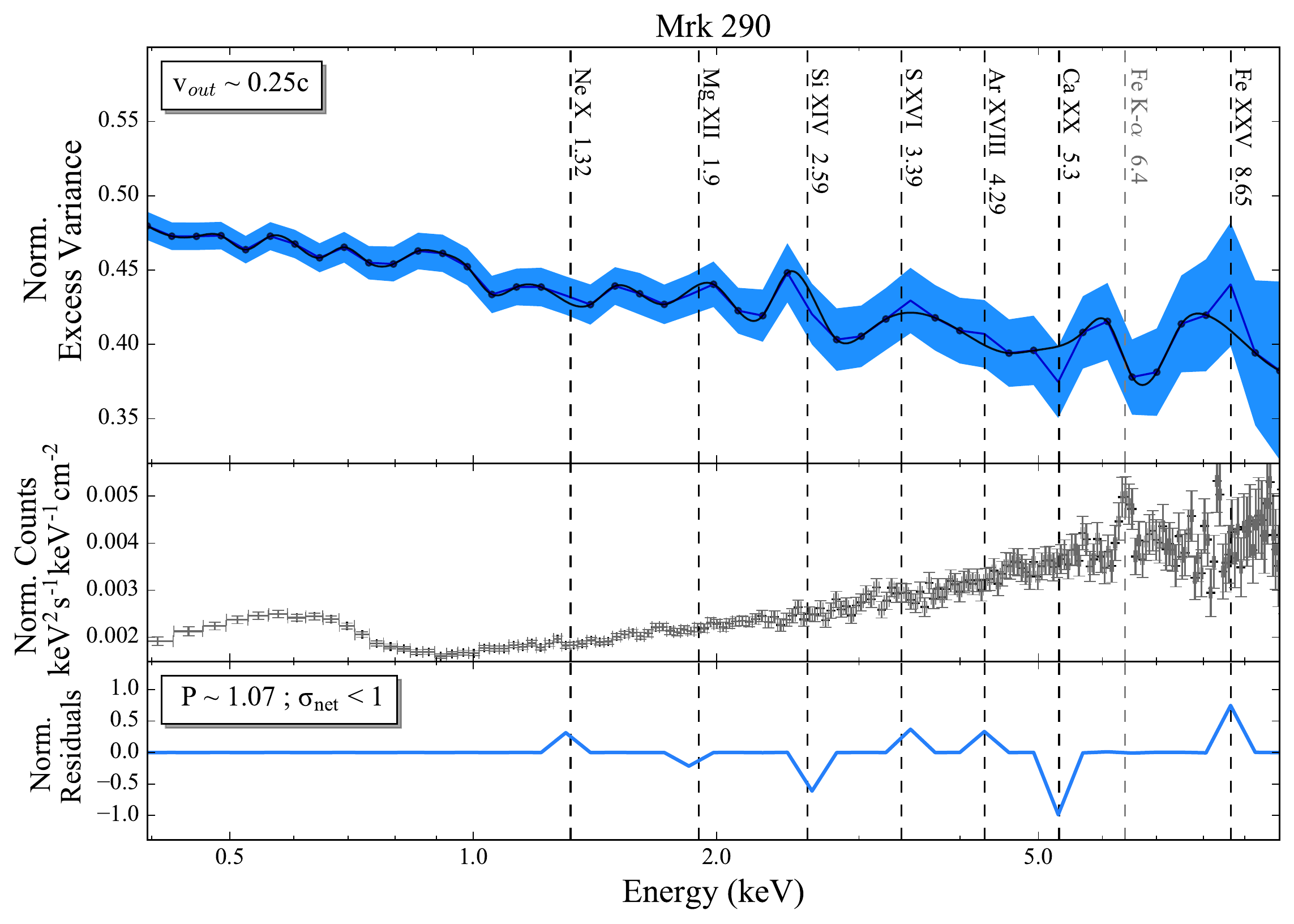}
    \caption{Examples of sources with no evidence for a UFO.}
\end{figure*}

\begin{figure*}
    \centering
    \includegraphics[width=85mm]{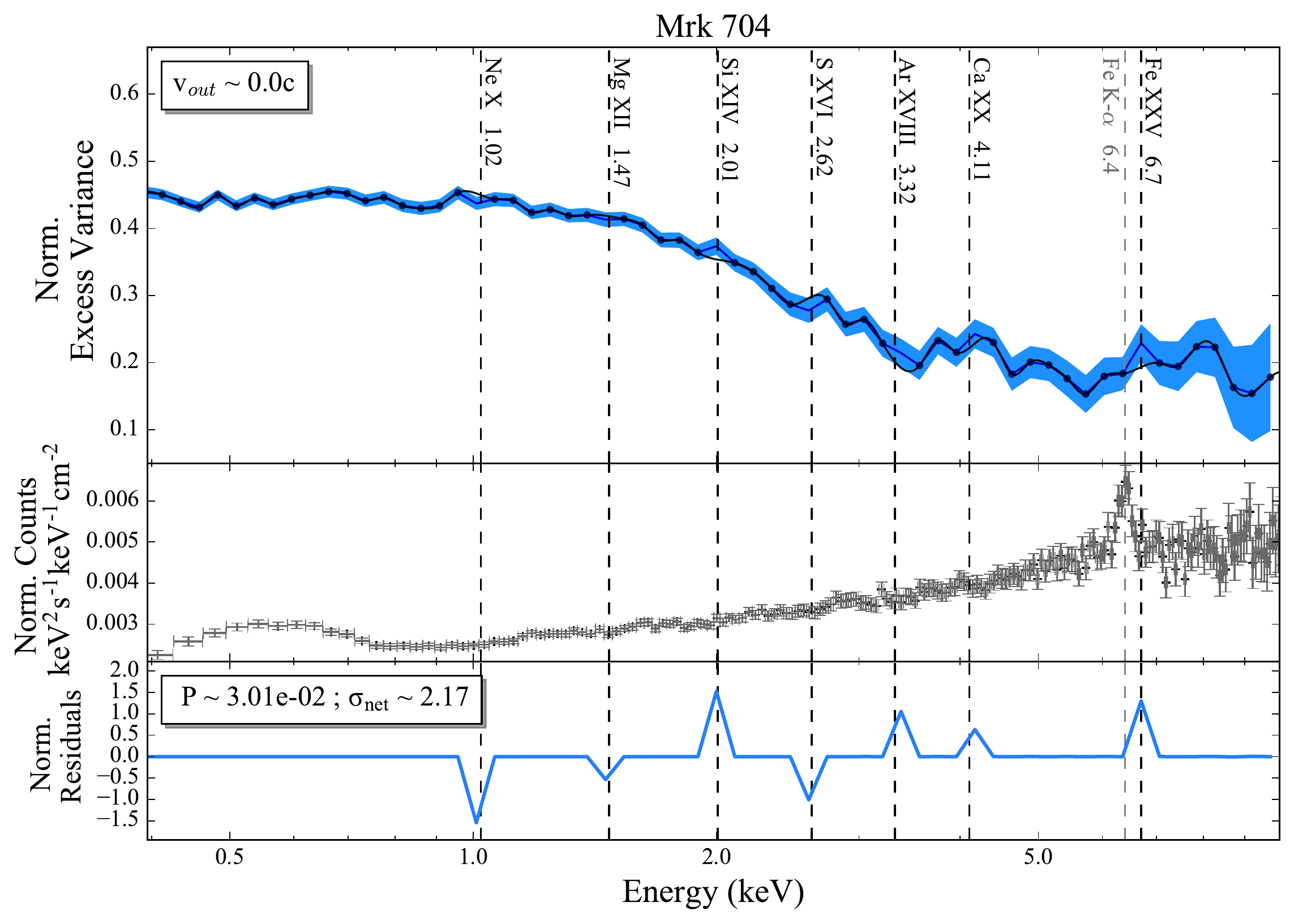}
    \includegraphics[width=85mm]{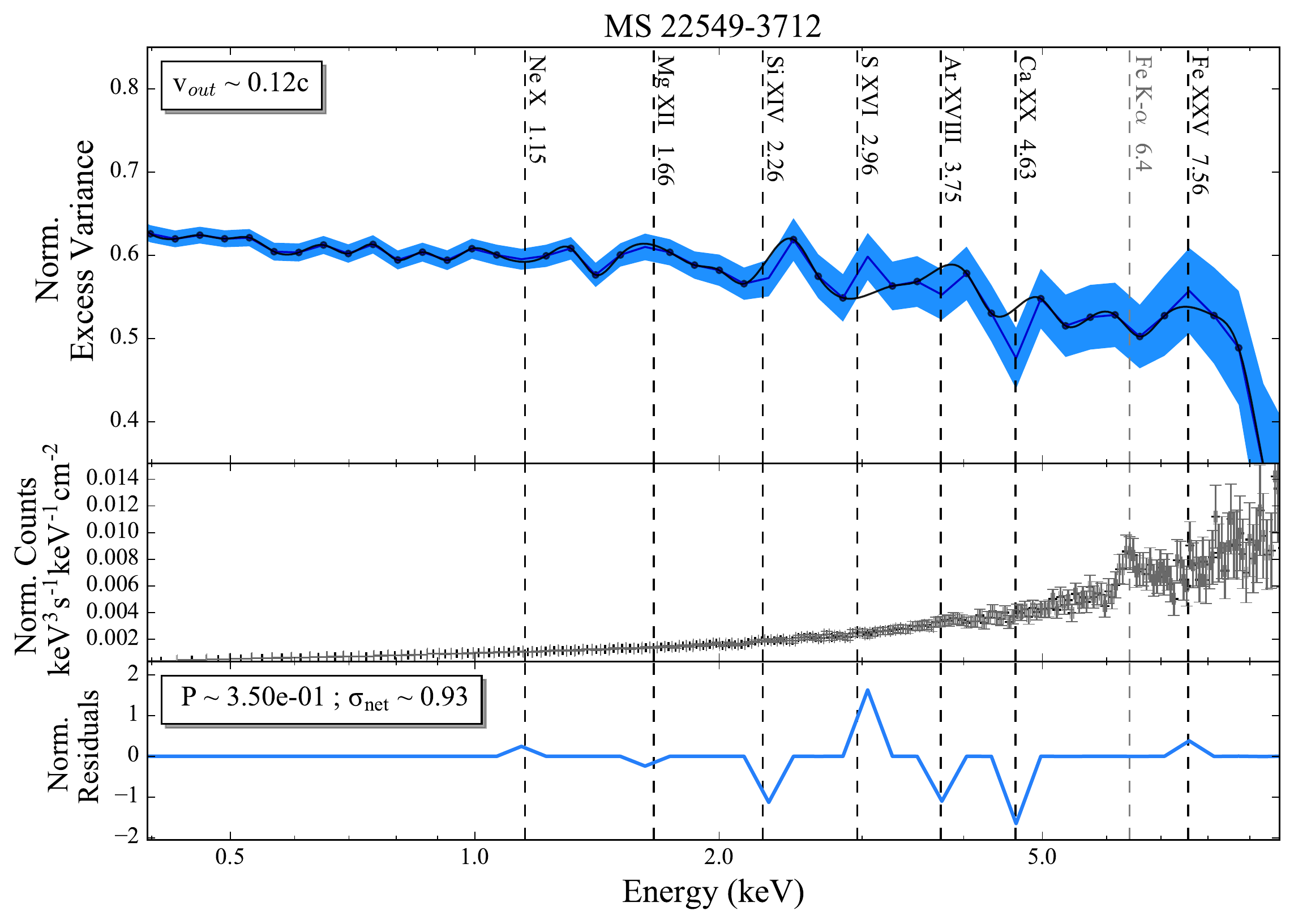}
    \includegraphics[width=85mm]{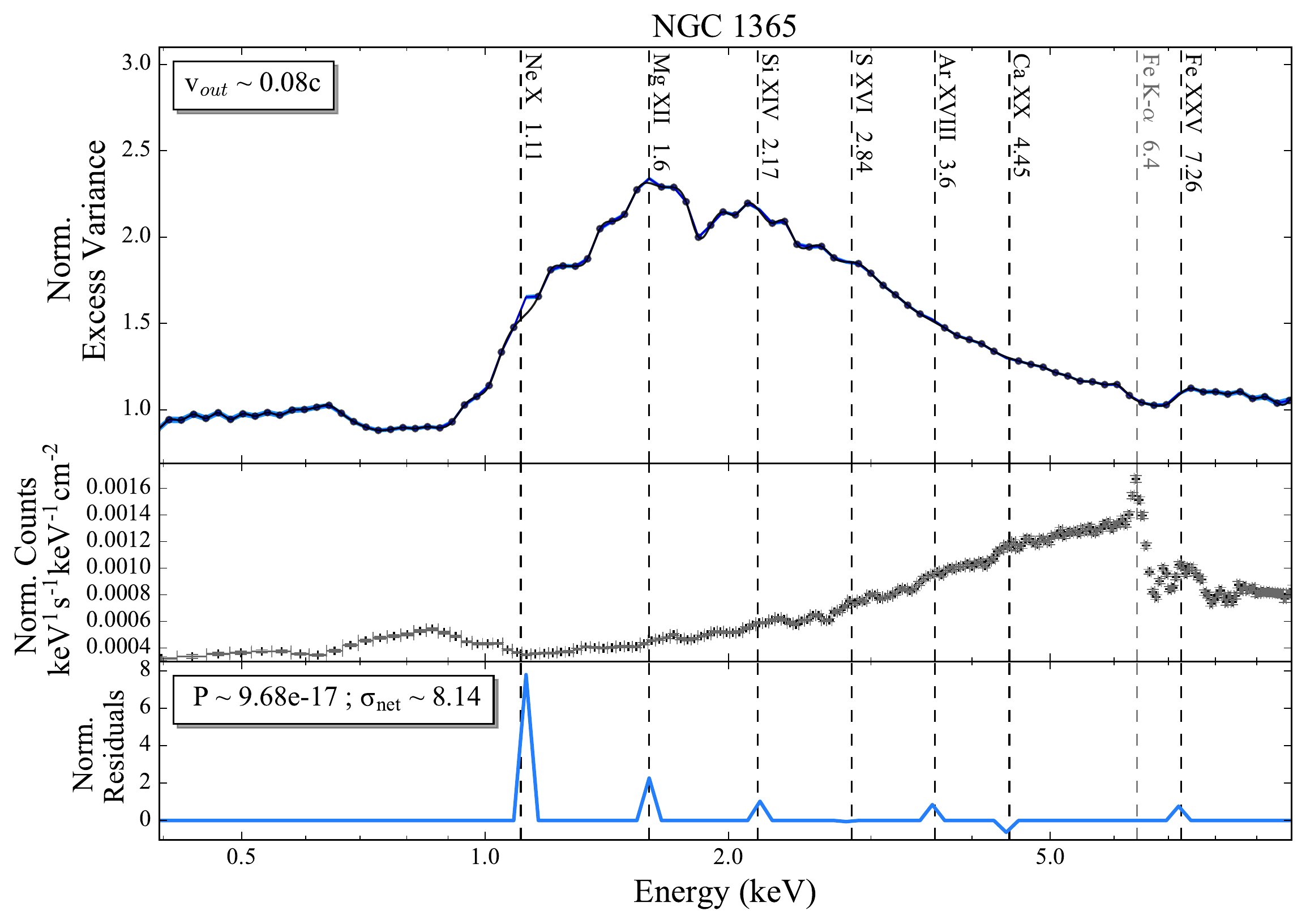}
    \includegraphics[width=85mm]{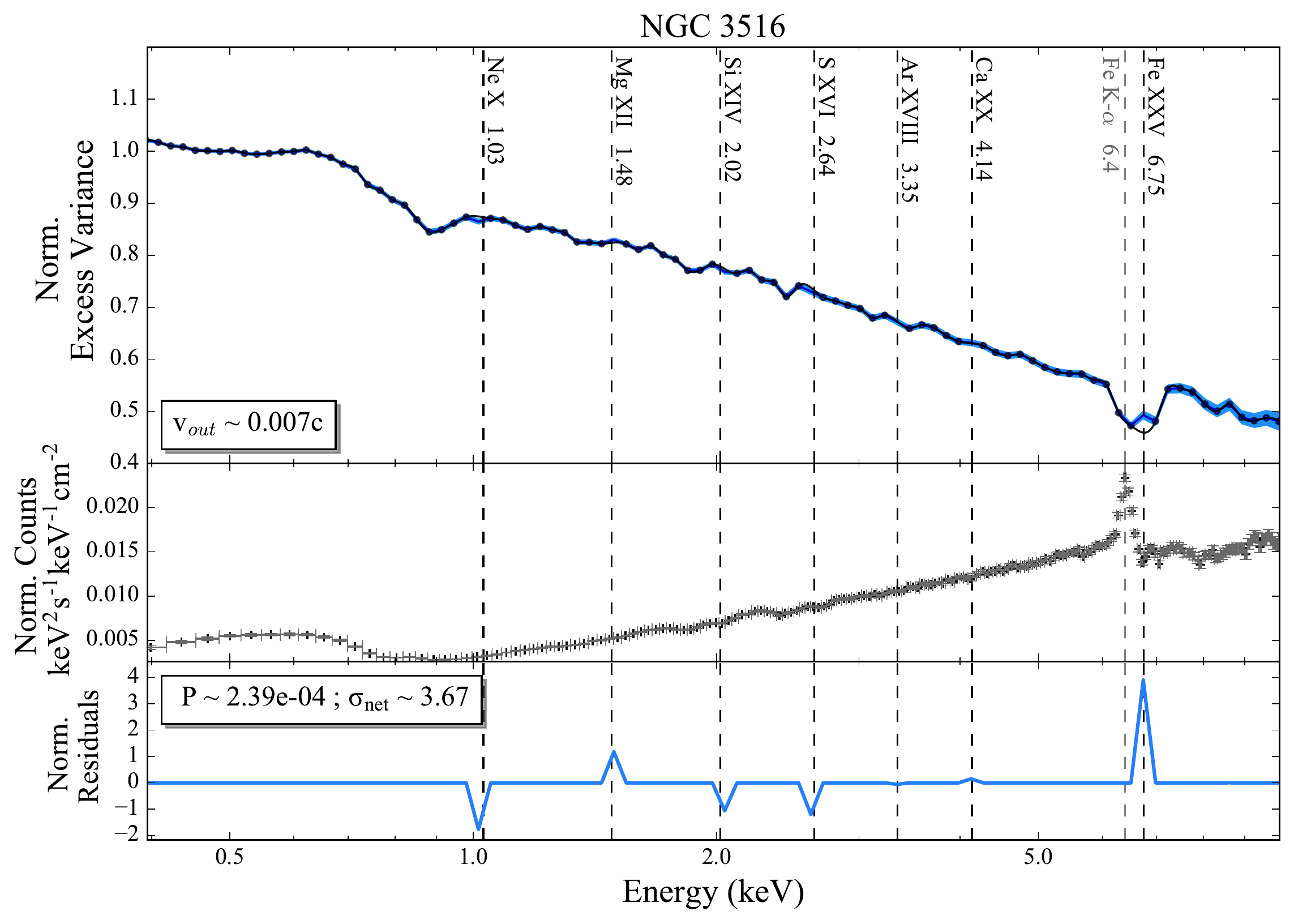}
    \includegraphics[width=85mm]{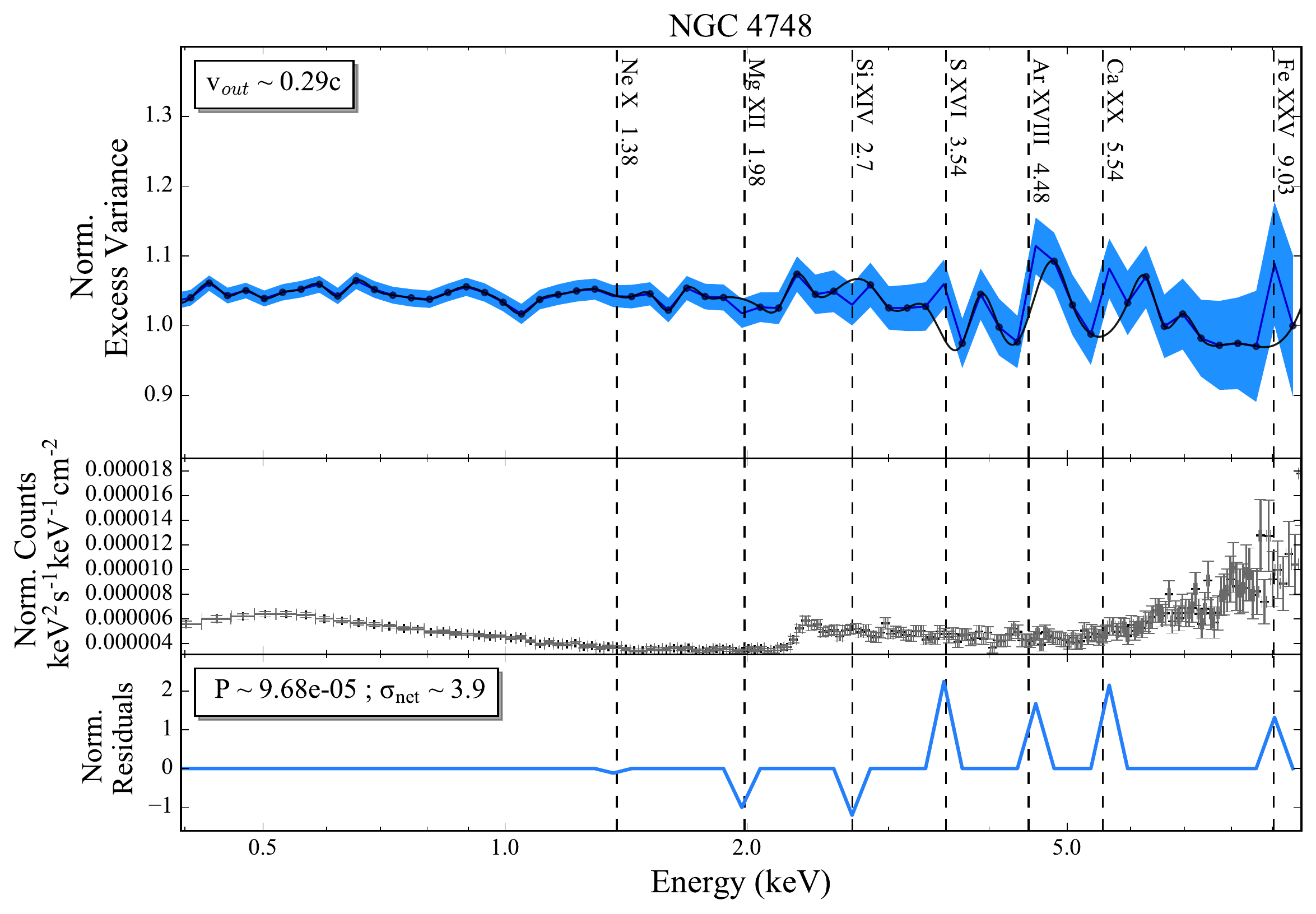}
    \includegraphics[width=85mm]{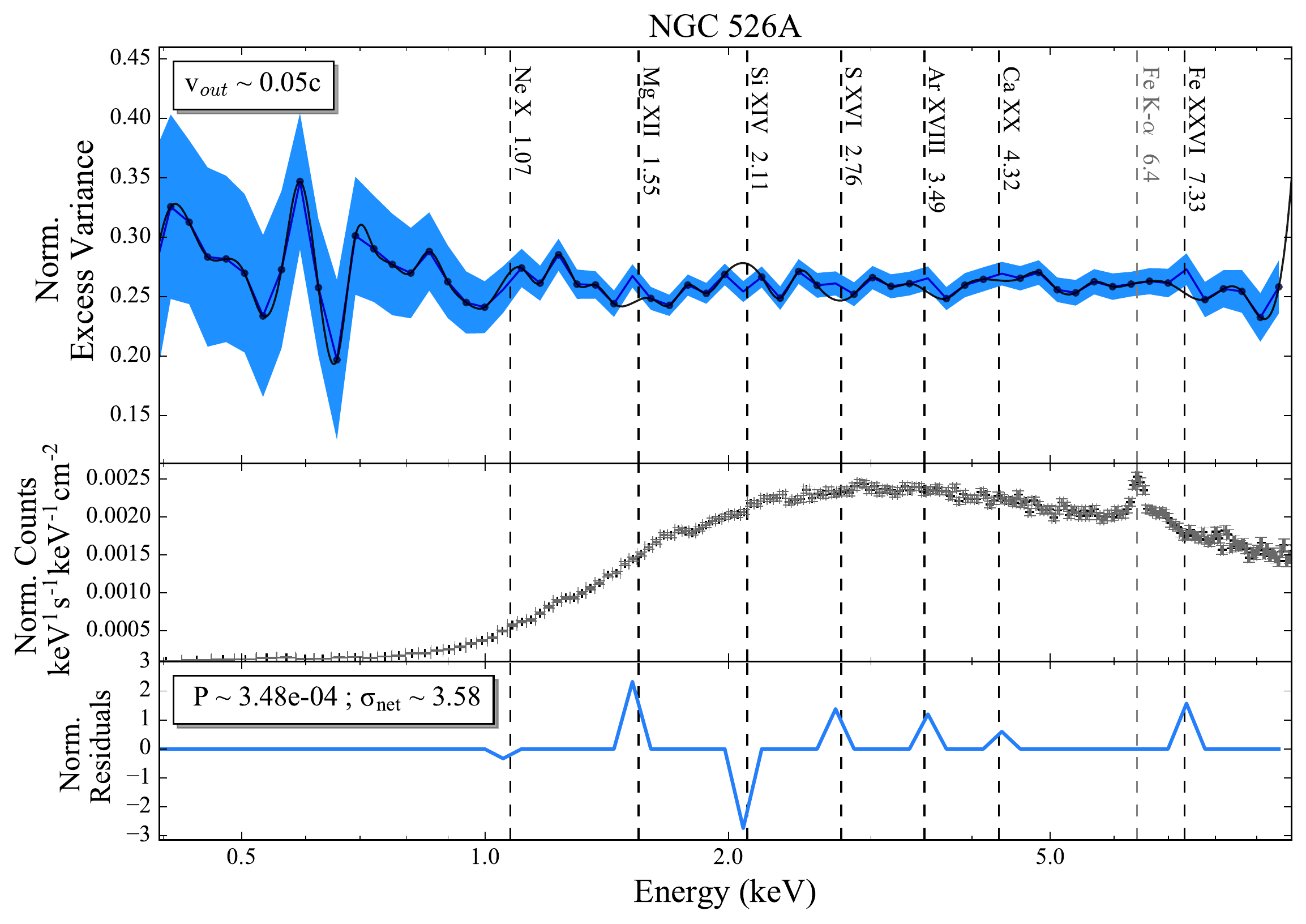}
    \caption{Examples of sources with no evidence for a UFO.}
\end{figure*}

\begin{figure*}
    \centering
    \includegraphics[width=85mm]{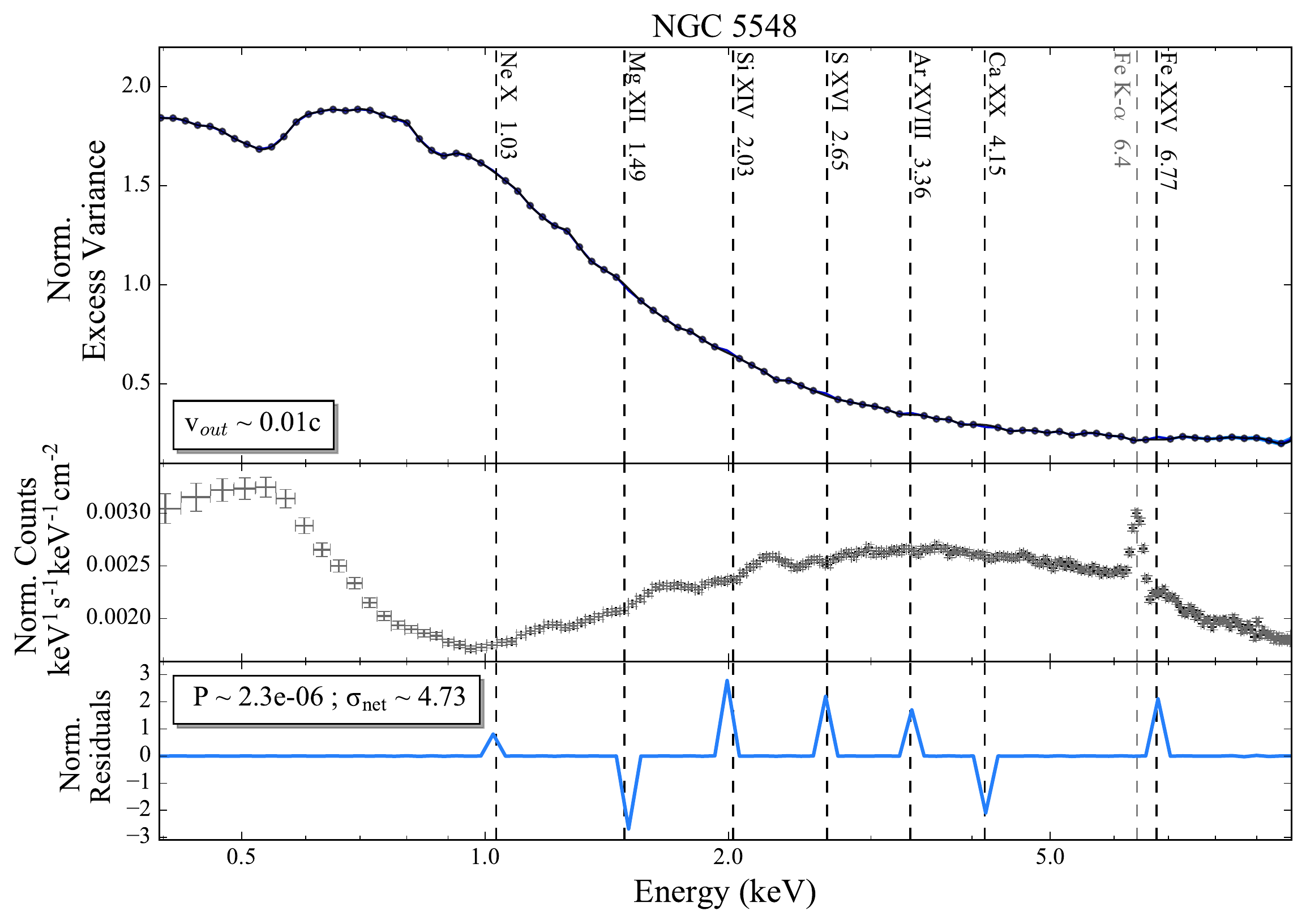}
    \includegraphics[width=85mm]{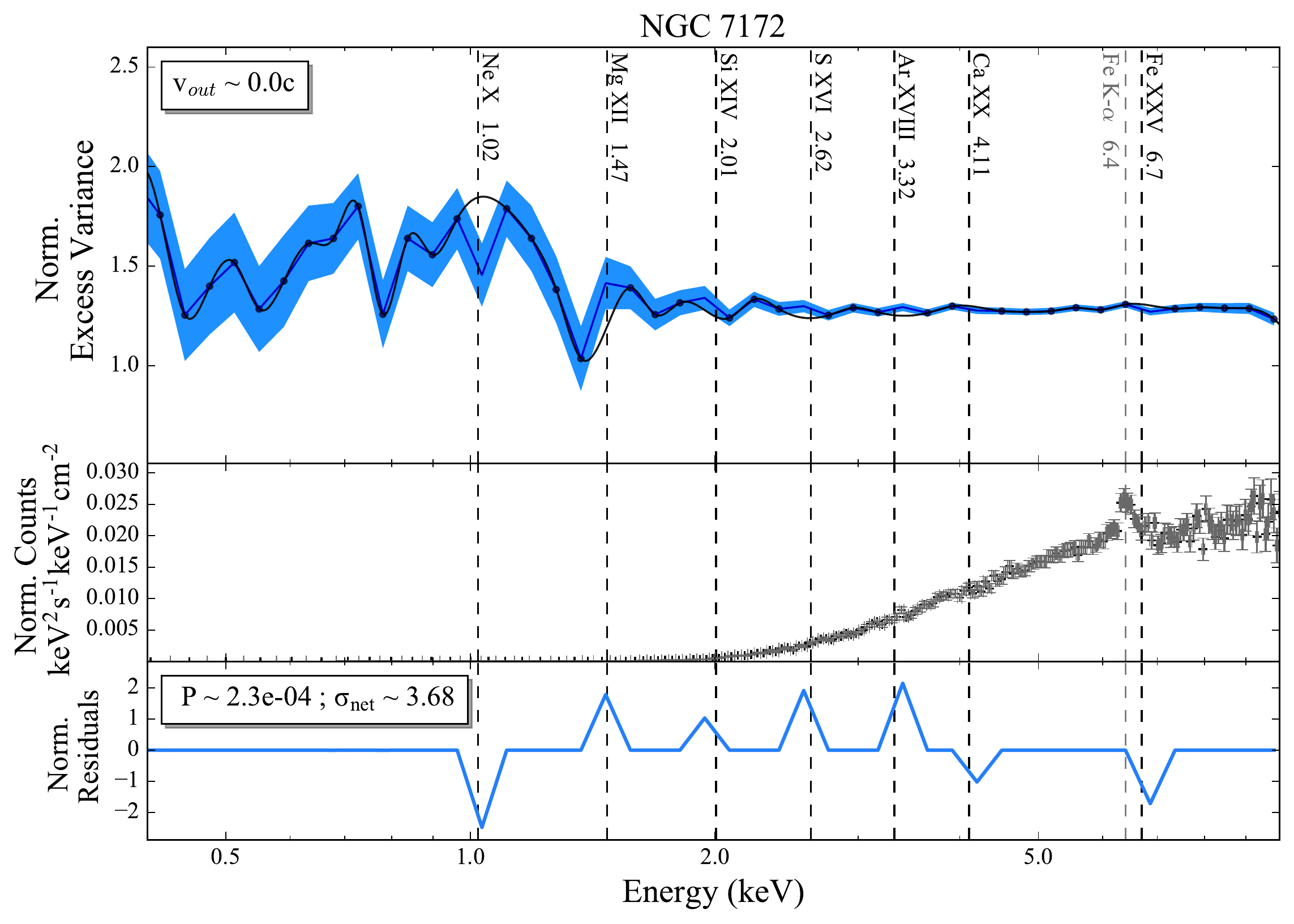}
    \includegraphics[width=85mm]{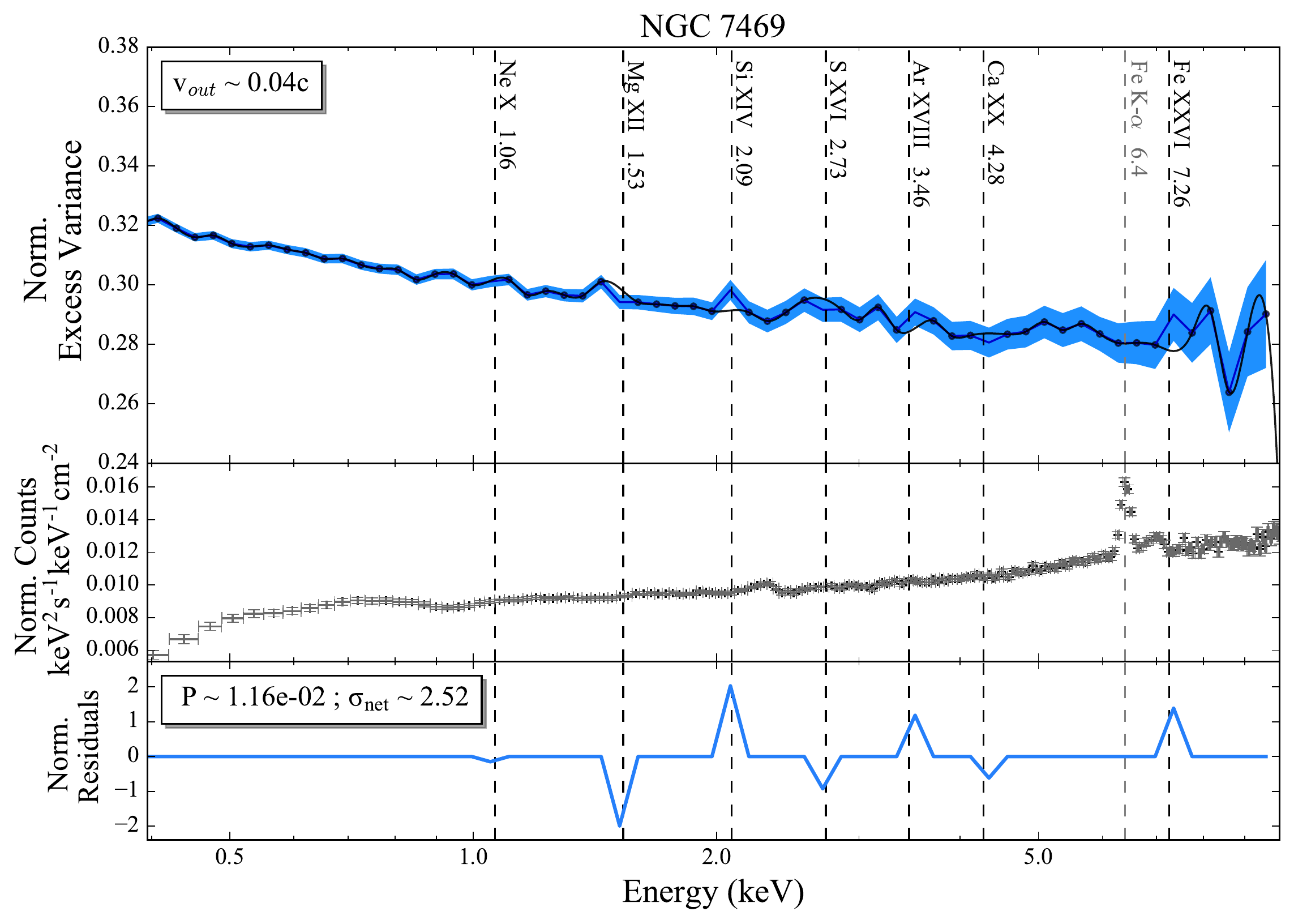}
    \includegraphics[width=85mm]{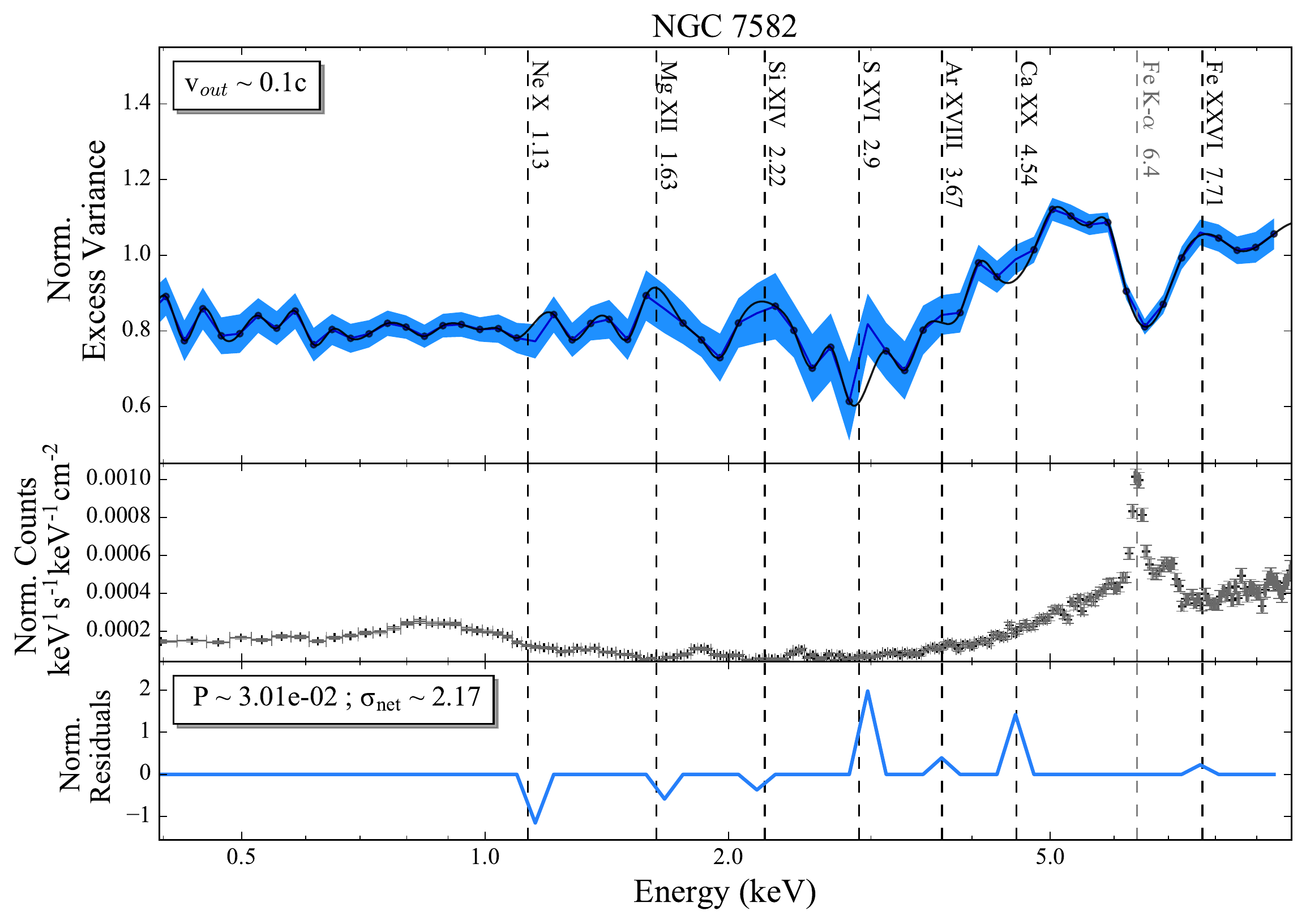}
    \caption{Examples of sources with no evidence for a UFO.}
\end{figure*}


\bsp	
\label{lastpage}
\end{document}